# Efficient, High Accuracy ADER-WENO Schemes for Hydrodynamics and Divergence-Free Magnetohydrodynamics

By

Dinshaw S. Balsara[1], Tobias Rumpf[1,3], Michael Dumbser[2,3] & Claus-Dieter Munz[3]

(dbalsara@nd.edu, iagrumpf@iag.uni-stuttgart.de, michael.dumbser@iag.uni-stuttgart.de, munz@iag.uni-stuttgart.de )

[1] Physics Department, University of Notre Dame, 225 Nieuwland Science Hall, Notre Dame, IN, 46556, USA

[2] Laboratory of Applied Mathematics, University of Trento, Via Mesiano 77, I-38100 Trento, Italy

[3] Institut für Aerodynamik und Gasdynamik, University of Stuttgart, Pfaffenwaldring 21, D-70550, Stuttgart, Germany

**Abstract**

The present paper introduces a class of finite volume schemes of increasing order of accuracy in space and time for hyperbolic systems that are in conservation form. The methods are specially suited for efficient implementation on structured meshes. The hyperbolic system is required to be non-stiff. This paper specifically focuses on Euler system that is used for modeling the flow of neutral fluids and the divergence-free, ideal magnetohydrodynamics (MHD) system that is used for large scale modeling of ionized plasmas.

Efficient techniques for weighted essentially non-oscillatory (WENO) interpolation have been developed for finite volume reconstruction on structured meshes. We have shown that the most elegant and compact formulation of WENO reconstruction obtains when the interpolating functions are expressed in modal space. Explicit formulae have been provided for schemes having up to fourth order of spatial accuracy. Divergence-free evolution of magnetic fields requires the magnetic field components and their moments to be defined in the zone faces. We draw on a reconstruction strategy developed recently by the first author to show that a high order specification of the magnetic field components in zone-faces naturally furnishes an appropriately high order representation of the magnetic field within the zone.

We also present a new formulation of the ADER (for Arbitrary Derivative Riemann Problem) schemes that relies on a local continuous space-time Galerkin formulation instead of the usual Cauchy-Kovalewski procedure. We call such schemes ADER-CG and show that a very elegant and compact formulation results when the scheme is formulated in modal space. Explicit formulae have been provided on structured meshes for ADER-CG schemes in three dimensions for all orders of accuracy that extend up to fourth order. Such ADER schemes have been used to temporally evolve the WENO-based spatial reconstruction. The resulting ADER-WENO schemes provide temporal accuracy that matches the spatial accuracy of the underlying WENO reconstruction.

In this paper we have also provided a point-wise description of ADER-WENO schemes for divergence-free MHD in a fashion that facilitates computer implementation. The schemes reported here have all been implemented in the RIEMANN framework for computational astrophysics. All the methods presented have a one-step update, making them low-storage alternatives to the usual Runge-Kutta time-discretization. Their one-step update also makes them suitable building blocks for adaptive mesh refinement (AMR) calculations.

We demonstrate that the ADER-WENO meet their design accuracies. Several stringent test problems of Euler flows and MHD flows are presented in one, two and three dimensions. Many of our test problems involve near infinite shocks in multiple dimensions and the higher order schemes are shown to perform very robustly and accurately under all conditions. It is shown that the increasing computational complexity with increasing order is handily offset by the increased accuracy of the scheme. The resulting ADER-WENO schemes are, therefore, very worthy alternatives to the standard second order schemes for compressible Euler and MHD flow.

# 1) Introduction

The accurate simulation of hydrodynamical and magnetohydrodynamical (MHD) flows is an important topic in several areas of science and engineering. Much progress has been made towards that goal. While second order accurate simulations have been carried out for a while, recent advances have made it possible to go beyond second order accuracy. Early attempts to go beyond second order have been catalogued in Harten et al [36], Shu & Osher [50] and [51], Barth & Frederickson [12] and Suresh & Huynh [53]. Liu, Osher & Chan [42], Jiang & Shu [40] and Balsara & Shu [9] presented finite difference weighted essentially non-oscillatory (WENO) schemes for hydrodynamics. The WENO interpolation used in such schemes is usually coupled with a Runge-Kutta (RK) time update strategy from Shu & Osher [50] to yield schemes that have spatial and temporal accuracies that are well-matched. While the finite difference WENO schemes handily meet their design accuracies, they do not take well to non-uniform or hierarchical meshes. For that reason it is advantageous to have finite-volume WENO schemes which can be seamlessly used as building blocks for adaptive mesh refinement (AMR) calculations, see Berger & Colella [13] and Balsara [3]. Higher order accurate schemes that have a finite volume-like structure have been designed for structured meshes, see Balsara et al [7] and Balsara [6] and also for unstructured meshes, see Hu & Shu [37], Zhang & Shu [60], Dumbser & Käser [28] and Dumbser et al [29]. The purpose of this paper is to catalogue finite-volume WENO schemes that go beyond second order accuracy on structured meshes.

As shown by Colella [17], it is also very advantageous to use schemes that have a one-step temporal update as building blocks for AMR calculations. All of the RK schemes from Shu & Osher [50] lack such a one-step structure. While dense output RK schemes can be devised, it is still very desirable to have schemes that retain a simple one-step time update. ADER (for Arbitrary Derivative Riemann Problem) schemes have seen a fair bit of recent evolution, see Titarev & Toro [55] and [56], Toro & Titarev [57], Dumbser, Enaux & Toro [27] and Dumbser et al [26]. Recent versions of ADER schemes, see Dumbser et al [26], have the right kind of one-step temporal update that makes them very convenient for higher order AMR work. Thus the further goal of this paper is to present finite-volume ADER-WENO schemes that have a one-step temporal update. In order to achieve balanced performance, all of the schemes presented here have increasing spatial accuracy that is matched by a corresponding increase in temporal accuracy. The resulting schemes are eminently well-suited for high accuracy hydrodynamical calculations and can serve as good building blocks for block structured AMR calculations.

Numerical magnetohydrodynamics (MHD) plays an important role in astrophysics, aerospace, space physics and plasma physics applications. It is therefore very interesting to develop highly accurate methods for simulating MHD phenomena. The structure of the compressible MHD eigensystem is well-understood, Jeffrey & Taniuti [39], Roe and Balsara [47], making it possible to develop high resolution shock-capturing methods for this system. Most of the early work was focused on developing higher order Godunov schemes with second order of accuracy, Dai & Woodward [20], Ryu & Jones [48], Balsara [1,2], Falle, Komissarov & Joarder [32] and Crockett et al [19]. The magnetic field in MHD obeys the following evolutionary equation

$$\frac{\partial \mathbf{B}}{\partial t} = \nabla \times (\mathbf{v} \times \mathbf{B}) \tag{1}$$

where $\mathbf{v}$ is the fluid velocity and $\mathbf{B}$ is the magnetic field. The structure of eqn. (1) is such that the magnetic field remains divergence-free in its time-evolution, i.e. it satisfies the constraint

$$\nabla \cdot \mathbf{B} = 0 \tag{2}$$

Retaining a divergence-free aspect in the evolution of the magnetic field has been a design goal in devising methods for numerical MHD, see Yee [59], Brackbill & Barnes [14], Brecht et al [15] and DeVore [24]. Higher order Godunov schemes that provide divergence-free evolution of magnetic fields have been available, see Dai & Woodward [21], Ryu et al [49], Balsara & Spicer [10] and [11], Balsara [5] and Londrillo & DelZanna [43]. Such schemes keep the magnetic field divergence free throughout its evolution while offering the stability and robustness of a total variation diminishing (TVD) scheme. Other formulations are also available that try to advect any magnetic field divergence that might form out of the computational domain, Powell [45], Dedner et al [22].

In his study of AMR-MHD Balsara [3] invented a divergence-free reconstruction strategy for the magnetic field. The method was based on realizing that the magnetic field in the interior of a zone is fully furnished by specifying its field components and their variation within the zone-faces and imposing the divergence-free constraint from eqn. (2). Balsara [3] also used a one-step temporal update strategy as a building block for AMR calculations. Balsara [5] showed that the same divergence-free reconstruction is also useful in designing very high quality second order accurate schemes for numerical MHD. The same divergence-free reconstruction has been extended to higher orders by Balsara [6] who used it along with an RK time-update strategy to obtain MHD schemes that were better than second order accurate. An early version of an ADER scheme for MHD was also presented by us in Taube et al [54]. The goal of the present paper is to present modern ADER-WENO schemes for divergence-free MHD that have a one-step temporal update and could serve as building blocks for AMR-MHD with spatial and temporal accuracy that goes beyond second order. The schemes mentioned in this paragraph have all been implemented in the RIEMANN code for astrophysical fluid dynamics and have been successfully applied to numerous astrophysical applications.

The paper is organized as follows. In Section 2 we present the WENO interpolation used here. Section 3 contains a description of the ADER scheme as well as its instantiation at third order. Section 4 very briefly describes the flux calculation and the time-update steps. Section 4 also provides a point-wise description of the whole ADER-WENO scheme as it is implemented on a computer. Section 5 catalogues the order property of the schemes that have been designed. Section 6 presents several stringent hydrodynamical test problems while Section 7 does the same for MHD test problems.

## 2) Efficient, Multi-Dimensional WENO Reconstruction on Structured Meshes

The first step in designing a high order scheme consists of reconstructing the flow variables within all zones in the computational domain to the desired order of accuracy. Several good options exist in one dimension, see Jiang & Shu [40] and Balsara & Shu [9]. The problem of carrying out a multidimensional reconstruction has been treated in Friedrich [33], Zhang & Shu [60], Dumbser & Käser [28] and Balsara et al [7]. We assume that each zone has a local set of coordinates given by $(x,y,z) \in [-1/2,1/2] \times [-1/2,1/2] \times [-1/2,1/2]$. The Legendre polynomials, suitably modified for the domain $[-1/2,1/2]$, are given by:

$$P_0(x) = 1 \ ; \ P_1(x) = x \ ; \ P_2(x) = x^2 - \frac{1}{12} \ ; \ P_3(x) = x^3 - \frac{3}{20} x \ ;$$
$$P_4(x) = x^4 - \frac{3}{14} x^2 + \frac{3}{560} \tag{3}$$

The polynomial basis set given in eqn. (3) is orthogonal and has a diagonal mass matrix. Tensor products of these polynomials yield an orthogonal, modal basis set in multiple dimensions. A variable "u" is, therefore, reconstructed to appropriate order in the zone being considered when one has all the coefficients of the polynomial

$$\begin{aligned}
u(x,y,z) = &\ u_0 + u_x\, P_1(x) + u_y\, P_1(y) + u_z\, P_1(z) && \leftarrow \text{second order} \\
& + u_{xx}\, P_2(x) + u_{yy}\, P_2(y) + u_{zz}\, P_2(z) \\
& + u_{xy}\, P_1(x) P_1(y) + u_{yz}\, P_1(y) P_1(z) + u_{xz}\, P_1(x) P_1(z) && \leftarrow \text{third order} \\
& + u_{xxx}\, P_3(x) + u_{yyy}\, P_3(y) + u_{zzz}\, P_3(z) + u_{xxy}\, P_2(x) P_1(y) + u_{xyy}\, P_1(x) P_2(y) \\
& + u_{yyz}\, P_2(y) P_1(z) + u_{yzz}\, P_1(y) P_2(z) + u_{xxz}\, P_2(x) P_1(z) + u_{xzz}\, P_1(x) P_2(z) \\
& + u_{xyz}\, P_1(x) P_1(y) P_1(z) && \leftarrow \text{fourth order}
\end{aligned} \tag{4}$$

The arrows in eqn. (4) show us the minimum sub-set of terms that are needed for achieving the desired order of accuracy. The variable $u_0$ is the zone-averaged value of the variable and is evolved using the governing equations. In a WENO scheme, the remaining moments in eqn. (4) above are obtained by examining the smoothness properties of the neighboring zones. In a pointwise WENO scheme, see Jiang & Shu [40] and Balsara & Shu [9] the cross-terms in eqn. (4) are not needed. Since we wish to build a finite volume scheme, we have to reconstruct all the terms including the cross terms in eqn. (4). Several of the strategies catalogued above for carrying out a multidimensional reconstruction can be used to obtain the moments in eqn. (4). However, as shown in Balsara et al [7], for structured meshes it is possible to simplify the reconstruction problem. In that paper we showed that the modes along each coordinate direction in eqn. (4) can be obtained by using the dimension-by-dimension formulation from Jiang & Shu [40] and Balsara & Shu [9]. In this paper we show that the expressions obtained in Jiang & Shu [40] and Balsara & Shu [9] can be substantially simplified if cased in a modal formulation and our goal in Sub-Section 2.1 is to catalogue that simplification. Balsara et al [7] also presented an inexpensive strategy for obtaining the remaining cross-terms in eqn. (4). In this paper we present an even less expensive strategy for obtaining the cross-terms and such an advance is catalogued in Sub-Section 2.2. Sub-Section 2.3 catalogues the divergence-free reconstruction of magnetic fields. In the vicinity of strong shocks it is also useful to flatten the interpolated profiles, as shown by Colella & Woodward [18]. In Appendix A we provide a simple and serviceable flattening algorithm that works with multi-dimensional reconstruction.

**2.1) One-Dimensional WENO Formulation**

The formulation presented here can be used along each dimension to limit the modes in eqn. (4) that do not contain cross-terms. Casting the problem in a modal basis enables us to obtain expressions that are even more compact than those in Jiang & Shu [40] and Balsara & Shu [9].

*2.1.1) Third Order Reconstruction in One Dimension*

Consider the reconstruction problem in a zone labeled by a subscript "0". We start with the neighboring zone-averaged variables $\{u_{-2}, u_{-1}, u_0, u_1, u_2\}$. A third order reconstruction over the zone labeled "0" can be carried out by using three stencils $S_1$, $S_2$ and $S_3$ that rely on the variables $\{u_{-2}, u_{-1}, u_0\}$, $\{u_{-1}, u_0, u_1\}$ and $\{u_0, u_1, u_2\}$ respectively. The reconstructed polynomial is then expressed as

$$u(x) = u_0 + u_x \, P_1(x) + u_{xx} \, P_2(x) \tag{5}$$

The left-biased stencil $S_1$ gives

$$u_x = -2\, u_{-1} + u_{-2}/2 + 3\, u_0/2 \quad , \quad u_{xx} = (u_{-2} - 2\, u_{-1} + u_0)/2 \tag{6}$$

The central stencil $S_2$ gives

$$u_x = (u_1 - u_{-1})/2 \quad , \quad u_{xx} = (u_{-1} - 2\, u_0 + u_1)/2 \tag{7}$$

The right-biased stencil $S_3$ gives

$$u_x = -3\, u_0/2 + 2\, u_1 - u_2/2 \quad , \quad u_{xx} = (u_0 - 2\, u_1 + u_2)/2 \tag{8}$$

Eqns. (6) to (8) show a clear analogy to finite difference approximations. The smoothness measure for each of the three stencils can then be written as

$$IS = u_x^2 + \frac{13}{3} u_{xx}^2 \tag{9}$$

In keeping with the philosophy of Dumbser & Käser [28] we do not strive to achieve fifth order accuracy by using the optimal linear weights of Jiang & Shu [40]. Instead, we seek out stability in our reconstruction by giving the central stencil $S_2$ a linear weight that is 100 times larger than its one-sided partners, i.e. $S_1$ and $S_3$. Also keeping with Dumbser & Käser [28], we raise the smoothness measures to the fourth power when constructing non-linear stencil weights. This choice of preferring stability over an increase in accuracy is also made for all the other WENO schemes in this section.

*2.1.2) Fourth Order Reconstruction in One Dimension*

Consider the reconstruction problem in a zone labeled by a subscript "0". We start with the neighboring zone-averaged variables $\{u_{-3}, u_{-2}, u_{-1}, u_0, u_1, u_2, u_3\}$. A fourth order reconstruction over the zone labeled "0" can be carried out by using four stencils $S_1$, $S_2$, $S_3$ and $S_4$ that rely on the variables $\{u_{-3}, u_{-2}, u_{-1}, u_0\}$, $\{u_{-2}, u_{-1}, u_0, u_1\}$ $\{u_{-1}, u_0, u_1, u_2\}$ and $\{u_0, u_1, u_2, u_3\}$ respectively. The reconstructed polynomial is then expressed as

$$u(x) = u_0 + u_x\, P_1(x) + u_{xx}\, P_2(x) + u_{xxx}\, P_3(x) \tag{10}$$

The stencil $S_1$ gives

$$\begin{aligned}
u_x &= (-177\,u_{-1} + 87\,u_{-2} - 19\,u_{-3} + 109\,u_0)/60, \\
u_{xx} &= -5u_{-1}/2 + 2u_{-2} - u_{-3}/2 + u_0, \\
u_{xxx} &= (-3u_{-1} + 3u_{-2} - u_{-3} + u_0)/6
\end{aligned} \tag{11}$$

The stencil $S_2$ gives

$$\begin{aligned}
u_x &= (-63\,u_{-1} + 11\,u_{-2} + 33\,u_0 + 19\,u_1)/60, \\
u_{xx} &= u_{-1}/2 - u_0 + u_1/2, \\
u_{xxx} &= (3u_{-1} - u_{-2} - 3u_0 + u_1)/6
\end{aligned} \tag{12}$$

The stencil $S_3$ gives

$$u_x = (-19u_{-1} - 33u_0 + 63u_1 - 11u_2)/60,$$
$$u_{xx} = u_{-1}/2 - u_0 + u_1/2,  \quad (13)$$
$$u_{xxx} = (-u_{-1} + 3u_0 - 3u_1 + u_2)/6$$

The stencil $S_4$ gives

$$u_x = (-109u_0 + 177u_1 - 87u_2 + 19u_3)/60,$$
$$u_{xx} = u_0 - 5u_1/2 + 2u_2 - u_3/2, \quad (14)$$
$$u_{xxx} = (-u_0 + 3u_1 - 3u_2 + u_3)/6$$

The smoothness measure for each of the four stencils can then be written as

$$IS = (u_x + u_{xxx}/10)^2 + \frac{13}{3}u_{xx}^2 + \frac{781}{20}u_{xxx}^2 \quad (15)$$

Eqn. (15) makes the positivity of the smoothness measure very apparent.

*2.1.3) Fifth Order Reconstruction in One Dimension*

Though we do not present a fifth order scheme in this paper, the one-dimensional WENO reconstruction presented in this sub-section was used as a building block for a very elegant ninth order pointwise WENO scheme in Balsara & Shu [9]. Because of the utility of that scheme, it is worthwhile presenting the simple and compact expressions for implementing that scheme in this sub-section. Thus consider the reconstruction problem in a zone labeled by a subscript "0". We start with the neighboring zone-averaged variables $\{u_{-4}, u_{-3}, u_{-2}, u_{-1}, u_0, u_1, u_2, u_3, u_4\}$. A fifth order reconstruction over the zone labeled "0" can be carried out by using five stencils $S_1$, $S_2$, $S_3$, $S_4$ and $S_5$ that rely on the variables $\{u_{-4}, u_{-3}, u_{-2}, u_{-1}, u_0\}$, $\{u_{-3}, u_{-2}, u_{-1}, u_0, u_1\}$ $\{u_{-2}, u_{-1}, u_0, u_1, u_2\}$, $\{u_{-1}, u_0, u_1, u_2, u_3\}$ and $\{u_0, u_1, u_2, u_3, u_4\}$ respectively. The reconstructed polynomial is then expressed as

$$u(x) = u_0 + u_x\, P_1(x) + u_{xx}\, P_2(x) + u_{xxx}\, P_3(x) + u_{xxxx}\, P_4(x) \tag{16}$$

The stencil $S_1$ gives

$$\begin{aligned}
u_x &= (-462 u_{-1} + 336 u_{-2} - 146 u_{-3} + 27 u_{-4} + 245 u_0)/120, \\
u_{xx} &= (-240 u_{-1} + 262 u_{-2} - 128 u_{-3} + 25 u_{-4} + 81 u_0)/56, \\
u_{xxx} &= (-18 u_{-1} + 24 u_{-2} - 14 u_{-3} + 3 u_{-4} + 5 u_0)/12, \\
u_{xxxx} &= (-4 u_{-1} + 6 u_{-2} - 4 u_{-3} + u_{-4} + u_0)/24
\end{aligned} \tag{17}$$

The stencil $S_2$ gives

$$\begin{aligned}
u_x &= (-192 u_{-1} + 66 u_{-2} - 11 u_{-3} + 110 u_0 + 27 u_1)/120, \\
u_{xx} &= (10 u_{-1} + 12 u_{-2} - 3 u_{-3} - 44 u_0 + 25 u_1)/56, \\
u_{xxx} &= (12 u_{-1} - 6 u_{-2} + u_{-3} - 10 u_0 + 3 u_1)/12, \\
u_{xxxx} &= (6 u_{-1} - 4 u_{-2} + u_{-3} - 4 u_0 + u_1)/24
\end{aligned} \tag{18}$$

The stencil $S_3$ gives

$$\begin{aligned}
u_x &= (-82 u_{-1} + 11 u_{-2} + 82 u_1 - 11 u_2)/120, \\
u_{xx} &= (40 u_{-1} - 3 u_{-2} - 74 u_0 + 40 u_1 - 3 u_2)/56, \\
u_{xxx} &= (2 u_{-1} - u_{-2} - 2 u_1 + u_2)/12, \\
u_{xxxx} &= (-4 u_{-1} + u_{-2} + 6 u_0 - 4 u_1 + u_2)/24
\end{aligned} \tag{19}$$

The stencil $S_4$ gives

$$\begin{aligned}
u_x &= (-27 u_{-1} - 110 u_0 + 192 u_1 - 66 u_2 + 11 u_3)/120, \\
u_{xx} &= (25 u_{-1} - 44 u_0 + 10 u_1 + 12 u_2 - 3 u_3)/56, \\
u_{xxx} &= (-3 u_{-1} + 10 u_0 - 12 u_1 + 6 u_2 - u_3)/12, \\
u_{xxxx} &= (u_{-1} - 4 u_0 + 6 u_1 - 4 u_2 + u_3)/24
\end{aligned} \tag{20}$$

The stencil $S_5$ gives

$$u_x = (-245 u_0 + 462 u_1 - 336 u_2 + 146 u_3 - 27 u_4)/120,$$
$$u_{xx} = (81 u_0 - 240 u_1 + 262 u_2 - 128 u_3 + 25 u_4)/56,$$
$$u_{xxx} = (-5 u_0 + 18 u_1 - 24 u_2 + 14 u_3 - 3 u_4)/12, \quad (21)$$
$$u_{xxxx} = (u_0 - 4 u_1 + 6 u_2 - 4 u_3 + u_4)/24$$

The smoothness measure for each of the five stencils can then be written as

$$IS = (u_x + u_{xxx}/10)^2 + \frac{13}{3}\left(u_{xx} + \frac{123}{455} u_{xxxx}\right)^2 + \frac{781}{20} u_{xxx}^2 + \frac{1421461}{2275} u_{xxxx}^2 \quad (22)$$

Eqn. (22) makes the positivity of the smoothness measure very apparent.

**2.2) WENO Formulation for the Cross-Terms**

Notice that a majority of the terms in eqn. (4) can be evaluated by dimension-by-dimension limiting. Balsara et al [7] therefore realized that once those terms have been obtained, the remaining cross-terms can be gathered quite efficiently by invoking smaller stencils. While this process is not generally extensible to all orders, it also yields an efficient strategy on structured meshes up to fourth order. In this section we catalogue a strategy for obtaining the cross-terms in eqn. (4) that is even more efficient than the one in Balsara et al [7] by virtue of the fact that it uses smaller stencils to gather up the cross-terms.

*2.2.1) Third Order Reconstruction of the Cross-Terms*

Consider a sub-set of the full polynomial in eqn. (4) given by

$$u(x,y,z) = u_{0,0} + u_x P_1(x) + u_y P_1(y) + u_{xx} P_2(x) + u_{yy} P_2(y) + u_{xy} P_1(x) P_1(y) \quad (23)$$

All the modes in eqn. (23) except for the cross-term $u_{xy}$ can be obtained by using the dimension-by-dimension reconstruction catalogued in the previous Sub-Section. Thus the modes $u_x, u_y$, $u_{xx}$ and $u_{yy}$ as well as the zone-averaged value $u_{0,0}$ are all known in the zone of interest, which we label with a two-index subscript "0,0". Four possible evaluations of the $u_{xy}$ term can then be obtained by taking all the known moments in the zone "0,0" and including any one of the four zone averaged values $u_{1,1}, u_{-1,1}, u_{1,-1}$ and $u_{-1,-1}$ from the zones that lie along the diagonals of the zone of interest. We catalogue the four possible evaluations of the cross-term $u_{xy}$ below:

$$\begin{aligned}
u_{xy} &= u_{1,1} - u_{0,0} - u_x - u_y - u_{xx} - u_{yy} \\
u_{xy} &= -u_{1,-1} + u_{0,0} + u_x - u_y + u_{xx} + u_{yy} \\
u_{xy} &= -u_{-1,1} + u_{0,0} - u_x + u_y + u_{xx} + u_{yy} \\
u_{xy} &= u_{-1,-1} - u_{0,0} + u_x + u_y - u_{xx} - u_{yy}
\end{aligned} \quad (24)$$

The smoothness measure for each of those stencils is given by

$$IS = 4\,u_{xx}^2 + 4\,u_{yy}^2 + u_{xy}^2 \quad (25)$$

and can be used in the usual way to obtain a non-linearly weighted value for $u_{xy}$. By viewing the problem in the yz-plane and the xz-plane it is possible to use the formulae developed here to obtain $u_{yz}$ and $u_{xz}$ respectively. This completes our description of third order WENO interpolation on structured meshes.

*2.2.2) Fourth Order Reconstruction of the Cross-Terms*

Consider a sub-set of the full polynomial in eqn. (4) given by

$$\begin{aligned}
u(x,y,z) = &\,u_{0,0} + u_x\,P_1(x) + u_y\,P_1(y) + u_{xx}\,P_2(x) + u_{yy}\,P_2(y) + u_{xy}\,P_1(x)\,P_1(y) \\
&+ u_{xxx}\,P_3(x) + u_{yyy}\,P_3(y) + u_{xxy}\,P_2(x)\,P_1(y) + u_{xyy}\,P_1(x)\,P_2(y)
\end{aligned} \quad (26)$$

All the modes in eqn. (26) except for the cross-terms $u_{xy}, u_{xxy}$ and $u_{xyy}$ can be obtained by using the dimension-by-dimension reconstruction catalogued in the previous Sub-Section. Thus the modes $u_x, u_y, u_{xx}, u_{yy}, u_{xxx}$ and $u_{yyy}$ as well as the zone-averaged value $u_{0,0}$ are all known in the zone of interest, which we label with a two-index subscript "0,0". Fig. 1 shows us five possible stencils that can each be used to evaluate the $u_{xy}, u_{xxy}$ and $u_{xyy}$ cross-terms. The central stencil was added for stability reasons and has a linear weight that is hundred times larger than the linear weights of the other four directionally-biased stencils. Our choice of five stencils in Fig. 1 gives us five possible evaluations of the cross-terms $u_{xy}, u_{xxy}$ and $u_{xyy}$ which we catalogue below:

For stencil $S_1$ we obtain

$$\begin{aligned}
u_{xy} &= (60 u_{1,1} - 10 u_{1,2} - 10 u_{2,1} - 40 u_{0,0} - 30 u_x - 30 u_y \\
&\quad - 10 u_{xx} - 10 u_{yy} + 27 u_{xxx} + 27 u_{yyy})/20 \\
u_{xxy} &= (-20 u_{1,1} + 10 u_{2,1} + 10 u_{0,0} + 10 u_y \\
&\quad - 20 u_{xx} + 10 u_{yy} - 60 u_{xxx} + 11 u_{yyy})/20 \\
u_{xyy} &= (-20 u_{1,1} + 10 u_{1,2} + 10 u_{0,0} + 10 u_x \\
&\quad + 10 u_{xx} - 20 u_{yy} + 11 u_{xxx} - 60 u_{yyy})/20
\end{aligned} \quad (27)$$

For stencil $S_2$ we obtain

$$\begin{aligned}
u_{xy} &= (-60 u_{-1,1} + 10 u_{-1,2} + 10 u_{-2,1} + 40 u_{0,0} - 30 u_x \\
&\quad + 30 u_y + 10 u_{xx} + 10 u_{yy} + 27 u_{xxx} - 27 u_{yyy})/20 \\
u_{xxy} &= (-20 u_{-1,1} + 10 u_{-2,1} + 10 u_{0,0} + 10 u_y - 20 u_{xx} \\
&\quad + 10 u_{yy} + 60 u_{xxx} + 11 u_{yyy})/20 \\
u_{xyy} &= (20 u_{-1,1} - 10 u_{-1,2} - 10 u_{0,0} + 10 u_x - 10 u_{xx} \\
&\quad + 20 u_{yy} + 11 u_{xxx} + 60 u_{yyy})/20
\end{aligned} \quad (28)$$

For stencil $S_3$ we obtain

$$u_{xy} = (-60u_{1,-1} + 10u_{1,-2} + 10u_{2,-1} + 40u_{0,0} + 30u_x$$
$$-30u_y + 10u_{xx} + 10u_{yy} - 27u_{xxx} + 27u_{yyy})/20$$
$$u_{xxy} = (20u_{1,-1} - 10u_{2,-1} - 10u_{0,0} + 10u_y + 20u_{xx}$$
$$-10u_{yy} + 60u_{xxx} + 11u_{yyy})/20 \quad (29)$$
$$u_{xyy} = (-20u_{1,-1} + 10u_{1,-2} + 10u_{0,0} + 10u_x + 10u_{xx}$$
$$-20u_{yy} + 11u_{xxx} + 60u_{yyy})/20$$

For stencil $S_4$ we obtain

$$u_{xy} = (60u_{-1,-1} - 10u_{-2,-1} - 10u_{-1,-2} - 40u_{0,0} + 30u_x$$
$$+30u_y - 10u_{xx} - 10u_{yy} - 27u_{xxx} - 27u_{yyy})/20$$
$$u_{xxy} = (20u_{-1,-1} - 10u_{-2,-1} - 10u_{0,0} + 10u_y + 20u_{xx}$$
$$-10u_{yy} - 60u_{xxx} + 11u_{yyy})/20 \quad (30)$$
$$u_{xyy} = (20u_{-1,-1} - 10u_{-1,-2} - 10u_{0,0} + 10u_x - 10u_{xx}$$
$$+20u_{yy} + 11u_{xxx} - 60u_{yyy})/20$$

For the central stencil $S_5$ we obtain

$$u_{xy} = (u_{-1,-1} - u_{-1,1} - u_{1,-1} + u_{1,1})/4$$
$$u_{xxy} = (-5u_{-1,-1} + 5u_{-1,1} - 5u_{1,-1} + 5u_{1,1} - 22u_{yyy} - 20u_y)/20 \quad (31)$$
$$u_{xyy} = (-5u_{-1,-1} - 5u_{-1,1} + 5u_{1,-1} + 5u_{1,1} - 22u_{xxx} - 20u_x)/20$$

The smoothness measure for each of the $u_{xy}$ terms in eqns. (27) to (31) is obtained by taking the square of all possible second derivatives of eqn. (26) and integrating them over the zone of interest, see Balsara et al [7]. It is given by

$$IS = 3(u_{xxx}^2 + u_{yyy}^2) + 4(u_{xx}^2 + u_{yy}^2) + u_{xy}^2 + \frac{2}{3}(u_{xxy}^2 + u_{xyy}^2) \quad (32)$$

The smoothness measure for each of the $u_{xxy}$ and $u_{xyy}$ terms in eqns. (27) to (31) is obtained by taking the square of all possible third derivatives of eqn. (26) and integrating them over the zone of interest, see Balsara et al [7]. It is given by

$$IS = 36\,(u_{xxx}^2 + u_{yyy}^2) + 4\,(u_{xxy}^2 + u_{xyy}^2) \tag{33}$$

Both smoothness measures can be used in the usual way to obtain a non-linearly weighted value for $u_{xy}$ or $u_{xxy}$ and $u_{xyy}$ respectively. It is also acceptable to sum eqns. (32) and (33) to obtain a single smoothness measure for all the cross-terms. By viewing the problem in the yz-plane and the xz-plane it is possible to use the formulae developed here to obtain $u_{yz}, u_{yyz}, u_{yzz}, u_{xz}, u_{xxz}$ and $u_{xzz}$.

The remaining $u_{xyz}$ term in eqn. (4) can now be obtained using a strategy that is similar to the one used for obtaining the $u_{xy}$ cross-term at third order. Instead of the four stencils using diagonal zones in the plane we build the eight stencils using the diagonal zones in space. Thus for each of the eight stencils we use one of the values in the set $\{u_{1,1,1},\ u_{-1,1,1},\ u_{1,-1,1},\ u_{1,1,-1},\ u_{-1,1,1},\ u_{-1,1,-1},\ u_{1,-1,-1},\ u_{-1,-1,-1}\}$ to get $u_{xyz}$ in the element of interest indexed by "0,0,0".

For stencil $S_1$ we obtain

$$\begin{aligned}u_{xyz} = u_{1,1,1} &- \frac{11}{10}u_{zzz} - u_{xzz} - u_{zz} - u_{yzz} - u_{yyz} - u_{xxz} - u_{xz} - u_z - u_{yz}\\ &- u_{yy} - \frac{11}{10}u_{yyy} - u_{xyy} - u_{xy} - u_{xxy} - u_y - u_x - u_{0,0,0} - \frac{11}{10}u_{xxx} - u_{xx}\end{aligned} \tag{34}$$

For stencil $S_2$ we obtain

$$u_{xyz} = -u_{-1,1,1} + \frac{11}{10}u_{zzz} - u_{xzz} + u_{zz} + u_{yzz} + u_{yyz} + u_{xxz} - u_{xz} + u_z + u_{yz}$$
$$+ u_{yy} + \frac{11}{10}u_{yyy} - u_{xyy} - u_{xy} + u_{xxy} + u_y - u_x + u_{0,0,0} - \frac{11}{10}u_{xxx} + u_{xx} \tag{35}$$

For stencil $S_3$ we obtain

$$u_{xyz} = -u_{1,-1,1} + \frac{11}{10}u_{zzz} + u_{xzz} + u_{zz} - u_{yzz} + u_{yyz} + u_{xxz} + u_{xz} + u_z - u_{yz}$$
$$+ u_{yy} - \frac{11}{10}u_{yyy} + u_{xyy} - u_{xy} - u_{xxy} - u_y + u_x + u_{0,0,0} + \frac{11}{10}u_{xxx} + u_{xx} \tag{36}$$

For stencil $S_4$ we obtain

$$u_{xyz} = -u_{1,1,-1} - \frac{11}{10}u_{zzz} + u_{xzz} + u_{zz} + u_{yzz} - u_{yyz} - u_{xxz} - u_{xz} - u_z - u_{yz}$$
$$+ u_{yy} + \frac{11}{10}u_{yyy} + u_{xyy} + u_{xy} + u_{xxy} + u_y + u_x + u_{0,0,0} + \frac{11}{10}u_{xxx} + u_{xx} \tag{37}$$

For stencil $S_5$ we obtain

$$u_{xyz} = u_{-1,-1,1} - \frac{11}{10}u_{zzz} + u_{xzz} - u_{zz} + u_{yzz} - u_{yyz} - u_{xxz} + u_{xz} - u_z + u_{yz}$$
$$- u_{yy} + \frac{11}{10}u_{yyy} + u_{xyy} - u_{xy} + u_{xxy} + u_y + u_x - u_{0,0,0} + \frac{11}{10}u_{xxx} - u_{xx} \tag{38}$$

For stencil $S_6$ we obtain

$$u_{xyz} = u_{-1,1,-1} + \frac{11}{10}u_{zzz} + u_{xzz} - u_{zz} - u_{yzz} + u_{yyz} + u_{xxz} - u_{xz} + u_z + u_{yz}$$
$$- u_{yy} - \frac{11}{10}u_{yyy} + u_{xyy} + u_{xy} - u_{xxy} - u_y + u_x - u_{0,0,0} + \frac{11}{10}u_{xxx} - u_{xx} \tag{39}$$

For stencil $S_7$ we obtain

$$u_{xyz} = u_{1,-1,-1} + \frac{11}{10}u_{zzz} - u_{xzz} - u_{zz} + u_{yzz} + u_{yyz} + u_{xxz} + u_{xz} + u_z - u_{yz}$$
$$- u_{yy} + \frac{11}{10}u_{yyy} - u_{xyy} + u_{xy} + u_{xxy} + u_y - u_x - u_{0,0,0} - \frac{11}{10}u_{xxx} - u_{xx}$$
(40)

For stencil $S_8$ we obtain

$$u_{xyz} = -u_{-1,-1,-1} - \frac{11}{10}u_{zzz} - u_{xzz} + u_{zz} - u_{yzz} - u_{yyz} - u_{xxz} + u_{xz} - u_z + u_{yz}$$
$$+ u_{yy} - \frac{11}{10}u_{yyy} - u_{xyy} + u_{xy} - u_{xxy} - u_y - u_x + u_{0,0,0} - \frac{11}{10}u_{xxx} + u_{xx}$$
(41)

The smoothness measure for each of the eight stencils is given by

$$IS = 36(u_{xxx}^2 + u_{yyy}^2 + u_{zzz}^2) + 4(u_{xxy}^2 + u_{xxz}^2 + u_{xyy}^2 + u_{yyz}^2 + u_{xzz}^2 + u_{yzz}^2) + u_{xyz}^2 \quad (42)$$

The smoothness measures can be used in the usual way to obtain a non-linearly weighted value for $u_{xyz}$. This completes our description of fourth order WENO interpolation on structured meshes.

**2.3) Cataloguing Divergence-Free Reconstruction of the Magnetic Field**

Divergence-free reconstruction for MHD has been detailed in Balsara [3] and [5] for second order schemes and in Balsara [6] for higher order schemes. We therefore present only as much detail here as is needed for understanding ADER-WENO schemes for MHD. Consequently, the method consists of realizing that the moments of the face-centered magnetic field components can be obtained by using the reconstruction techniques given in the previous two Sub-Sections. Assuming a zone to be a unit cube, the x-components of the magnetic field in the upper and lower x-faces of a zone are then given by

$$B_x(x = \pm 1/2, y, z) = B_0^{x\pm} + B_y^{x\pm} P_1(y) + B_z^{x\pm} P_1(z) \quad \leftarrow \text{ second order}$$
$$+ B_{yy}^{x\pm} P_2(y) + B_{yz}^{x\pm} P_1(y) P_1(z) + B_{zz}^{x\pm} P_2(z) \quad \leftarrow \text{ third order}$$
$$+ B_{yyy}^{x\pm} P_3(y) + B_{yyz}^{x\pm} P_2(y) P_1(z) + B_{yzz}^{x\pm} P_1(y) P_2(z) + B_{zzz}^{x\pm} P_3(z) \quad \leftarrow \text{ fourth order}$$

(43)

The arrows in eqn. (43) show us the minimum sub-set of terms that are needed for achieving the desired order of accuracy. Thus for a second order scheme we would only use the first line in eqn. (43). For a third order scheme we would need the first and second lines in eqn. (43). For a fourth order scheme we would use all three lines in eqn. (43). Similar expressions for the y and z-components of the field in the appropriate zone faces can be written as:

$$B_y(x, y = \pm 1/2, z) = B_0^{y\pm} + B_x^{y\pm} P_1(x) + B_z^{y\pm} P_1(z) \quad \leftarrow \text{ second order}$$
$$+ B_{xx}^{y\pm} P_2(x) + B_{xz}^{y\pm} P_1(x) P_1(z) + B_{zz}^{y\pm} P_2(z) \quad \leftarrow \text{ third order}$$
$$+ B_{xxx}^{y\pm} P_3(x) + B_{xxz}^{y\pm} P_2(x) P_1(z) + B_{xzz}^{y\pm} P_1(x) P_2(z) + B_{zzz}^{y\pm} P_3(z) \quad \leftarrow \text{ fourth order}$$

(44)

$$B_z(x, y, z = \pm 1/2) = B_0^{z\pm} + B_x^{z\pm} P_1(x) + B_y^{z\pm} P_1(z) \quad \leftarrow \text{ second order}$$
$$+ B_{xx}^{z\pm} P_2(x) + B_{xy}^{z\pm} P_1(x) P_1(y) + B_{yy}^{z\pm} P_2(y) \quad \leftarrow \text{ third order}$$
$$+ B_{xxx}^{z\pm} P_3(x) + B_{xxy}^{z\pm} P_2(x) P_1(y) + B_{xyy}^{z\pm} P_1(x) P_2(y) + B_{yyy}^{z\pm} P_3(y) \quad \leftarrow \text{ fourth order}$$

(45)

The moments in eqn. (43) can be obtained by limiting the x-component of the magnetic field in the yz-plane. Similarly, the moments in eqns. (44) and (45) can be obtained by limiting in the xz-plane and xy-plane respectively. The WENO limiting strategies catalogued in the previous two Sub-Sections can be used to carry out the limiting. To reconstruct the field in the interior of the zone we pick the following functional forms for the fields:

$$B_x(x, y, z) = a_0 + a_x P_1(x) + a_y P_1(y) + a_z P_1(z)$$
$$+ a_{xx} P_2(x) + a_{xy} P_1(x) P_1(y) + a_{xz} P_1(x) P_1(z) \qquad \leftarrow \text{second order}$$
$$+ a_{yy} P_2(y) + a_{xyy} P_1(x) P_2(y) + a_{zz} P_2(z) + a_{xzz} P_1(x) P_2(z) + a_{yz} P_1(y) P_1(z) + a_{xyz} P_1(x) P_1(y) P_1(z)$$
$$+ a_{xxx} P_3(x) + a_{xxy} P_2(x) P_1(y) + a_{xxz} P_2(x) P_1(z) \qquad \leftarrow \text{third order}$$
$$+ a_{yyy} P_3(y) + a_{xyyy} P_1(x) P_3(y) + a_{yyz} P_2(y) P_1(z) + a_{xyyz} P_1(x) P_2(y) P_1(z)$$
$$+ a_{yzz} P_1(y) P_2(z) + a_{xyzz} P_1(x) P_1(y) P_2(z) + a_{zzz} P_3(z) + a_{xzzz} P_1(x) P_3(z)$$
$$+ a_{xxxx} P_4(x) + a_{xxxy} P_3(x) P_1(y) + a_{xxxz} P_3(x) P_1(z)$$
$$+ a_{xxyy} P_2(x) P_2(y) + a_{xxzz} P_2(x) P_2(z) \qquad \leftarrow \text{fourth order}$$

$$(46)$$

$$B_y(x, y, z) = b_0 + b_x P_1(x) + b_y P_1(y) + b_z P_1(z)$$
$$+ b_{yy} P_2(y) + b_{xy} P_1(x) P_1(y) + b_{yz} P_1(y) P_1(z) \qquad \leftarrow \text{second order}$$
$$+ b_{xx} P_2(x) + b_{xxy} P_2(x) P_1(y) + b_{zz} P_2(z) + b_{yzz} P_1(y) P_2(z) + b_{xz} P_1(x) P_1(z) + b_{xyz} P_1(x) P_1(y) P_1(z)$$
$$+ b_{yyy} P_3(y) + b_{xyy} P_1(x) P_2(y) + b_{yyz} P_2(y) P_1(z) \qquad \leftarrow \text{third order}$$
$$+ b_{xxx} P_3(x) + b_{xxxy} P_3(x) P_1(y) + b_{xxz} P_2(x) P_1(z) + b_{xxyz} P_2(x) P_1(y) P_1(z)$$
$$+ b_{xzz} P_1(x) P_2(z) + b_{xyzz} P_1(x) P_1(y) P_2(z) + b_{zzz} P_3(z) + b_{yzzz} P_1(y) P_3(z)$$
$$+ b_{yyyy} P_4(y) + b_{xyyy} P_1(x) P_3(y) + b_{yyyz} P_3(y) P_1(z)$$
$$+ b_{xxyy} P_2(x) P_2(y) + b_{yyzz} P_2(y) P_2(z) \qquad \leftarrow \text{fourth order}$$

$$(47)$$

$$B_z(x, y, z) = c_0 + c_x P_1(x) + c_y P_1(y) + c_z P_1(z)$$
$$+ c_{zz} P_2(z) + c_{xz} P_1(x) P_1(z) + c_{yz} P_1(y) P_1(z) \qquad \leftarrow \text{second order}$$
$$+ c_{xx} P_2(x) + c_{xxz} P_2(x) P_1(z) + c_{yy} P_2(y) + c_{yyz} P_2(y) P_1(z) + c_{xy} P_1(x) P_1(y) + c_{xyz} P_1(x) P_1(y) P_1(z)$$
$$+ c_{zzz} P_3(z) + c_{xzz} P_1(x) P_2(z) + c_{yzz} P_1(y) P_2(z) \qquad \leftarrow \text{third order}$$
$$+ c_{xxx} P_3(x) + c_{xxxz} P_3(x) P_1(z) + c_{xxy} P_2(x) P_1(y) + c_{xxyz} P_2(x) P_1(y) P_1(z)$$
$$+ c_{xyy} P_1(x) P_2(y) + c_{xyyz} P_1(x) P_2(y) P_1(z) + c_{yyy} P_3(y) + c_{yyyz} P_3(y) P_1(z)$$
$$+ c_{zzzz} P_4(z) + c_{xzzz} P_1(x) P_3(z) + c_{yzzz} P_1(y) P_3(z)$$
$$+ c_{xxzz} P_2(x) P_2(z) + c_{yyzz} P_2(y) P_2(z) \qquad \leftarrow \text{fourth order}$$

$$(48)$$

The rationale for picking this set of moments follows from Balsara [3]. As shown in Balsara [6], eqns. (43) to (45) can be used along with the divergence-free condition in eqn. (2) to completely specify the coefficients in eqns. (46) to (48). Observe that eqns. (43) to (45) only hold in the zone-faces while eqns. (46) to (48) are divergence-free to all orders and hold at all points within the zone being considered. As a result, our evaluation of the coefficients in eqns. (46) to (48) reconstructs the magnetic field in the whole zone.

It is worthwhile to consider eqn. (46) at third order to make two important points. First, notice that all the linear and quadratic variations that one would require for the third order accurate reconstruction within the unit cube are all present. As a result, although we started with just the facial moments in eqns. (43) to (45), the divergence-free reconstruction has enabled us to fully specify all the requisite moments for third order accuracy within the unit cube's interior. This observation extends to all orders. Second, notice that the coefficients $a_{xyy}$, $a_{xzz}$, $a_{xyz}$, $a_{xxx}$, $a_{xxy}$ and $a_{xxz}$ correspond to variations that are only needed for fourth order accuracy and yet they are present in the third order reconstruction. Their presence is mandated by the divergence-free condition, not by accuracy conditions. As a result, these terms do not need to participate in the third order ADER time-evolution. Their contribution does, however, need to be included in the flux evaluation in the ADER scheme as well as in the construction of Riemann solvers at zone boundaries. We therefore say that these coefficients provide non-evolutionary terms in the ADER update. Their contribution needs to be included wherever possible but there are no time-evolving terms associated with them in the ADER formulation.

**3) ADER-CG Formulation**

In contrast to the classical ADER schemes of Titarev & Toro [55] and [56], which needed many analytical algebraic manipulations due to the underlying Cauchy-Kovalewski procedure, the new formulation of ADER schemes recently proposed in Dumbser et al. [26] is based on a local weak formulation of the governing PDE in space-time and only needs flux evaluations at point values. These new ADER schemes rely on an iterative convergence to the actual space-time representation of the solution within each zone. In Sub-Section 3.1 we provide the general formulation of ADER-CG schemes (where CG stands for continuous Galerkin representation in space and time) and describe one iteration of the ADER scheme. In Sub-Section 3.2 we describe in detail the implementation of the third order ADER-CG scheme in an effort to make the ADER method easily accessible to all readers. In Appendices B and C we provide the most essential formulae that are needed for making implementations of the second and fourth order ADER-CG schemes respectively.

**3.1) General Formulation of ADER-CG Schemes for Structured Meshes**

Say we want to evolve the nonlinear time-dependent hyperbolic system of conservation laws given by

$$\frac{\partial U}{\partial t} + \frac{\partial F}{\partial x} + \frac{\partial G}{\partial y} + \frac{\partial H}{\partial z} = S \qquad (49)$$

where $u$ is an $n$-component vector of conserved variables and $F = F(U)$, $G = G(U)$ and $H = H(U)$ are flux vectors and $S = S(U)$ is a non-stiff source term. We wish to take a time step of size $\Delta t$ on a mesh having zones of size $\Delta x$, $\Delta y$ and $\Delta z$ in each of the three directions. Each zone can be mapped to a unit cube in space. Since ADER schemes operate in space and time, we consider a four dimensional reference element in space-time given by $[-1/2, 1/2] \times [-1/2, 1/2] \times [-1/2, 1/2] \times [0, 1]$ where the first three coordinates span the unit cube and the fourth coordinate represents the time axis. In this space-time element we set up the coordinates $(\xi, \eta, \zeta, \tau)$ and make the transcriptions $u = U$, $f = \Delta t\, F/\Delta x$, $g = \Delta t\, G/\Delta y$, $h = \Delta t\, H/\Delta z$ and $s = \Delta t\, S$. This allows us to write eqn. (49) in the reference element as

$$\frac{\partial u}{\partial \tau} + \frac{\partial f}{\partial \xi} + \frac{\partial g}{\partial \eta} + \frac{\partial h}{\partial \zeta} = s \qquad (50)$$

The ADER scheme that we describe here is a modal variant of the ADER scheme with a continuous Galerkin representation in time (also known as ADER-CG) described in Dumbser et al [26]. Such ADER-CG schemes are very efficient because they minimize the number of flux evaluations though they have the drawback that they are not well-suited for handling stiff source terms. We now specify a set of $L$ basis functions $\{\theta_l = \theta_l(\xi,\eta,\zeta,\tau), l=1,L\}$ in the reference element. For a general Galerkin formulation, any reasonable set of basis functions would suffice. For an ADER-CG scheme we make the further requirement that the first $L_S$ elements in the set of basis functions only have a spatial dependence and lack any dependence on time $\tau$. The solution vector $u$ can now be represented in this basis space as

$$u(\xi,\eta,\zeta,\tau) = \sum_{l=1}^{L} \hat{u}_l \, \theta_l(\xi,\eta,\zeta,\tau) \tag{51}$$

where $\hat{u} \equiv \left(\hat{u}_1,..,\hat{u}_{L_S},\hat{u}_{L_S+1},...,\hat{u}_L\right)^T$ is a vector of modes. The first $L_S$ elements of this vector of modes lack time-dependence so that only the last $L-L_S$ of these modes carry the time-evolution of the solution $u$. Equations similar to eqn. (51) can be formulated for the flux components as well as the source terms. Thus the $\xi$, $\eta$ and $\zeta$-directional fluxes in space-time reference element are, therefore, completely specified by providing the modal vectors given by $\hat{f} \equiv \left(\hat{f}_1,..,\hat{f}_{L_S},\hat{f}_{L_S+1},...,\hat{f}_L\right)^T$, $\hat{g} \equiv \left(\hat{g}_1,..,\hat{g}_{L_S},\hat{g}_{L_S+1},...,\hat{g}_L\right)^T$ and $\hat{h} \equiv \left(\hat{h}_1,..,\hat{h}_{L_S},\hat{h}_{L_S+1},...,\hat{h}_L\right)^T$ respectively. Likewise, the source terms are specified by providing $\hat{s} \equiv \left(\hat{s}_1,...,\hat{s}_{L_S},\hat{s}_{L_S+1},...,\hat{s}_L\right)^T$. These flux terms and source terms can be obtained by using $\hat{u}$ from the previous iteration. The method for doing so is illustrated at third order in the next Sub-Section. The ADER-CG formulation consists of making a further *essential* simplification that at $\tau = 0$ the solution $u(\xi,\eta,\zeta,\tau)$ is continuous with the initial condition $w(\xi,\eta,\zeta)$. This simplification is advantageous because one needs to evaluate $\hat{f}_l$, $\hat{g}_l$, $\hat{h}_l$ and $\hat{s}_l$ for $l=1,..,L_S$ only once at $\tau=0$, resulting in a substantial savings in computational complexity. Thus if $w$ is written in a modal space as

$$w(\xi,\eta,\zeta) = \sum_{l=1}^{L_S} \hat{w}_l \; \theta_l(\xi,\eta,\zeta,\tau=0) \tag{52}$$

then the ADER-CG simplification consists of asserting that $\hat{u}_l = \hat{w}_l$ for $l=1,..,L_S$. Notice that this assertion simultaneously relinquishes the prospect of obtaining a weak formulation in time as well as the scheme's ability to handle stiff source terms. (In Balsara et al [8] and Dumbser et al [30] we present ADER schemes that retain the weak formulation in time and can, therefore, handle stiff source terms. We refer to those schmes as ADER-DG to show their discontinuous Galerkin aspect.)

Applying the Galerkin approach to eqn. (50) then gives us

$$\left\langle \theta_j, \frac{\partial \theta_l}{\partial \tau}\right\rangle \hat{u}_l + \left\langle \theta_j, \frac{\partial \theta_l}{\partial \xi}\right\rangle \hat{f}_l + \left\langle \theta_j, \frac{\partial \theta_l}{\partial \eta}\right\rangle \hat{g}_l + \left\langle \theta_j, \frac{\partial \theta_l}{\partial \zeta}\right\rangle \hat{h}_l = \left\langle \theta_j, \theta_l \right\rangle \hat{s}_l \tag{53}$$

The angled brackets in the above equation denote space-time integration over the reference element. Eqn. (53) can then be written as

$$K_\tau \hat{u} + K_\xi \hat{f} + K_\eta \hat{g} + K_\zeta \hat{h} = M\hat{s} \tag{54}$$

where, in keeping with the usual terminology of Galerkin schemes, $M$ is the mass matrix, $K_\tau$ is the time-stiffness matrix and $K_\xi, K_\eta$ and $K_\zeta$ are the flux-stiffness matrices. The $(j,l)^{th}$ elements of these matrices can be made explicit as follows

$$K_{\tau;j,l} = \left\langle \frac{\partial \theta_j}{\partial \tau}, \theta_l \right\rangle \; ; \; K_{\xi;j,l} = \left\langle \theta_j, \frac{\partial \theta_l}{\partial \xi}\right\rangle \; ; \; K_{\eta;j,l} = \left\langle \theta_j, \frac{\partial \theta_l}{\partial \eta}\right\rangle \; ; \; K_{\zeta;j,l} = \left\langle \theta_j, \frac{\partial \theta_l}{\partial \zeta}\right\rangle \; ; M_{j,l} = \left\langle \theta_j, \theta_l \right\rangle \tag{55}$$

Notice from the structure of the vector $\hat{u}$ that only the last $L$–$L_S$ components are unknowns to be obtained from the ADER-CG iteration. We thus write $\hat{u}$ as $\hat{u} = \left(\hat{u}^0, \hat{u}^1\right)^T$ where $\hat{u}^0$ has the first $L_S$ components of $\hat{u}$ and $\hat{u}^1$ has the last $L$–$L_S$ components of $\hat{u}$. A similar split can be effected for $\hat{f}, \hat{g}, \hat{h}$ and $\hat{s}$. The mass matrix and the stiffness matrices can now be written as

$$M = \begin{bmatrix} M^{00} & M^{01} \\ M^{10} & M^{11} \end{bmatrix}, \quad K_\alpha = \begin{bmatrix} K_\alpha^{00} & K_\alpha^{01} \\ K_\alpha^{10} & K_\alpha^{11} \end{bmatrix} \tag{56}$$

where $\alpha$ can be $\xi, \eta, \zeta$ or $\tau$ in the above equation. Only the last $L-L_S$ components of eqn. (54) are useful and yield the equation

$$\hat{u}^1 + \hat{K}_\xi \hat{f}^1 + \hat{K}_\eta \hat{g}^1 + \hat{K}_\zeta \hat{h}^1 = \hat{M}\hat{s}^1 + \hat{M}^0 \hat{s}^0 - \hat{K}_\xi^0 \hat{f}^0 - \hat{K}_\eta^0 \hat{g}^0 - \hat{K}_\zeta^0 \hat{h}^0 \tag{57}$$

Where the matrices in the above equation are given by

$$\hat{K}_\xi = \left(K_\tau^{11}\right)^{-1} K_\xi^{11}, \; \hat{K}_\eta = \left(K_\tau^{11}\right)^{-1} K_\eta^{11}, \; \hat{K}_\zeta = \left(K_\tau^{11}\right)^{-1} K_\zeta^{11}, \; \hat{M} = \left(K_\tau^{11}\right)^{-1} M^{11},$$
$$\hat{K}_\xi^0 = \left(K_\tau^{11}\right)^{-1} K_\xi^{10}, \; \hat{K}_\eta^0 = \left(K_\tau^{11}\right)^{-1} K_\eta^{10}, \; \hat{K}_\zeta^0 = \left(K_\tau^{11}\right)^{-1} K_\zeta^{10}, \; \hat{M}^0 = \left(K_\tau^{11}\right)^{-1} M^{10} \tag{58}$$

Thus a specification of the matrices in eqn. (58) along with eqn. (57) furnishes the entire ADER-CG scheme. In the next Sub-Section we will explicitly show the third order ADER-CG scheme that results from using these matrices.

Notice that the matrices $\hat{M}$ and $\hat{K}_\alpha$ are square matrices with a rank of $L-L_S$ while the matrices $\hat{M}^0$ and $\hat{K}_\alpha^0$ are rectangular with a dimension $(L-L_S)\times L_S$. It is interesting to remark that while the $K_\alpha^{10}$ and $K_\alpha^{11}$ matrices in eqn. (58) are non-sparse, the matrices $\hat{K}_\alpha$ and $\hat{K}_\alpha^0$ are sparse at all orders. As a result, the form presented in eqn. (57) is also the one in which the equations are most elegant. This is true both for the tensor product basis functions that are used for logically rectilinear meshes and also for the Dubiner [25] basis functions that are used for unstructured meshes. At an intuitive level, the sparsity of $\hat{K}_\xi$, $\hat{K}_\eta$ and $\hat{K}_\zeta$ stems from the fact that in Legendre basis as well as in Dubiner basis the derivative operator only couples one basis function to two other basis functions.

In an ADER-CG scheme eqn. (57) is made to converge via iteration. Our experience has shown that we only require "M" iterations of eqn. (57) to achieve the requisite accuracy of an $M^{th}$ order scheme. Dumbser et al [26] provide an intuitive explanation, based on contractive mappings, for this rapid convergence. There also exists formal theory based on the Picard iteration which supports the claim that "M" iterations are adequate for an $M^{th}$ order scheme. As a result, while the ADER-CG schemes do iterate on eqn. (57), the iteration is not very expensive. Even the most stringent test problems presented here were always run with the minimum requisite number of ADER-CG iterations.

**3.2) Implementation of the ADER-CG Scheme at Third Order**

We start with the initial condition at $\tau=0$ which is given by expressing $w(\xi,\eta,\zeta)$ in terms of the $L_S = 10$ spatial basis functions as follows

$$\begin{aligned} w(\xi,\eta,\zeta) &= \hat{w}_1 P_0(\xi) P_0(\eta) P_0(\zeta) \\ &+ \hat{w}_2 P_1(\xi) P_0(\eta) P_0(\zeta) + \hat{w}_3 P_0(\xi) P_1(\eta) P_0(\zeta) + \hat{w}_4 P_0(\xi) P_0(\eta) P_1(\zeta) \\ &+ \hat{w}_5 P_2(\xi) P_0(\eta) P_0(\zeta) + \hat{w}_6 P_0(\xi) P_2(\eta) P_0(\zeta) + \hat{w}_7 P_0(\xi) P_0(\eta) P_2(\zeta) \\ &+ \hat{w}_8 P_1(\xi) P_1(\eta) P_0(\zeta) + \hat{w}_9 P_0(\xi) P_1(\eta) P_1(\zeta) + \hat{w}_{10} P_1(\xi) P_0(\eta) P_1(\zeta) \end{aligned}$$

(59)

We can now define a space-time solution $u(\xi,\eta,\zeta,\tau)$ in the reference space-time element by forming tensor products of the spatial basis set with the temporal basis set. The temporal basis set has to be specially chosen in an ADER-CG scheme so that the first $L_S$ basis functions match up with those in eqn. (59). Thus our temporal basis functions are taken to be

$$Q_0(\tau)=1 \quad, \quad Q_1(\tau)=\tau \quad, \quad Q_2(\tau)=\tau^2 \quad, \quad Q_3(\tau)=\tau^3 \qquad (60)$$

The first three basis functions in eqn. (60) are needed for the third order scheme; the last basis function in eqn. (60) is only needed for fourth order schemes. To obtain full third order accuracy in space-time we use a total of $L=15$ basis functions. The conserved variables $u(\xi,\eta,\zeta,\tau)$ can be expressed in terms of the degrees of freedom, i.e. the modes, and the basis functions as

$$
\begin{aligned}
u(\xi,\eta,\zeta,\tau) &= \hat{w}_1 P_0(\xi)P_0(\eta)P_0(\zeta)Q_0(\tau) \\
&+ \hat{w}_2 P_1(\xi)P_0(\eta)P_0(\zeta)Q_0(\tau) + \hat{w}_3 P_0(\xi)P_1(\eta)P_0(\zeta)Q_0(\tau) + \hat{w}_4 P_0(\xi)P_0(\eta)P_1(\zeta)Q_0(\tau) \\
&+ \hat{w}_5 P_2(\xi)P_0(\eta)P_0(\zeta)Q_0(\tau) + \hat{w}_6 P_0(\xi)P_2(\eta)P_0(\zeta)Q_0(\tau) + \hat{w}_7 P_0(\xi)P_0(\eta)P_2(\zeta)Q_0(\tau) \\
&+ \hat{w}_8 P_1(\xi)P_1(\eta)P_0(\zeta)Q_0(\tau) + \hat{w}_9 P_0(\xi)P_1(\eta)P_1(\zeta)Q_0(\tau) + \hat{w}_{10} P_1(\xi)P_0(\eta)P_1(\zeta)Q_0(\tau) \\
&+ \hat{u}_{11} P_0(\xi)P_0(\eta)P_0(\zeta)Q_1(\tau) + \hat{u}_{12} P_0(\xi)P_0(\eta)P_0(\zeta)Q_2(\tau) \\
&+ \hat{u}_{13} P_1(\xi)P_0(\eta)P_0(\zeta)Q_1(\tau) + \hat{u}_{14} P_0(\xi)P_1(\eta)P_0(\zeta)Q_1(\tau) + \hat{u}_{15} P_0(\xi)P_0(\eta)P_1(\zeta)Q_1(\tau)
\end{aligned}
$$
(61)

Notice that eqn. (61) already incorporates the essential simplification that is built into an ADER-CG scheme because we have set $\hat{u}_l = \hat{w}_l$ for $l=1,..,L_S$.

While it is always possible to explicitly write down all the matrices from eqn. (58), it is much easier to write down the iterative scheme that they give rise to. The resultant ADER-CG iteration at third order is therefore given by

$$
\begin{aligned}
\hat{u}_{11} &= -\hat{f}_2 - \hat{g}_3 - \hat{h}_4 + \hat{s}_1 - \frac{3}{10}\hat{s}_{12} \\
\hat{u}_{12} &= -\frac{\hat{f}_{13}}{2} - \frac{\hat{g}_{14}}{2} - \frac{\hat{h}_{15}}{2} + \frac{\hat{s}_{11}}{2} + \frac{3}{5}\hat{s}_{12} \\
\hat{u}_{13} &= -2\hat{f}_5 - \hat{g}_8 - \hat{h}_{10} + \hat{s}_2 + \frac{2}{3}\hat{s}_{13} \\
\hat{u}_{14} &= -\hat{f}_8 - 2\hat{g}_6 - \hat{h}_9 + \hat{s}_3 + \frac{2}{3}\hat{s}_{14} \\
\hat{u}_{15} &= -\hat{f}_{10} - \hat{g}_9 - 2\hat{h}_7 + \hat{s}_4 + \frac{2}{3}\hat{s}_{15}
\end{aligned}
$$
(62)

The set of equations provided in eqn. (62) completely describe one iteration the ADER-CG scheme at third order on structured meshes.

Now that the ADER-CG iteration has been described in eqn. (62), we only need to specify a strategy for obtaining the vectors $\hat{f}$, $\hat{g}$, $\hat{h}$ and $\hat{s}$ from the vector $\hat{u}$. To accomplish that, we establish a set of nodal points in space-time on the reference element. Several choices of nodal points are possible. Realize that we have $L$ modes so that we could define a minimal set of $L$ nodal points that allow us to make a one-to-one transcription from the nodal space to the modal space. This would yield the most economical ADER-CG scheme. Choosing a tensor product set of Gaussian quadrature points might yield the most accurate transcription from nodal to modal space. It would also be computationally expensive because for an $M^{th}$ order scheme, this choice would call for $M^4$ quadrature points. We prefer an intermediate strategy where we choose a set of $L_n$ set of nodal points where $L_n$ is slightly larger than $L$. The node placement in this choice has the special property that it yields compact, finite-difference like formulae for transcribing from nodal space to modal space. For third order we have $L_n = 22$ and the nodes are chosen to have geometric symmetries which yield expected cancellations in problems that have a great deal of symmetry. We have found such symmetrical node placements even for second and fourth order ADER-CG. For third order ADER-CG the $L_n$ nodes are given by the ordered set

$$\begin{aligned}
\{ & (0,0,0,0); (1/2,0,0,0); (-1/2,0,0,0); (0,1/2,0,0); (0,-1/2,0,0); \\
& (0,0,1/2,0); (0,0,-1/2,0); (1/2,1/2,1/2,0); (-1/2,1/2,1/2,0); \\
& (1/2,-1/2,1/2,0); (-1/2,-1/2,1/2,0); (1/2,1/2,-1/2,0); \\
& (-1/2,1/2,-1/2,0); (1/2,-1/2,-1/2,0); (-1/2,-1/2,-1/2,0); \\
& (1/2,0,0,1/2); (-1/2,0,0,1/2); (0,1/2,0,1/2); (0,-1/2,0,1/2); \\
& (0,0,1/2,1/2); (0,0,-1/2,1/2); (0,0,0,1)\}
\end{aligned} \quad (63)$$

Using the ordered set of nodal points we can then define an $L_n$ component vector $\bar{u}$ which contains the nodal values of the conserved variables. The ordering of the components of $\bar{u}$ follows that of the nodal points. Note that the first 15 elements of $\bar{u}$ have to be evaluated only once. Using the $L_n$ component vector $\bar{u}$ we can now define an an $L_n$ component vector $\bar{f}$ which contains the x-directional fluxes from the hyperbolic system in eqn. (50). The ordering of the components of $\bar{f}$ also follows that of the nodal points. As a result the first 15 elements of $\bar{f}$ have to be evaluated only once, leading to some of the computational efficiency of the ADER-CG scheme. The process of obtaining the vector $\hat{f}$ from the vector $\bar{f}$ is just a matter of transcribing from nodal to modal space and is given below.

$$\begin{aligned}
\hat{f}_1 &= (\bar{f}_2 + \bar{f}_3 + \bar{f}_4 + \bar{f}_5 + \bar{f}_6 + \bar{f}_7)/6 \\
\hat{f}_2 &= \bar{f}_2 - \bar{f}_3 \\
\hat{f}_3 &= \bar{f}_4 - \bar{f}_5 \\
\hat{f}_4 &= \bar{f}_6 - \bar{f}_7 \\
\hat{f}_5 &= 2\bar{f}_2 - 4\bar{f}_1 + 2\bar{f}_3 \\
\hat{f}_6 &= 2\bar{f}_4 - 4\bar{f}_1 + 2\bar{f}_5 \\
\hat{f}_7 &= 2\bar{f}_6 - 4\bar{f}_1 + 2\bar{f}_7 \\
\hat{f}_8 &= (\bar{f}_8 - \bar{f}_9 - \bar{f}_{10} + \bar{f}_{11} + \bar{f}_{12} - \bar{f}_{13} - \bar{f}_{14} + \bar{f}_{15})/2 \\
\hat{f}_9 &= (\bar{f}_8 + \bar{f}_9 - \bar{f}_{10} - \bar{f}_{11} - \bar{f}_{12} - \bar{f}_{13} + \bar{f}_{14} + \bar{f}_{15})/2 \\
\hat{f}_{10} &= (\bar{f}_8 - \bar{f}_9 + \bar{f}_{10} - \bar{f}_{11} - \bar{f}_{12} + \bar{f}_{13} - \bar{f}_{14} + \bar{f}_{15})/2
\end{aligned} \tag{64}$$

$$\begin{aligned}
\hat{t}_1 &= (\bar{f}_{16} + \bar{f}_{17} + \bar{f}_{18} + \bar{f}_{19} + \bar{f}_{20} + \bar{f}_{21})/3 - 2\hat{f}_1 \\
\hat{f}_{11} &= 2\hat{t}_1 - \bar{f}_{22} + \bar{f}_1 \\
\hat{f}_{12} &= 2\hat{t}_1 - 2\hat{f}_{11} \\
\hat{f}_{13} &= 2(\bar{f}_{16} - \bar{f}_{17} - \bar{f}_2 + \bar{f}_3) \\
\hat{f}_{14} &= 2(\bar{f}_{18} - \bar{f}_{19} - \bar{f}_4 + \bar{f}_5) \\
\hat{f}_{15} &= 2(\bar{f}_{20} - \bar{f}_{21} - \bar{f}_6 + \bar{f}_7)
\end{aligned} \tag{65}$$

We point out that $\hat{t}_1$ is a temporary variable. Notice from eqn. (64) that the first 10 components of $\hat{f}$ have to be evaluated only once and are completely specified by the first fifteen components of $\bar{f}$. The first fifteen components of $\bar{f}$ are, in turn, evaluated only once at τ=0 before starting the ADER-CG iterations. Notice too from eqn. (65) that the last 5 components of $\hat{f}$ will have to be re-evaluated in every ADER-CG iteration and only require a re-evaluation of the last seven components of $\bar{f}$ at τ>0. This clear separation between the fluxes that have to be evaluated only once at τ=0 and the much smaller number of fluxes that have to be evaluated at τ>0 yields even further computational efficiency. A similar approach is followed for the other fluxes and the source terms in eqn. (50). This completes our description of the ADER-CG scheme at third order.

**4) Flux Calculation , Time-Update and a Step-by-step Description of the ADER-WENO Scheme**

In Sub-Section 4.1 we describe the one-step time update and the flux calculation. This includes the electric field calculation that is needed for MHD. In Sub-Section 4.2 we provide a step-by-step description of the ADER-WENO scheme.

**4.1) Flux Calculation and Time-Update**

The MHD system can be described in conservation form of the form shown in eqn. (49) by writing it as

$$\frac{\partial}{\partial t}\begin{pmatrix} \rho \\ \rho v_x \\ \rho v_y \\ \rho v_z \\ \varepsilon \\ B_x \\ B_y \\ B_z \end{pmatrix} + \frac{\partial}{\partial x}\begin{pmatrix} \rho v_x \\ \rho v_x^2 + P + \mathbf{B}^2/8\pi - B_x^2/4\pi \\ \rho v_x v_y - B_x B_y/4\pi \\ \rho v_x v_z - B_x B_z/4\pi \\ (\varepsilon+P+\mathbf{B}^2/8\pi)v_x - B_x(\mathbf{v}\cdot\mathbf{B})/4\pi \\ 0 \\ (v_x B_y - v_y B_x) \\ -(v_z B_x - v_x B_z) \end{pmatrix}$$

$$+ \frac{\partial}{\partial y}\begin{pmatrix} \rho v_y \\ \rho v_x v_y - B_x B_y/4\pi \\ \rho v_y^2 + P + \mathbf{B}^2/8\pi - B_y^2/4\pi \\ \rho v_y v_z - B_y B_z/4\pi \\ (\varepsilon+P+\mathbf{B}^2/8\pi)v_y - B_y(\mathbf{v}\cdot\mathbf{B})/4\pi \\ -(v_x B_y - v_y B_x) \\ 0 \\ (v_y B_z - v_z B_y) \end{pmatrix} + \frac{\partial}{\partial z}\begin{pmatrix} \rho v_z \\ \rho v_x v_z - B_x B_z/4\pi \\ \rho v_y v_z - B_y B_z/4\pi \\ \rho v_z^2 + P + \mathbf{B}^2/8\pi - B_z^2/4\pi \\ (\varepsilon+P+\mathbf{B}^2/8\pi)v_z - B_z(\mathbf{v}\cdot\mathbf{B})/4\pi \\ (v_z B_x - v_x B_z) \\ -(v_y B_z - v_z B_y) \\ 0 \end{pmatrix} = 0$$

(66)

where $\varepsilon = \rho v^2/2 + P/(\gamma-1) + \mathbf{B}^2/8\pi$ is the total energy and $\gamma$ is the ratio of specific heats. The Euler equations can be obtained from eqn. (66) by setting the magnetic fields to zero.

The first five components of eqn. (66) follow a straightforward conservation form and their one-step update from a time $t^n$ to a time $t^{n+1} = t^n + \Delta t$ in a zone labeled by subscripts "$i,j,k$" is given by

$$\bar{U}_{i,j,k}^{n+1} = \bar{U}_{i,j,k}^{n} - \frac{\Delta t}{\Delta x}\left(\bar{F}_{i+1/2,j,k} - \bar{F}_{i-1/2,j,k}\right) - \frac{\Delta t}{\Delta y}\left(\bar{G}_{i,j+1/2,k} - \bar{G}_{i,j-1/2,k}\right) - \frac{\Delta t}{\Delta z}\left(\bar{H}_{i,j,k+1/2} - \bar{H}_{i,j,k-1/2}\right)$$

(67)

The overbars in eqn. (67) denote suitable averagings as will be detailed below. For eqn. (67) to be a high order update, the fluxes in eqn. (67) have to be averaged in space and time at the zone faces. These averages have to be obtained using quadratures having the appropriate accuracy. Traditionally, this has been obtained by solving a large number of Riemann problems at a large number of quadrature points, see Cockburn & Shu [16]. This makes the time-update very expensive. A substantially simpler strategy was presented by Dumbser et al [29] which views the flux at a face as being a linear combination of four vectors. The four vectors are : a) the conserved variables to the left of the zone boundary given by $U_{L;\,i+1/2,j,k}(y,z,t) = u_{i,j,k}(\xi=1/2,\eta,\zeta,\tau)$, b) the conserved variables to the right of the zone boundary given by $U_{R;\,i+1/2,j,k}(y,z,t) = u_{i+1,j,k}(\xi=-1/2,\eta,\zeta,\tau)$, c) the flux to the left of the zone boundary given by $F_{L;\,i+1/2,j,k}(y,z,t) = f_{i,j,k}(\xi=1/2,\eta,\zeta,\tau)\,\Delta x/\Delta t$ and d) the flux to the right of the zone boundary given by $F_{R;\,i+1/2,j,k}(y,z,t) = f_{i+1,j,k}(\xi=-1/2,\eta,\zeta,\tau)\,\Delta x/\Delta t$. Let us illustrate this for the HLL flux at any general point on the boundary "$i+1/2,j,k$". Consider a situation where the fastest left-going and right-going signal speeds at that boundary are $\lambda_L$ and $\lambda_R$ respectively. In the usual way, we reset $\lambda_L = \min(\lambda_L, 0)$ and $\lambda_R = \max(\lambda_R, 0)$. The HLL flux at any general point on the top x-face of the zone being considered is then given by

$$F_{i+1/2,j,k}(y,z,t) = \left[\frac{\lambda_R}{\lambda_R - \lambda_L}\right] F_{L;\,i+1/2,j,k}(y,z,t) - \left[\frac{\lambda_L}{\lambda_R - \lambda_L}\right] F_{R;\,i+1/2,j,k}(y,z,t)$$
$$+ \left[\frac{\lambda_R \lambda_L}{\lambda_R - \lambda_L}\right] \left(U_{R;\,i+1/2,j,k}(y,z,t) - U_{L;\,i+1/2,j,k}(y,z,t)\right)$$

(68)

The flux $\bar{F}_{i+1/2,j,k}$ that is used in eqn. (67) is an average of the flux in eqn. (68) where the averaging process is applied to the whole zone boundary being considered. The central idea of Dumbser et al [29] consists of freezing $\lambda_L$ and $\lambda_R$ to equal their values evaluated at the space-time barycenters of the face under consideration. As a result, the square brackets in eqn. (68) also become constants. This is tantamount to assuming that the same dissipation model holds at all space-time points at the face being considered. With that assumption, eqn. (68) becomes a linear function in the four vectors $U_{L;\,i+1/2,j,k}$, $U_{R;\,i+1/2,j,k}$

, $F_{L;\,i+1/2,j,k}$ and $F_{R;\,i+1/2,j,k}$. Notice from eqn. (51) as well as its explicit instantiation at third order in eqn. (61) that a space-time averaging of $U_{L;\,i+1/2,j,k}$ and $U_{R;\,i+1/2,j,k}$ is easily done by using the ADER scheme's space-time representation of $u$ in the two zones that abut the boundary "$i+1/2,j,k$". The ADER scheme also provides a space-time representation of the fluxes, making it possible to obtain the space-time averages of $F_{L;\,i+1/2,j,k}$ and $F_{R;\,i+1/2,j,k}$. Consequently, eqn. (68) can be averaged over the upper x-face of the zone "$i,j,k$" by integrating over the limits $[-\Delta y/2, \Delta y/2] \times [-\Delta z/2, \Delta z/2] \times [0, \Delta t]$ and dividing the integral by $\Delta y\, \Delta z\, \Delta t$. Please recall that the non-evolutionary terms in the magnetic field reconstruction (see last paragraph in Sub-Section 2.3) also contribute to $U_{L;\,i+1/2,j,k}$, $U_{R;\,i+1/2,j,k}$, $F_{L;\,i+1/2,j,k}$ and $F_{R;\,i+1/2,j,k}$. This completes our description of the one-step update for the mass, momentum and energy densities in eqn. (66).

As first shown by Yee [59], a divergence-free evolution of the magnetic field requires that one has a face-centered representation of the magnetic fields that is updated using an edge-centered representation of the electric fields. As shown by Balsara & Spicer [11], setting $\mathbf{E} = -\mathbf{v} \times \mathbf{B}$ shows us that specific components of the fluxes in eqn. (66) are indeed the electric fields that one seeks. This enables us to use the upwinded fluxes evaluated at the zone edges to obtain those components of the electric fields, as shown in Balsara [5]. Thus we have a one-step update for the magnetic fields given by

$$\overline{B}_{x;\,i+1/2,j,k}^{n+1} = \overline{B}_{x;\,i+1/2,j,k}^{n} - \frac{\Delta t}{2\Delta y \Delta z}\left(\Delta z \overline{E}_{z;\,i+1/2,j+1/2,k} - \Delta z \overline{E}_{z;\,i+1/2,j-1/2,k} + \Delta y \overline{E}_{y;\,i+1/2,j,k-1/2} - \Delta y \overline{E}_{y;\,i+1/2,j,k+1/2}\right)$$

$$\overline{B}_{y;\,i,j-1/2,k}^{n+1} = \overline{B}_{y;\,i,j-1/2,k}^{n} - \frac{\Delta t}{2\Delta x \Delta z}\left(\Delta x \overline{E}_{x;\,i,j-1/2,k+1/2} - \Delta x \overline{E}_{x;\,i,j-1/2,k-1/2} + \Delta z \overline{E}_{z;\,i-1/2,j-1/2,k} - \Delta z \overline{E}_{z;\,i+1/2,j-1/2,k}\right)$$

$$\overline{B}_{z;\,i,j,k+1/2}^{n+1} = \overline{B}_{z;\,i,j,k+1/2}^{n} - \frac{\Delta t}{2\Delta x \Delta y}\left(\Delta x \overline{E}_{x;\,i,j-1/2,k+1/2} - \Delta x \overline{E}_{x;\,i,j+1/2,k+1/2} + \Delta y \overline{E}_{y;\,i+1/2,j,k+1/2} - \Delta y \overline{E}_{y;\,i-1/2,j,k+1/2}\right)$$

(69)

Just as the fluxes in eqn. (67) are space-time averages over the zone faces, the electric fields to be used in eqn. (69) are space-time averages over the zone edges. As before, the ADER formulation can be used to obtain these averages. Notice that four faces come together at each zone edge. The Riemann problems that furnish the electric fields at the space-time center of the edge of interest are solved at space-time points within each face that are closest to the edge center, see Fig. 1 from Balsara [5]. The actual electric field at each edge is the arithmetic average of the electric field contributions from each of the four faces that come together at that edge. This completes our description of the one-step update for the magnetic fields in eqn. (1).

Balsara & Spicer [11] realized that the correct amount of upwinding for the electric fields in (69) could be a delicate issue, a topic that has also been addressed by Londrillo & DelZanna [43]. Notice that the electric fields are picked out by examining the last three components of the flux in eqn. (68). When HLL fluxes are used, an extremely simple solution to this issue is obtained by doubling the value of the third square bracket in eqn. (68). Such a doubling should only be done when using eqn. (68) to evaluate electric fields and, that too, only when $\lambda_L < 0 < \lambda_R$. In all other instances, eqn. (68) can be used straightforwardly to obtain the electric fields in eqn. (69).

**4.2) Step-by-step Description of the ADER-WENO Scheme**

Here we provide a step-by-step description of one time step of the ADER-WENO scheme presented in this paper.

1) Use the WENO formulae presented in Sub-Section 2.1 and 2.2 to obtain the moments of the face-centered magnetic field components in eqns. (43), (44) and (45). Do this without recourse to characteristic interpolation. Once the facial moments are obtained, use them to reconstruct the magnetic field within all the zones in of the mesh. This can be accomplished using the formulae in Balsara [6]. This step also gives us a zone-centered mean magnetic field evaluated with the requisite accuracy.

2) Use the WENO reconstruction formulae presented in Sub-Section 2.1 and 2.2 to obtain the moments of eqn. (4) for each of the zone-centered quantities. For obtaining the moments in each dimension, as described in Sub-Section 2.1, we used reconstruction in characteristic space. The cross-term reconstruction, described in Sub-Section 2.2, was carried out directly in the space of conserved variables to keep the scheme inexpensive. For some of the most stringent test problems we also followed the suggestion of Dumbser & Käser (2007) and reconstructed the moments from Sub-Section 2.1 twice, once in characteristic space and once in the space of conserved variables, and took the smaller of those moments. This helps stability without damaging the order property. The double reconstruction, applied only to the moments that are reconstructed in a dimension-by-dimension fashion, adds very little to the computational complexity of the scheme.

3) Use the WENO formulae presented in Sub-Section 2.1 and 2.2 to obtain the moments of the face-centered magnetic field components in eqns. (43), (44) and (45). We are now in a position to carry out this reconstruction in characteristic space. We now use the facial moments to reconstruct the magnetic field within all the zones in of the mesh using the formulae in Balsara [6].

4) Apply the flattening algorithm from Appendix A if that is desired.

5) Use the ADER-CG scheme that is detailed in Sub-Section 3.2 and Appendices B and C to obtain the space-time representation of the flow variables within each zone. The number of ADER iterations was always equated to the order of the scheme, i.e. we used the minimum permissible number of ADER iterations for the time-update. As a result, we used two, three and four ADER iterations for the second, third and fourth order schemes respectively. We followed this practice for all the test problems presented in this paper. We have never seen the need for using more than the minimum number of ADER iterations in our simulations and a good reason for that, based on the Picard iteration, was presented in Dumbser et al [26].

6) Obtain the space-time averaged values of the fluxes in eqn. (67). Similarly, obtain the space-time averaged values of the electric fields in eqn. (69).

7) Make the one-step updates described in eqns. (67) and (69).

Notice that after step 5) above we obtain not just the space-time representation of the conserved variable but also all the fluxes. We have the option of storing all the flux information. That option does add to the memory usage, but yields a faster scheme. One also has the option of discarding the flux information and rebuilding it for step 6) as and when it is needed. This yields an ADER scheme that uses memory much more economically. We have chosen the latter approach in the schemes presented here.

The second order ADER-WENO scheme for MHD simulations uses characteristic reconstruction and updates ~31,000 zones per second in three dimensions on a single core mid-grade Intel processor. This makes it very cost-effective relative to modern, sophisticated second order TVD schemes which also use characteristic reconstruction. The third order ADER-WENO scheme has a computational complexity that is 2.5 times that of the second order scheme. Likewise, the fourth order ADER-WENO scheme has a computational complexity that is only 3 times that of the third order scheme. The examples presented in this paper will show that the increased computational complexity of higher order schemes is easily offset by their increased accuracy. It is also worth pointing out that all the schemes presented here use the ADER time update and are considerably less expensive than their counterparts that use a Runge-Kutta time update strategy, see Balsara [6].

**5) Order Property**

The schemes presented here easily pass all the standard one dimensional tests for demonstrating order of accuracy. Thus we prefer to focus on two and three dimensional demonstrations of the order of accuracy in this section. All the two dimensional tests were run with a CFL number of 0.45 and all the three dimensional tests were run with a CFL number of 0.3 . A linearized Riemann solver for MHD, of the type presented in Balsara [1,4] was used for all the tests in this section. All of the results presented in this section use Balsara's RIEMANN code for astrophysical fluid dynamics.

It is also worthwhile making a note of the reconstruction used for the second order scheme that we present in this paper. Following Balsara [5] we used the slopes from the r=3 WENO reconstruction of Jiang & Shu [40] for our second order scheme. As a result, the slopes have one more order of accuracy than the accuracy that would be furnished by a TVD-preserving limiter. This yields a very superior second order scheme. It would be very difficult for a basic second order scheme to obtain the same accuracies as the second order scheme presented here.

**5.1) Unmagnetized Isentropic Vortex in Two Dimensions**

In the unmagnetized vortex problem, presented by Balsara & Shu [9], an isentropic vortex propagates at 45° to the grid lines in a domain with periodic boundaries given by [-5, 5] x [-5, 5]. As the original test problem was set up for the Euler equations, the magnetic field in all three directions is initialized to zero. The unperturbed flow at the initialial time can be written as $(\rho, P, v_x, v_y, B_x, B_y, B_z) = (1, 1, 1, 1, 0, 0, 0)$. The ratio of the specific heats is given by $\gamma = 1.4$. The entropy and the temperature are defined as $S = P/\rho^\gamma$ and $T = P/\rho$. The vortex is set up as a fluctuation of the unperturbed flow with the fluctuations given by

$$(\delta v_x, \delta v_y) = \frac{\varepsilon}{2\pi} e^{0.5(1-r^2)}(-y, x)$$
$$\delta T = -\frac{(\gamma-1)\varepsilon^2}{8\gamma\pi^2} e^{(1-r^2)}$$
$$\delta S = 0$$

Its strength is controlled by the parameter $\varepsilon$ which we set to $\varepsilon = 5$ according to Balsara & Shu [9]. r is the radius from the origin of the domain and can be written as $r^2 = x^2 + y^2$. Please note that the problem has to be initialized in each zone using numerical quadrature and that the accuracy of the quadrature formulae should match that of the numerical scheme being used. Also notice that the exponential function in the velocity and temperature fluctuations above ensures that the fluctuations are quite close to zero at the domain boundaries. However, for the fourth order scheme the domain is increased to [- 10, 10] x [-10, 10] due to the fact that the nonzero values of the exponential function at the boundaries are picked up by the fourth order scheme on the smaller domain. The stopping time was set to 10 time units for the second and third order schemes and to 20 time units for the fourth order scheme because of the bigger domain. The stopping time was chosen so that the vortex has completed one periodic passage through the computational domain.

TABLE I

| Method | Number of zones | $L_1$ error | $L_1$ order | $L_\infty$ error | $L_\infty$ order |
|---|---|---|---|---|---|
| $2^{nd}$ order ADER CG | 32×32 | $5.1124900 \times 10^{-3}$ | | $1.1677400 \times 10^{-1}$ | |
| | 64×64 | $1.0527400 \times 10^{-3}$ | 2.28 | $2.3322500 \times 10^{-2}$ | 2.32 |
| | 128×128 | $2.2522500 \times 10^{-4}$ | 2.22 | $4.6105000 \times 10^{-3}$ | 2.34 |
| | 256×256 | $5.4364900 \times 10^{-5}$ | 2.05 | $1.0438700 \times 10^{-3}$ | 2.14 |
| $3^{rd}$ order ADER CG | 32×32 | $3.9555500 \times 10^{-3}$ | | $9.5757200 \times 10^{-2}$ | |
| | 64×64 | $6.4692800 \times 10^{-4}$ | 2.61 | $1.3762400 \times 10^{-2}$ | 2.80 |
| | 128×128 | $7.6747300 \times 10^{-5}$ | 3.08 | $1.9531200 \times 10^{-3}$ | 2.82 |
| | 256×256 | $9.3029100 \times 10^{-6}$ | 3.04 | $2.4996400 \times 10^{-4}$ | 2.97 |
| $4^{th}$ order ADER CG | 32×32 | $4.5318300 \times 10^{-3}$ | | $2.7546100 \times 10^{-1}$ | |
| | 64×64 | $4.7962700 \times 10^{-4}$ | 3.24 | $3.1474100 \times 10^{-2}$ | 3.13 |
| | 128×128 | $2.3561700 \times 10^{-5}$ | 4.35 | $1.6096600 \times 10^{-3}$ | 4.29 |
| | 256×256 | $8.7922100 \times 10^{-7}$ | 4.74 | $7.2832400 \times 10^{-5}$ | 4.47 |

Table I shows the accuracy analysis for the second, third and fourth order schemes presented here. The errors were measured using the density variable. All three methods meet the expected order of accuracy even for a small number of zones. The third order ADER-WENO scheme obtains an $L_1$ error norm at 128x128 zones which is comparable to the second order ADER-WENO at 256x256 zones, which demonstrates the advantage of a high order scheme. The fourth order scheme cannot be directly compared to the third and second order schemes because of our use of a much larger computational domain. We do see though that the fourth order scheme also meets its design accuracy.

**5.2) Magnetized Isodensity Vortex in Two Dimensions**

The magnetized isodensity vortex problem described in Balsara [5] consists of a magnetized vortex moving across a domain given by [-5, 5] x [-5, 5] at an angle of 45° for a time of 10 units. As before, for the fourth order scheme the domain is increased to [-10, 10] x [-10, 10] and the simulation time is increased to 20 units. Periodic boundaries are used for the domain and it is initialized with an unperturbed flow of $(\rho, P, v_x, v_y, B_x, B_y) = (1, 1, 1, 1, 0, 0)$. The ratio of the specific heat is set to $\gamma = 5/3$. The vortex is set up as a fluctuation of the unperturbed flow in the velocities and the magnetic field given by:

$$(\delta v_x, \delta v_y) = \frac{\kappa}{2\pi} e^{0.5(1-r^2)}(-y, x)$$

$$(\delta B_x, \delta B_y) = \frac{\mu}{2\pi} e^{0.5(1-r^2)}(-y, x)$$

According to Balsara [5] the pressure fluctuation can be written as

$$\delta P = \frac{1}{8\pi}(\frac{\mu}{2\pi})^2 (1-r^2) e^{(1-r^2)} - \frac{1}{2}(\frac{\kappa}{2\pi})^2 e^{(1-r^2)}$$

and the density is set to unity.

TABLE II

| Method | Number of zones | $L_1$ error | $L_1$ order | $L_\infty$ error | $L_\infty$ order |
|---|---|---|---|---|---|
| 2nd order ADER CG | 32×32 | $7.8294900 \times 10^{-3}$ | | $1.2119700 \times 10^{-1}$ | |
| | 64×64 | $2.2175500 \times 10^{-3}$ | 1.82 | $3.0823400 \times 10^{-2}$ | 1.98 |
| | 128×128 | $5.4236600 \times 10^{-4}$ | 2.03 | $6.8924200 \times 10^{-3}$ | 2.16 |
| | 256×256 | $1.3477000 \times 10^{-4}$ | 2.01 | $1.6531500 \times 10^{-3}$ | 2.06 |
| 3rd order ADER CG | 32×32 | $5.5966400 \times 10^{-3}$ | | $1.0136700 \times 10^{-1}$ | |
| | 64×64 | $9.7810500 \times 10^{-4}$ | 2.51 | $1.7964500 \times 10^{-2}$ | 2.50 |
| | 128×128 | $1.2692200 \times 10^{-4}$ | 2.95 | $2.3763700 \times 10^{-3}$ | 2.92 |
| | 256×256 | $1.5983600 \times 10^{-5}$ | 2.99 | $2.9869600 \times 10^{-4}$ | 2.99 |
| 4th order ADER CG | 32×32 | $5.3198700 \times 10^{-3}$ | | $2.8849000 \times 10^{-1}$ | |
| | 64×64 | $4.7436200 \times 10^{-4}$ | 3.48 | $3.1605200 \times 10^{-2}$ | 3.19 |
| | 128×128 | $1.7658600 \times 10^{-5}$ | 4.75 | $8.9386200 \times 10^{-4}$ | 5.14 |
| | 256×256 | $1.0736400 \times 10^{-6}$ | 4.04 | $5.4681200 \times 10^{-5}$ | 4.03 |

Table II shows the error measured in the x-component of the magnetic field. All three schemes meet the design order of accuracy even at a small number of zones. As in the previous test problem we see that the third order scheme at 128x128 zone resolution shows the same $L_1$ error as the second order scheme at 256x256 zone resolution. This illustrates the utility and cost-effectiveness of the higher order schemes because the third order scheme easily offsets its slightly greater computational complexity (relative to the second order scheme) by delivering a comparably accurate solution on a mesh that has half as many zones in each direction.

**5.3) Torsional Alfven Wave Propagation in Three Dimensions**

The previous test problems used flows that were exact, equilibrium structures of the governing equations. While torsional Alfven waves also satisfy the governing equations, they are susceptible to parametric instabilities. These instabilities exist at low values of plasma-β, see Goldstein [34] and Del Zanna et al [23], and also at high values of plasma-β, see Jayanti & Hollweg [38]. The instabilities can of course be suppressed by numerical dissipation and all schemes have such numerical dissipation. As a result, second order schemes do not show these instabilities till the Alfven wave is very highly resolved. However, higher order schemes can pick up on the slightest amount of numerical noise and propagate it as a true fluctuation. Since the torsional Alven waves are susceptible to physical growth of such fluctuations, they will be treated as such by the numerical scheme. To avoid such deleterious effects, we carry out this test problem at very high values of plasma-β where the growth of such instabilities is suppressed.

The problem consists of initializing a torsional Alfven wave along the $x'$ axis of an ($x'$, $y'$, $z'$) coordinate system with the following variables

$$\rho = 1 \; , \; P = 1000 \; , \; \Phi = \frac{2\pi}{\lambda}\left(x' - 2\,t\right)$$
$$v_{x'} = 1 \; , \; v_{y'} = \varepsilon \cos \Phi \; , \; v_{z'} = \varepsilon \sin \Phi$$
$$B_{x'} = \sqrt{4\pi\rho} \; , \; B_{y'} = -\varepsilon\sqrt{4\pi\rho} \cos \Phi \; , \; B_{z'} = -\varepsilon\sqrt{4\pi\rho} \sin \Phi$$

where we take $\varepsilon = 0.02$ and $\lambda = \sqrt{3}$. We utilize the magnetic vector potential when initializing the magnetic field in a divergence-free fashion on a three dimensional mesh. The magnetic vector potential is given by

$$A_{x'} = 0, \quad A_{y'} = \varepsilon \lambda \sqrt{\rho/\pi} \cos \Phi, \quad A_{z'} = \sqrt{4\pi\rho} \, y' + \varepsilon \lambda \sqrt{\rho/\pi} \sin \Phi$$

Please note that the magnetic vector potential has to be assigned to each zone's edges using numerical quadrature. Also note that the accuracy of the quadrature formula should match the accuracy of the scheme. An application of Stokes law in integral form at each face then yields the magnetic field component in that face.

The actual problem is solved on a unit cube with periodic boundary conditions in the (x,y,z) coordinate frame which is rotated relative to the (x', y', z') coordinate system. The rotation matrix is given by **A** so that we have

$$\mathbf{A} = \begin{bmatrix} \cos\psi \cos\phi - \cos\theta \sin\phi \sin\psi & \cos\psi \sin\phi + \cos\theta \cos\phi \sin\psi & \sin\psi \sin\theta \\ -\sin\psi \cos\phi - \cos\theta \sin\phi \cos\psi & -\sin\psi \sin\phi + \cos\theta \cos\phi \cos\psi & \cos\psi \sin\theta \\ \sin\theta \sin\phi & -\sin\theta \cos\phi & \cos\theta \end{bmatrix}$$

where $\phi = -\pi/4$, $\theta = \sin^{-1}\left(-\sqrt{2/3}\right)$ and $\psi = \sin^{-1}\left(\left(\sqrt{2}-\sqrt{6}\right)/4\right)$. As a result, the position vector $\mathbf{r}'$ in the primed frame transforms to the position vector $\mathbf{r}$ in the unprimed frame as $\mathbf{r} = \mathbf{A} \, \mathbf{r}'$. Other vectors transform similarly. The effect of the rotation is to make the wave propagate along the diagonal of the unit cube. The wave propagates at a speed of 2 units and the problem is stopped at a time of $\sqrt{3}/2$ by which time it has propagated once around the unit cube.

Table III

| Method | Number of zones | $L_1$ error | $L_1$ order | $L_\infty$ error | $L_\infty$ order |
|---|---|---|---|---|---|
| 2nd order ADER CG | 8×8×8 | $3.46872 \times 10^{-2}$ | | $5.17784 \times 10^{-2}$ | |
| | 16×16×16 | $2.13576 \times 10^{-2}$ | 0.70 | $3.37639 \times 10^{-2}$ | 0.62 |
| | 32×32×32 | $4.21518 \times 10^{-3}$ | 2.34 | $6.61278 \times 10^{-3}$ | 2.35 |
| | 48×48×48 | $1.42618 \times 10^{-3}$ | 2.67 | $2.23978 \times 10^{-3}$ | 2.67 |
| 3rd order ADER CG | 8×8×8 | $3.56780 \times 10^{-2}$ | | $5.33670 \times 10^{-2}$ | |
| | 16×16×16 | $1.67602 \times 10^{-2}$ | 1.09 | $2.58530 \times 10^{-2}$ | 1.05 |
| | 32×32×32 | $2.62382 \times 10^{-3}$ | 2.68 | $4.13702 \times 10^{-3}$ | 2.64 |
| | 48×48×48 | $7.92737 \times 10^{-4}$ | 2.95 | $1.25713 \times 10^{-3}$ | 2.94 |
| 4th order ADER CG | 8×8×8 | $2.64017 \times 10^{-2}$ | | $3.99367 \times 10^{-2}$ | |
| | 16×16×16 | $1.31645 \times 10^{-3}$ | 4.32 | $2.04612 \times 10^{-3}$ | 4.29 |
| | 32×32×32 | $6.06938 \times 10^{-5}$ | 4.44 | $1.02099 \times 10^{-4}$ | 4.33 |
| | 48×48×48 | $1.30835 \times 10^{-5}$ | 3.79 | $2.42201 \times 10^{-5}$ | 3.55 |

Table III shows the accuracy analysis for the second, third and fourth order schemes presented here. The errors were measured using the x-component of the magnetic field. All three methods meet the expected order of accuracy even for a small number of zones. Compared to the second and third order schemes we see that the fourth order scheme has reached a very high accuracy of one part in $10^5$ on the 48×48×48 zone mesh. The fourth order scheme shows a slight evidence for parametric instability at 48×48×48 zone resolution since it has picked up extremely tiny, numerically generated errors in the pressure and propagated them. The second and third order schemes never reach the same small value of the error on the meshes that are displayed but on very high resolution meshes we have been able to verify that they too pick up slight traces of the parametric instability.

Since all three schemes were run on the same problem, we can cross-compare the errors in the second, third and fourth order schemes using this accuracy analysis presented in Table III. Notice that on a resolution starved mesh, such as the 16x16x16 mesh in Table III the fourth order scheme offers almost an order of magnitude improvement in accuracy over the second and third order schemes. We also see that the fourth order scheme at 16x16x16 zone resolution is already as accurate as the second order scheme at 48x48x48 zone resolution. The 48x48x48 zone calculation at second order takes 12 times longer to complete than the 16x16x16 zone calculation at fourth order, thus illustrating the advantage of using a higher order scheme.

## 5.4) Density Wave Propagation in Three Dimensions

This test problem consists of propagating a density wave with a sinusoidal profile along the diagonal of the same unit cube that was described in the previous sub-section. Now the parameters in the $(x', y', z')$ coordinate system are given by

$$\rho = 1 + \varepsilon \sin \Phi \ , P = 1 \ , \ \Phi = \frac{2\pi}{\lambda}(x' - t) \ ,$$
$$v_{x'} = 1 \ , v_{y'} = 0 \ , v_{z'} = 0 \ , B_{x'} = 0 \ , B_{y'} = 0 \ , B_{z'} = 0$$

where we take $\varepsilon = 0.2$ and $\lambda = \sqrt{3}$. The density profile and velocities are then rotated into a periodic unit cube using the rotation matrix described in the previous Sub-section. The problem is stopped at a time of $\sqrt{3}$ by which time the density wave has propagated once around the unit cube.

Table IV

| Method | Number of zones | $L_1$ error | $L_1$ order | $L_\infty$ error | $L_\infty$ order |
|---|---|---|---|---|---|
| 2nd order ADER CG | 8×8×8 | 6.09811 × 10$^{-2}$ | | 9.64241 × 10$^{-2}$ | |
| | 16×16×16 | 1.58837 × 10$^{-2}$ | 1.94 | 2.43894 × 10$^{-2}$ | 1.98 |
| | 32×32×32 | 3.63924 × 10$^{-3}$ | 2.13 | 5.69284 × 10$^{-3}$ | 2.10 |
| | 48×48×48 | 1.58011 × 10$^{-3}$ | 2.06 | 2.47718 × 10$^{-3}$ | 2.05 |
| 3rd order ADER CG | 8×8×8 | 5.30213 × 10$^{-2}$ | | 8.25208 × 10$^{-2}$ | |
| | 16×16×16 | 9.48506 × 10$^{-3}$ | 2.48 | 1.37539 × 10$^{-2}$ | 2.59 |
| | 32×32×32 | 1.29720 × 10$^{-3}$ | 2.87 | 2.07369 × 10$^{-3}$ | 2.73 |
| | 48×48×48 | 3.95625 × 10$^{-4}$ | 2.93 | 5.80456 × 10$^{-4}$ | 3.14 |
| 4th order ADER CG | 8×8×8 | 1.76010 × 10$^{-2}$ | | 2.90944 × 10$^{-2}$ | |
| | 16×16×16 | 4.50487 × 10$^{-4}$ | 5.29 | 8.94523 × 10$^{-4}$ | 5.02 |
| | 32×32×32 | 1.56149 × 10$^{-5}$ | 4.85 | 3.61468 × 10$^{-5}$ | 4.63 |
| | 48×48×48 | 2.50965 × 10$^{-6}$ | 4.51 | 6.80014 × 10$^{-6}$ | 4.12 |

Table IV shows the accuracy analysis for the second, third and fourth order schemes presented here. The errors were measured using the density variable. All three methods meet the expected order of accuracy even for a small number of zones. Since all three schemes were run on the same problem, we can cross-compare the errors in the second, third and fourth order schemes using this accuracy analysis. We see that the fourth order scheme at 16x16x16 zone resolution provides a more accurate result that the second order scheme at 48x48x48 zone resolution, again underscoring the advantages of using higher order schemes.

**6) Hydrodynamical Test Problems**

In this section we present several stringent hydrodynamical test problems. The schemes we have presented operate accurately and robustly on all of these problems. This illustrates the utility of our methods for simulating hydrodynamical flows. The RIEMANN code was used for all these tests.

**6.1) Interacting Blast Problem in One Dimension**

The interacting blast problem was presented by Woodward and Colella [58]. We used the fourth order ADER-WENO scheme with the linearized Riemann solver to compute this problem using exactly the same parameters as the original authors of this problem. The CFL number was set to 0.8. Fig. 2 shows the density variable of a simulation with 400 zones as diamonds. The solid line is the converged density profile of a simulation using 1600 zones. We see that the left-going contact discontinuity is captured well in the simulation using 400 zones. We further note that all the flow structures in the 400 zone simulation are very close to the converged simulation.

**6.2) Shock-Entropy Wave Interaction in One Dimension**

The one dimensional shock-entropy wave interaction problem was first presented by Shu & Osher [51]. It consists of a Mach 3 shock interacting with a density disturbance. That generates a flow field that is a combination of discontinuities and smooth structures. Therefore the problem is a good model for the interactions occurring in simulations of compressible turbulences. Additionally it represents the amplification of entropy fluctuations as they pass through a shock. These interactions of smooth structures with shocks pose a problem for TVD schemes as the damaging effects of the TVD limiters are maximal in these cases. Jiang & Shu [40] made a detailed study showing that the r = 3 WENO scheme performs superior to a well-designed TVD scheme. They concluded that the r = 3 WENO scheme using 800 zones outperformed the TVD scheme using 2000 zones substantially. We computed the problem using several ADER-WENO schemes with the linearized Riemann solver at a CFL number of 0.8. To highlight the role of TVD limiters, we even ran a simulation using the MC limiter of vanLeer for the spatial interpolation and the ADER scheme for the time-evolution. The computational domain spans [-1, 1] and was set up with 200 zones. The initial condition is given by

$$(\rho_L, P_L, v_{x,L}) = (3.857143, 10.3333, 2.629369) \quad x < -0.8$$
$$(\rho_R, P_R, v_{x,R}) = (1 + 0.2\sin(5\pi x), 1, 0) \quad x > -0.8$$

The simulation was stopped at 0.47 time units.

Fig. 3 shows the density profile of the third and fourth order ADER-WENO schemes as well as the TVD scheme using 200 zones as diamonds and the reference solution, which was calculated on an 800 zones grid, as a solid line. We note that the density profile of the fourth order scheme has almost converged to the reference solution and shows all the extrema that are seen in the reference solution even though it uses a 200 zone grid. Furthermore we see that the scheme needs no more than 11 points between extrema in the density variable immediately after the shock. The third order scheme is quite close to the reference solution while the TVD scheme misses the reference solution by a wide margin. Therefore we note, that the solution of the ADER-WENO schemes converged to the reference solution using a small number of points. This shows that the third order ADER-WENO scheme converges faster to the reference solution than a lower order scheme and has the smaller error if the number of zones is kept constant.

**6.3) Resolution Study of the Forward Facing Step Problem in Two Dimensions**

This test problem was first presented by Woodward and Colella [58]. Cockburn and Shu [16] carried out a resolution study using schemes of increasing order of accuracy. Increasing the resolution enabled them to capture important details such as the roll up of the vortex sheet via Kelvin-Helmholtz instability. They also showed that more accurate schemes were able to capture the vortex sheet roll-up with smaller number of zones. Our purpose is to make a similar resolution study and to prove that the schemes are accurate and perform robustly on this stringent problem. We therefore simulated this test problem using the fourth order ADER-WENO scheme with increasing resolution as shown in Fig. 4.

The problem consists of a two-dimensional wind tunnel that spans a domain of [0, 3] x [0, 1]. A forward-facing step is set up at a location given by the coordinates (0.6,0.2). Inflow boundary conditions are applied at the left boundary, where the gas enters the wind tunnel at Mach 3.0 with a density of 1.4 and a pressure of unity. The right boundary is an outflow boundary. The walls are set to be reflective boundaries. The singularity at the corner was treated with the same technique that Woodward and Colella [58] suggested. The simulation was run until a time of 4.0 time units and the ratio of specific heats is given by 1.4.

Fig. 4 shows the density at the final time at a resolution of 240x80, 480x160 and 960x320. All of the three simulations were run with a fourth order scheme and a linearized Riemann solver. The CFL number was set to 0.4. All the shocks are properly captured on the computing grid and have sharp profiles. The vortex sheet that emanates from the Mach stem is correctly resolved with only a few zones across the sheet. We notice that the vortex sheet shows little or no spreading over the length of the computational domain. At a resolutions of 960x320 and 480x160 the roll up of the vortex sheet is clearly visible. An exceptionally good second order scheme would need at least a resolution of 960x320 zones to start showing evidence of the vortex sheet's roll-up. Such a second order scheme operating on this problem with a resolution of 960x320 zones would furnish the same solution quality as the fourth order scheme does at a resolution of 480x160. This demonstrates the ability of the high order schemes to provide a better resolution at a smaller number of zones.

## 6.4) Resolution Study of the Double Mach Reflection of a Strong Shock in Two Dimensions

This problem was presented by Woodward and Colella [58]. We use the exactly the same setup for the test problem as the authors did. A Mach 10 shock hits a reflecting wall which spreads from $x=1/6$ to $x=4$ at the bottom of the domain given by [0, 4] × [0, 1]. The angle between the shock and the wall is 60°. At the start of the computation the position of the shock is given by $(x, y) = (1/6, 0)$. The undisturbed fluid in front of the shock is initialized with a density of 1.4 and a pressure of 1. The exact post-shock condition is used for the bottom boundary from $x=0$ to $x=1/6$ to mimic an angled wedge. For the remaining boundary at the bottom of the domain we used a reflective boundary condition. The top boundary condition imposes the exact motion of a Mach 10 shock in the flow variables. The left and right boundaries are set to be inflow and outflow boundaries.

Fig. 5 shows the density variable at $t=0.2$ in [0, 3] × [0, 1] as in Woodward and Colella [58]. The upper panel shows a resolution of 960 × 240 zones, the second panel shows a resolution of 1920 × 480 zones. The two panels at the bottom show a blow-up of the region around the double Mach stem for both computations. All the plots show 30 contours equally distributed from $\rho=1.3965$ to $\rho=22.682$. We used the fourth order ADER-WENO scheme with an HLL Riemann solver for both simulations.

Notice that the fourth order ADER-WENO scheme resolves all the structures that are shown in Cockburn & Shu [16]. According to Cockburn & Shu [16] a second order scheme would need at least four times as many zones in each direction to resolve the instability and for such a simulation it would need more CPU time than the fourth order scheme. That demonstrates the efficiency of the higher order schemes presented here.

**7) MHD Test Problems**

We present several stringent MHD test problems in this section. The MHD schemes we have presented operate accurately and robustly on all of these problems showing the utility of our methods. The RIEMANN code for astrophysical simulations was used for all of these MHD tests.

**7.1) MHD Riemann Problems in One Dimension**

First we present one of the Riemann problems from Ryu & Jones [48]. It is set up on a 400-zone mesh spanning the domain given by [-0.5, 0.5]. The initial conditions are given by

$$(\rho_L, P_L, v_{x,L}, v_{y,L}, v_{z,L}, B_{y,L}, B_{z,L}) = (1.08, 0.95, 1.2, 0.01, 0.5, 3.6, 2.0) \quad x < 0$$
$$(\rho_R, P_R, v_{x,R}, v_{y,R}, v_{z,R}, B_{y,R}, B_{z,R}) = (1.0, 1.0, 0, 0, 0, 4.0, 2.0) \quad x > 0$$

The x-component of the magnetic field is given by $B_x = 2$. The simulation was stopped at a time of 0.2 and the ratio of specific heats was set to 5/3. As this is a non-coplanar problem it generates seven waves, which are a right-going fast shock, a right-going rotational discontinuity, a right-going slow shock, a contact discontinuity, a left-going slow-shock, a left-going rotational discontinuity and a left-going fast shock. Ryu & Jones [48] also provide the exact solution for this Riemann problem. We simulated the problem using the fourth order ADER-WENO scheme using an HLL Riemann solver and a CFL number of 0.8. Fig. 6 shows the density, the pressure, x-velocity, y-velocity, z-velocity and the y- and z-component of the magnetic field. All the shock profiles are properly captured within a few zones. We note that our high order scheme captures slow shocks with only a few zones across them. In Balsara [2] it was shown that profiles of slow shocks sometimes have a little more than the optimal number of zones across them if TVD schemes for MHD are used. In Fig. 6 we see that the fourth order ADER-WENO scheme resolves the slow shock as a sharp profile. Therefore we note that the representation of slow shocks is improved by high order schemes. We also see that the contact discontinuity and the rotational discontinuity profiles are captured with a few zones. Notice the small number of zones between the rotational discontinuity and the corresponding slow shock. The ability of the scheme to resolve every discontinuity as a sharp profile is necessary to distinguish the rotational discontinuity from the slow shock and to maintain a high accuracy.

Our next Riemann problem comes from Dai & Woodward [21]. It is set up on a 400-zone mesh spanning the domain given by [-0.5, 0.5]. The initial conditions are given by

$$(\rho_L, P_L, v_{x,L}, v_{y,L}, v_{z,L}, B_{y,L}, B_{z,L}) = (1.0, 1.0,\ 36.87,\ -0.155,\ -0.0386,\ 4.0,\ 1.0) \quad x < 0$$
$$(\rho_R, P_R, v_{x,R}, v_{y,R}, v_{z,R}, B_{y,R}, B_{z,R}) = (1.0,\ 1.0,\ -36.87,\ 0,\ 0,\ 4.0,\ 1.0) \qquad x > 0$$

The x-component of the magnetic field is given by $B_x = 4$. The simulation was stopped at a time of 0.03 and the ratio of specific heats was set to 5/3. The problem consists of two very high Mach number streams of magnetized fluid rushing towards each other. It can be thought of as the MHD equivalent of the Noh problem. The resolved state consists of two fast magnetosonic shocks of Mach number 25.5 propagating out of the interaction region. We simulated the problem using the fourth order ADER-WENO scheme using an HLL Riemann solver and a CFL number of 0.8. Fig. 7 shows the density, the pressure, x-velocity, y-velocity, z-velocity and the y- and z-component of the magnetic field. All the shock profiles are properly captured within a few zones and do not display any post-shock oscillations. This problem, along with a few other problems presented in this section, demonstrates that higher order schemes can successfully tackle problems with very strong shocks while simultaneously giving us the advantages of high resolution, high accuracy and low numerical dissipation.

## 7.2) Numerical Dissipation and Long-Term Decay of Alfven Waves in Two Dimensions

In several fields, like astrophysics or space physics, one is interested in the evolution of waves to simulate certain problems such as turbulence. The Alfven wave decay test problem first presented by Balsara [5] examines the dissipation of torsional Alfven waves in two dimensions. In this test problem torsional Alfven waves propagate at an angle of $\tan^{-1}(1/r) = \tan^{-1}(1/6) = 9.462°$ to the y-axis through a domain given by [-r/2, r/2] x [-r/2, r/2] with r = 6. The domain was set up with 120 x 120 zones and has periodic boundary conditions. The pressure and density are uniformly initialized as $P_0 = 1$ and $\rho_0 = 1$. The unperturbed velocity and unperturbed magnetic field are given by $v_0 = 0$ and $B_0 = 1$. The amplitude of the Alfven waves is parametrized by a velocity fluctuation $\varepsilon$, which is set to 0.2. Different test problems can be set up by changing these values. The simulation was stopped at 129 time units by which time the waves had crossed the domain several times. The CFL number was set to 0.4. The direction of the wave propagation along the unit vector can be written as

$$\hat{n} = n_x \hat{i} + n_y \hat{j} = \frac{1}{\sqrt{r^2+1}} \hat{i} + \frac{r}{\sqrt{r^2+1}} \hat{j}.$$

The phase of the waves is given by

$$\phi = \frac{2\pi}{n_y}(n_x x + n_y y - V_A t), \text{ where } V_A = \frac{B_0}{\sqrt{4\pi\rho_0}}.$$

The velocity is given by

$$\mathbf{v} = (v_0 n_x - \varepsilon n_y \cos\phi)\hat{i} + (v_0 n_y - \varepsilon n_x \cos\phi)\hat{j} + \varepsilon \sin\phi \hat{k}.$$

The magnetic field is given by

$$\mathbf{B} = (B_0 n_x + \varepsilon n_y \sqrt{4\pi\rho_0} \cos\phi)\hat{i} + (B_0 n_y - \varepsilon n_x \sqrt{4\pi\rho_0} \cos\phi)\hat{j} - \varepsilon\sqrt{4\pi\rho_0}\sin\phi\hat{k}.$$

The corresponding vector potential is given by

$$\mathbf{A} = -\frac{\varepsilon\sqrt{4\pi\rho_0}}{2\pi}\cos\phi\hat{i} + (-B_0 n_y x + B_0 n_x y + \frac{\varepsilon n_y \sqrt{4\pi\rho_0}}{2\pi}\sin\phi)\hat{k}$$

and it is used to initialize the magnetic field.

The dissipation of the numerical scheme can be measured in the decay of the maximum values of the z-component of the velocity and the magnetic field. The r.m.s. values of the velocity and the magnetic field decay in the same fashion as the maximal values of these quantities do. For this reason they are not presented here. Kim et al. [41] showed that these plots give a good qualitative understanding of numerical viscosities and resistivities in a numerical scheme. In Fig 8 the maximum z component of the velocity and of the magnetic field are plotted at every time step in a log-linear plot. We used the HLL and linearized Riemann solvers with the second, third and fourth order ADER-WENO schemes. It can be seen that with increasing order of accuracy the numerical dissipation of the scheme reduces significantly independent of the Riemann solver that is used. This makes the higher order schemes more favorable for simulations of complex phenomena that involve wave propagation. Further it can be noted that the linearized Riemann solver is substantially less dissipative than the HLL Riemann solver in both measured quantities. But this effect decreases with increasing order of accuracy because the improved reconstruction significantly reduces the difference in flow variables at the zone boundaries where the Riemann problem is solved. We therefore see that it is acceptable to use less expensive Riemann solvers as the order of the scheme is increased.

**7.3) The Rotor Problem in Two Dimensions**

The two dimensional rotor problem was presented in Balsara & Spicer [11] and in Balsara [5]. Here we describe a version of this test problem. The computational mesh has 200 × 200 zones and spans the domain [-0.5,0.5]×[-0.5,0.5]. A dense and rapidly spinning cylinder is set up in the center of an initially stationary, light ambient fluid. The ambient fluid is initially static. A uniform magnetic field initially threads the two fluids. Its value is set to 2.5 units. The pressure in both fluids is set to unity. The density in the ambient fluid is uniformly set to unity, while the constant density in the rotor is 10 units out to a radius of 0.1. A linear taper is applied to the density between a radius of 0.1 and 0.13 so that the density in the rotor decreases linearly to the value of the density in the ambient fluid. Therefore the taper needs six zones to join the density of the two fluids. That number should be kept fixed if the resolution is increased or decreased. The initial angular velocity of the rotor is uniform out to a radius of 0.1. At this radius the toroidal velocity has a value of one unit. The toroidal velocity decreases linearly from one unit to zero between a radius of 0.1 and 0.13 so that it joins the velocity of the ambient fluid at a radius of 0.13. The ratio of specific heats is given by 5/3. The Courant number was set to 0.4. The fourth order ADER-WENO scheme with the linearized Riemann solver was applied to this problem. In Fig. 9 the density, the pressure, the Mach number and the magnitude of the magnetic field are shown at a time of 0.29 units. Balsara & Spicer [11] provided a detailed physical description of this problem. The results presented in Fig. 9 show a good consistency with the descriptions in Balsara & Spicer [11]. We therefore conclude that the multidimensional limiting presented in this paper works well for numerical MHD.

**7.4) The Blast Problem in Two Dimensions**

Balsara & Spicer [11] first presented the two dimensional blast problem. It is set up by following the prescription in Balsara & Spicer [11]. The fourth order ADER-WENO scheme with the HLL Riemann solver was applied to a mesh having 200 × 200 zones and covering the domain [-0.5,0.5]×[-0.5,0.5]. In Fig. 10 the logarithm (base 10) of the density, the logarithm (base 10) of the pressure, the magnitude of the velocity and the magnitude of the magnetic field are shown at a time of 0.01. A detailed physical description of the problem is given in Balsara & Spicer [11]. We see that the results that we present are consistent with that description. Notice that the plasma-β is 0.000251 in the ambient medium. An almost circular, fast magnetosonic shock propagates through the ambient plasma and it is the fastest wave structure in this problem. The propagation of this extremely strong shock at all angles to the initial magnetic field in the low-β ambient plasma makes this a challenging test problem. In Fig. 10 we see that the positivity of the pressure variable is maintained even in regions where the strong shock propagates obliquely to the mesh. This is a direct result of using the divergence-free reconstruction to obtain the volume-averaged magnetic fields at the zone centers. Therefore we conclude that the divergence-free reconstruction presented in §2.3 provides a significant improvement in the simulation of low-β plasmas.

**7.5) The Blast Problem in Three Dimensions**

The present problem extends the previous problem to three dimensions. The problem is initialized on the domain given by [-0.5,0.5]×[-0.5,0.5]×[-0.5,0.5] using a 151×151×151 zone mesh. The primitive variables are specified by

$$(\rho, P, v_x, v_y, v_z, B_x, B_y, B_z) = (1, 1000, 0, 0, 0, 1000/\sqrt{3}, 1000/\sqrt{3}, 1000/\sqrt{3}) \text{ for } r < 0.1$$
$$= (1, 0.1, 0, 0, 0, 1000/\sqrt{3}, 1000/\sqrt{3}, 1000/\sqrt{3}) \text{ for } r > 0.1$$

The ratio of specific heats is given by 1.4. The Courant number was set to 0.3 and the problem was run to a time of 0.01. The problem was run using a fourth order ADER-WENO scheme with an HLL Riemann solver. Please note that the present blast problem in three dimensions is substantially more stringent than similar blast problems that have been presented in the literature.

Fig. 11 shows the logarithm (base 10) of the density, the logarithm (base 10) of the pressure, the magnitude of the velocity and the magnitude of the magnetic field in the central xy-plane at the final time of the simulation. We see that an almost spherical fast magnetosonic shock propagates through the low-β ambient plasma. While this shock is a nearly infinite shock, the fourth order scheme handles it without problem. All structures are crisply captured and there is no sign of undue oscillations anywhere on the computational mesh. The pressure remains positive throughout the simulation showing the utility of the divergence-free reconstruction in the simulation of low-β plasmas.

**8) Conclusions**

We have presented a new class of ADER-WENO schemes for high order evolution of hyperbolic systems of conservation laws. The methods are very general and can be used for several hyperbolic systems. In the present paper we have applied them with success to Euler and MHD flows. Below we make a point-wise catalogue of the advances reported in this paper:

1) A very efficient finite volume WENO reconstruction strategy has been presented for structured meshes. We have shown that the most elegant and compact formulation of WENO reconstruction obtains when the interpolating functions are expressed in modal space. Explicit formulae have been developed for spatial reconstruction that go up to fourth order of accuracy.

2) The most essential aspects of divergence-free reconstruction of magnetic fields have been discussed in this paper. Further details for carrying out such a reconstruction have been reported in Balsara [6]. It is shown here that the reconstruction naturally furnishes all the moments of the magnetic field within a zone consistent with retaining a specified order of accuracy.

3) A general purpose flattener algorithm has been presented in Appendix A. The algorithm detects regions with strong shocks and suitably stabilizes higher order schemes in those regions.

4) ADER-CG schemes, especially as they are compactly formulated in modal space, are reported here. Sub-Section 3.1 presents a general purpose formulation that makes it possible to design ADER-CG schemes in modal space for structured and unstructured meshes. For structured meshes we have explicitly demonstrated that the modal formulation yields the most compact and elegant formulation. It is also worth mentioning that on unstructured meshes the use of Dubiner [25] bases yields a similarly compact and elegant formulation of ADER-CG schemes.

5) Sub-section 3.2 presents a detailed instantiation of the third order ADER-CG scheme. This is done with the intent of facilitating its easy implementation by other practitioners. Appendices B and C present the most essential details for ADER-CG schemes at second and fourth orders respectively.

6) The one-step update of the resultant ADER-WENO schemes makes them lower storage alternatives to the multi-stage Runge-Kutta time discretizations that have been used in the past. The ADER-WENO schemes also bypass the reconstruction step that is needed in each stage of the multi-stage Runge-Kutta time discretization, making them the more efficient alternative. The ADER-WENO schemes are also free of the Butcher barriers that seem to occur in Runge-Kutta time discretizations, see Spiteri & Ruuth [52].

7) The one-step update of the ADER-WENO schemes makes them desirable building blocks for AMR calculations.

8) Section 5 presents several examples showing that the ADER-WENO schemes meet their design accuracies in two and three dimensions for Euler and MHD flows.

9) Sections 6 and 7 present several stringent test problems in one, two and three dimensions. The tests span Euler and MHD flows. Several of our test problems are very demanding on the numerical scheme because they require an ability to capture delicate flow structures accurately in the presence of almost infinite shocks. The higher order schemes along with the divergence-free reconstruction strategies for treating magnetic fields that we have presented here perform very well on all of those tests.

10) It is shown that the increasing computational complexity with increasing order is handily offset by the increased accuracy of the scheme. The resulting ADER-WENO schemes are, therefore, very worthy alternatives to the standard second order schemes for compressible Euler and MHD flow.

**Acknowledgements**

DSB acknowledges support via NSF grant AST-0607731. DSB also acknowledges NASA grants HST-AR-10934.01-A, NASA-NNX07AG93G and NASA-NNX08AG69G. The majority of simulations were performed on PC clusters at UND but a few initial simulations were also performed at NASA-NCCS. TR thanks the Erich-Becker-Stiftung for support. MD was funded by the Deutsche Forschungsgemeinschaft (DFG) in the framework of the DFG Forschungsstipendium (DU 1107/1-1). CDM acknowledges funding by Deutsche Forschungsgemeinschaft.

**Appendix A) Flattening Algorithm in the Vicinity of Strong Shocks**

In this appendix we describe the flattening algorithm used in the vicinity of strong MHD shocks. As shown by Colella & Woodward [18] and Balsara [4], such algorithms are useful for producing practical higher order schemes with a broad range of good operation. We construct the undivided divergence of the velocity in each zone and call it $\Delta x \left( \nabla \cdot \mathbf{v} \right)_{i,j,k}$. The zones are labeled by a subscript "i,j,k" on a three dimensional mesh. In each zone we also construct the largest magnetosonic speed of the MHD waves relative to the mean flow in that zone and call it $\lambda_{i,j,k} \equiv \sqrt{\left(\gamma\, P_{i,j,k} + \mathbf{B}^2_{i,j,k}/(4\pi)\right)/\rho_{i,j,k}}$. In each zone we construct $\lambda_{m;\,i,j,k}$ which is the minimum of $\lambda_{i,j,k}$ in the zone of interest and the neighboring zones that abut it. Thus in two dimensions we evaluate $\lambda_{m;\,i,j,k}$ by scanning nine zones and in three dimensions we scan twenty seven zones. Making a comparison between $\Delta x\,(\nabla\bullet\mathbf{v})_{i,j,k}$ and $\lambda_{m;\,i,j,k}$ then enables us to detect strong shocks. We therefore construct the detector function as

$$d_{i,j,k} = \min\left(1,\ \text{abs}\left(\Delta x\,(\nabla\bullet\mathbf{v})_{i,j,k} + \delta\,\lambda_{m;\,i,j,k}\right)\Big/\left(\delta\,\lambda_{m;\,i,j,k}\right)\right)\,\mathrm{H}\!\left(-\left(\Delta x\,(\nabla\bullet\mathbf{v})_{i,j,k} + \delta\,\lambda_{m;\,i,j,k}\right)\right)$$

where H(x) is the Heaviside function and is unity for x>0 and zero for x<0. $\delta$ is a positive number that is set to be of order unity. Notice that the detector function $d_{i,j,k}$ is zero in the vicinity of smooth flow or even in the presence of moderately compressive shocks. It only deviates from 0 and goes smoothly to unity only in the vicinity of strongly compressive shocks. This threshold is important for retaining the order property. In some problems there might be a tendency for generating strong rarefactions which can also become problematical. In that case we can modify the above detector function to include rarefactions as

$$d_{i,j,k} = \min\left(1,\ \text{abs}\left(-\left|\Delta x\,(\nabla\bullet\mathbf{v})_{i,j,k}\right| + \delta\,\lambda_{m;\,i,j,k}\right)\Big/\left(\delta\,\lambda_{m;\,i,j,k}\right)\right)\,\mathrm{H}\!\left(-\left(-\left|\Delta x\,(\nabla\bullet\mathbf{v})_{i,j,k}\right| + \delta\,\lambda_{m;\,i,j,k}\right)\right)$$

In zones with a non-zero strong shock detector function we modify the modes in eqns. (4), (43), (44) and (45). Except for the piecewise linear variation in eqn. (4), we prefer to multiply all the higher moments (i.e. the ones with quadratic or cubic variation) by $(1-d_{i,j,k})$ so that those zone-centered moments are effectively zero in the vicinity of strong shocks. Likewise, when limiting the x-component of the magnetic field in eqn. (43) for the face $(i+1/2, j, k)$ we multiply the higher moments with $(1-d_{i,j,k})(1-d_{i+1,j,k})/(2-d_{i,j,k}-d_{i+1,j,k})$, i.e. the reciprocal sum derived from the two abutting zones. A similar approach can be taken for limiting the y and z-components of the magnetic fields in eqns. (44) and (45).

It was felt that even in the vicinity of strong shocks one should not obliterate structure altogether. For that reason, we felt that terms that have linear variations in eqns. (4), (43), (44) and (45) should be blended with some fraction of a slope limiter. The detector function should also be active in a zone that is about to be run over by a strong shock in the next step. For such reasons, our treatment of the first moments is modified a little. In a dimension-by-dimension fashion we make the modification:

$$if\left(\left(d_{i,j,k}>0\right)and\left(d_{i+1,j,k}=0\right)and\left(P_{i,j,k}>P_{i+1,j,k}\right)\right)then\ d_{i+1,j,k}=d_{i,j,k}$$
$$if\left(\left(d_{i,j,k}>0\right)and\left(d_{i-1,j,k}=0\right)and\left(P_{i,j,k}>P_{i-1,j,k}\right)\right)then\ d_{i-1,j,k}=d_{i,j,k}$$

Then say that $u_x$ is the slope from eqn. (4) evaluated from a WENO scheme and $\tilde{u}_x$ is the slope evaluated using a MinMod limiter. We then reset $u_x$ in the vicinity of a strong shock as follows

$$u_x \leftarrow \left(1-d_{i,j,k}\right)u_x + \chi\, d_{i,j,k}\, \tilde{u}_x$$

where $\chi \leq 1$. A similar flattening algorithm for treating the first moments can be instituted in the other two directions.

For second and third order schemes we use $\delta = 1.5$ and $\chi = 1$. For fourth order schemes we usually use $\delta = 0.75$ and $\chi = 0.5$.

The detector function described here can also play an important role when using a linearized Riemann solver. As shown by Quirk [46], linearized Riemann solvers are susceptible to a carbuncle instability when grid-aligned strong shocks are present. Einfeldt et al [31] also showed that linearized Riemann solvers do not function well in the presence of strong rarefactions. In both situations, a simple solution consists of blending in some fraction of an HLL flux and this is the approach we have used here. As a result, at strong shocks or rarefactions, the flux function consists of just the HLL flux while in weak shocks or rarefactions, the flux function is given entirely by the linearized Riemann solver. Using the detector function we provide a linear blend of the two in intermediate situations. Other approaches for curing the linearized Riemann solvers have been presented in Pandolfini & D'Ambrosio [44], Hanawa et al [35] and references therein but we have not explored them here. When building a detector function for linearized Riemann solvers one has the option of using not just the undivided divergence of the velocity but also the difference in wave speeds of any given wave family on either side of the Riemann problem.

**Appendix B) Implementation of the ADER-CG Scheme at Second Order of Accuracy**

For second order ADER-CG schemes we start with

$$
\begin{aligned}
u(\xi,\eta,\zeta,\tau) = & \hat{w}_1 P_0(\xi) P_0(\eta) P_0(\zeta) Q_0(\tau) \\
& + \hat{w}_2 P_1(\xi) P_0(\eta) P_0(\zeta) Q_0(\tau) + \hat{w}_3 P_0(\xi) P_1(\eta) P_0(\zeta) Q_0(\tau) + \hat{w}_4 P_0(\xi) P_0(\eta) P_1(\zeta) Q_0(\tau) \\
& + \hat{u}_5 P_0(\xi) P_0(\eta) P_0(\zeta) Q_1(\tau)
\end{aligned}
$$

The resultant ADER-CG iteration at second order is therefore given by

$$
\hat{u}_5 = -\hat{f}_2 - \hat{g}_3 - \hat{h}_4 + \hat{s}_1 + \frac{2}{3} \hat{s}_5
$$

The nodal to modal transcription can be carried out by picking a small number of symmetrically placed nodes in the reference element. We pick the nodal points

$$
\{(1/2,0,0,0); (-1/2,0,0,0); (0,1/2,0,0); (0,-1/2,0,0); \\
(0,0,1/2,0); (0,0,-1/2,0); (0,0,0,1)\}
$$

The nodal to modal transcription of the fluxes at $\tau=0$ is given by

$$\hat{f}_1 = \left( \overline{f}_1 + \overline{f}_2 + \overline{f}_3 + \overline{f}_4 + \overline{f}_5 + \overline{f}_6 \right) / 6$$
$$\hat{f}_2 = \overline{f}_1 - \overline{f}_2$$
$$\hat{f}_3 = \overline{f}_3 - \overline{f}_4$$
$$\hat{f}_4 = \overline{f}_5 - \overline{f}_6$$

The nodal to modal transcription of the fluxes at $\tau>0$ is given by

$$\hat{f}_5 = \overline{f}_7 - \hat{f}_1$$

**Appendix C) Implementation of the ADER-CG Scheme at Fourth Order of Accuracy**

For fourth order ADER-CG schemes we start with

$$\begin{aligned}
u(\xi,\eta,\zeta,\tau) = &\ \hat{w}_1 P_0(\xi) P_0(\eta) P_0(\zeta) Q_0(\tau) \\
&+ \hat{w}_2 P_1(\xi) P_0(\eta) P_0(\zeta) Q_0(\tau) + \hat{w}_3 P_0(\xi) P_1(\eta) P_0(\zeta) Q_0(\tau) + \hat{w}_4 P_0(\xi) P_0(\eta) P_1(\zeta) Q_0(\tau) \\
&+ \hat{w}_5 P_2(\xi) P_0(\eta) P_0(\zeta) Q_0(\tau) + \hat{w}_6 P_0(\xi) P_2(\eta) P_0(\zeta) Q_0(\tau) + \hat{w}_7 P_0(\xi) P_0(\eta) P_2(\zeta) Q_0(\tau) \\
&+ \hat{w}_8 P_1(\xi) P_1(\eta) P_0(\zeta) Q_0(\tau) + \hat{w}_9 P_0(\xi) P_1(\eta) P_1(\zeta) Q_0(\tau) + \hat{w}_{10} P_1(\xi) P_0(\eta) P_1(\zeta) Q_0(\tau) \\
&+ \hat{w}_{11} P_3(\xi) P_0(\eta) P_0(\zeta) Q_0(\tau) + \hat{w}_{12} P_0(\xi) P_3(\eta) P_0(\zeta) Q_0(\tau) + \hat{w}_{13} P_0(\xi) P_0(\eta) P_3(\zeta) Q_0(\tau) \\
&+ \hat{w}_{14} P_2(\xi) P_1(\eta) P_0(\zeta) Q_0(\tau) + \hat{w}_{15} P_2(\xi) P_0(\eta) P_1(\zeta) Q_0(\tau) \\
&+ \hat{w}_{16} P_1(\xi) P_2(\eta) P_0(\zeta) Q_0(\tau) + \hat{w}_{17} P_0(\xi) P_2(\eta) P_1(\zeta) Q_0(\tau) \\
&+ \hat{w}_{18} P_1(\xi) P_0(\eta) P_2(\zeta) Q_0(\tau) + \hat{w}_{19} P_0(\xi) P_1(\eta) P_2(\zeta) Q_0(\tau) \\
&+ \hat{w}_{20} P_1(\xi) P_1(\eta) P_1(\zeta) Q_0(\tau) \\
&+ \hat{u}_{21} P_0(\xi) P_0(\eta) P_0(\zeta) Q_1(\tau) + \hat{u}_{22} P_0(\xi) P_0(\eta) P_0(\zeta) Q_2(\tau) + \hat{u}_{23} P_0(\xi) P_0(\eta) P_0(\zeta) Q_3(\tau) \\
&+ \hat{u}_{24} P_1(\xi) P_0(\eta) P_0(\zeta) Q_1(\tau) + \hat{u}_{25} P_0(\xi) P_1(\eta) P_0(\zeta) Q_1(\tau) + \hat{u}_{26} P_0(\xi) P_0(\eta) P_1(\zeta) Q_1(\tau) \\
&+ \hat{u}_{27} P_1(\xi) P_0(\eta) P_0(\zeta) Q_2(\tau) + \hat{u}_{28} P_0(\xi) P_1(\eta) P_0(\zeta) Q_2(\tau) + \hat{u}_{29} P_0(\xi) P_0(\eta) P_1(\zeta) Q_2(\tau) \\
&+ \hat{u}_{30} P_2(\xi) P_0(\eta) P_0(\zeta) Q_1(\tau) + \hat{u}_{31} P_0(\xi) P_2(\eta) P_0(\zeta) Q_1(\tau) + \hat{u}_{32} P_0(\xi) P_0(\eta) P_2(\zeta) Q_1(\tau) \\
&+ \hat{u}_{33} P_1(\xi) P_1(\eta) P_0(\zeta) Q_1(\tau) + \hat{u}_{34} P_0(\xi) P_1(\eta) P_1(\zeta) Q_1(\tau) + \hat{u}_{35} P_1(\xi) P_0(\eta) P_1(\zeta) Q_1(\tau)
\end{aligned}$$

The resultant ADER-CG iteration at fourth order is therefore given by

$$\hat{u}_{21} = -\frac{\hat{f}_{11}}{10} - \hat{f}_2 - \frac{\hat{g}_{12}}{10} - \hat{g}_3 - \frac{\hat{h}_{13}}{10} - \hat{h}_4 + \hat{s}_1 + \frac{8}{70}\hat{s}_{23}$$

$$\hat{u}_{22} = -\frac{\hat{f}_{24}}{2} - \frac{\hat{g}_{25}}{2} - \frac{\hat{h}_{26}}{2} + \frac{\hat{s}_{21}}{2} - \frac{3}{7}\hat{s}_{23}$$

$$\hat{u}_{23} = -\frac{\hat{f}_{27}}{3} - \frac{\hat{g}_{28}}{3} - \frac{\hat{h}_{29}}{3} + \frac{\hat{s}_{22}}{3} + \frac{4}{7}\hat{s}_{23}$$

$$\hat{u}_{24} = -2\hat{f}_5 - \hat{g}_8 - \hat{h}_{10} + \hat{s}_2 - \frac{3}{10}\hat{s}_{27}$$

$$\hat{u}_{25} = -\hat{f}_8 - 2\hat{g}_6 - \hat{h}_9 + \hat{s}_3 - \frac{3}{10}\hat{s}_{28}$$

$$\hat{u}_{26} = -\hat{f}_{10} - \hat{g}_9 - 2\hat{h}_7 + \hat{s}_4 - \frac{3}{10}\hat{s}_{29}$$

$$\hat{u}_{27} = -\hat{f}_{30} - \frac{\hat{g}_{33}}{2} - \frac{\hat{h}_{35}}{2} + \frac{\hat{s}_{24}}{2} + \frac{3}{5}\hat{s}_{27}$$

$$\hat{u}_{28} = -\frac{\hat{f}_{33}}{2} - \hat{g}_{31} - \frac{\hat{h}_{34}}{2} + \frac{\hat{s}_{25}}{2} + \frac{3}{5}\hat{s}_{28}$$

$$\hat{u}_{29} = -\frac{\hat{f}_{35}}{2} - \frac{\hat{g}_{34}}{2} - \hat{h}_{32} + \frac{\hat{s}_{26}}{2} + \frac{3}{5}\hat{s}_{29}$$

$$\hat{u}_{30} = -3\hat{f}_{11} - \hat{g}_{14} - \hat{h}_{15} + \hat{s}_5 + \frac{2}{3}\hat{s}_{30}$$

$$\hat{u}_{31} = -\hat{f}_{16} - 3\hat{g}_{12} - \hat{h}_{17} + \hat{s}_6 + \frac{2}{3}\hat{s}_{31}$$

$$\hat{u}_{32} = -\hat{f}_{18} - \hat{g}_{19} - 3\hat{h}_{13} + \hat{s}_7 + \frac{2}{3}\hat{s}_{32}$$

$$\hat{u}_{33} = -2\hat{f}_{14} - 2\hat{g}_{16} - \hat{h}_{20} + \hat{s}_8 + \frac{2}{3}\hat{s}_{33}$$

$$\hat{u}_{34} = -\hat{f}_{20} - 2\hat{g}_{17} - 2\hat{h}_{19} + \hat{s}_9 + \frac{2}{3}\hat{s}_{34}$$

$$\hat{u}_{35} = -2\hat{f}_{15} - \hat{g}_{20} - 2\hat{h}_{18} + \hat{s}_{10} + \frac{2}{3}\hat{s}_{35}$$

The nodal to modal transcription can be carried out by picking a small number of symmetrically placed nodes in the reference element. We pick the nodal points

{ (0,0,0,0); (1/2,0,0,0); (1/4,0,0,0); (−1/4,0,0,0); (−1/2,0,0,0); (0,1/2,0,0);
(0,1/4,0,0); (0,−1/4,0,0); (0,−1/2,0,0); (0,0,1/2,0); (0,0,1/4,0); (0,0,−1/4,0);
(0,0,−1/2,0); (1/2,1/2,1/2,0); (0,1/2,1/2,0); (−1/2,1/2,1/2,0); (1/2,−1/2,1/2,0);
(0,−1/2,1/2,0); (−1/2,−1/2,1/2,0); (1/2,1/2,−1/2,0); (0,1/2,−1/2,0); (−1/2,1/2,−1/2,0);
(1/2,−1/2,−1/2,0); (0,−1/2,−1/2,0); (−1/2,−1/2,−1/2,0); (1/2,0,1/2,0); (−1/2,0,1/2,0);
(1/2,0,−1/2,0); (−1/2,0,−1/2,0); (1/2,1/2,0,0); (−1/2,1/2,0,0); (1/2,−1/2,0,0);
(−1/2,−1/2,0,0);
(0,0,0,1/3); (1/2,0,0,1/3); (−1/2,0,0,1/3); (0,1/2,0,1/3); (0,−1/2,0,1/3);
(0,0,1/2,1/3); (0,0,−1/2,1/3); (1/2,1/2,1/2,1/3); (−1/2,1/2,1/2,1/3);
(1/2,−1/2,1/2,1/3); (−1/2,−1/2,1/2,1/3); (1/2,1/2,−1/2,1/3); (−1/2,1/2,−1/2,1/3);
(1/2,−1/2,−1/2,1/3); (−1/2,−1/2,−1/2,1/3);
(1/2,0,0,2/3); (−1/2,0,0,2/3); (0,1/2,0,2/3); (0,−1/2,0,2/3); (0,0,1/2,2/3);
(0,0,−1/2,2/3);
(0,0,0,1);}

The nodal to modal transcription of the fluxes at τ=0 is given by

$$\hat{f}_5 = 2(\bar{f}_2 - 2\bar{f}_1 + \bar{f}_5)$$

$$\hat{f}_6 = 2(\bar{f}_6 - 2\bar{f}_1 + \bar{f}_9)$$

$$\hat{f}_7 = 2(\bar{f}_{10} - 2\bar{f}_1 + \bar{f}_{13})$$

$$\hat{f}_8 = (\bar{f}_{14} - \bar{f}_{16} - \bar{f}_{17} + \bar{f}_{19} + \bar{f}_{20} - \bar{f}_{22} - \bar{f}_{23} + \bar{f}_{25}) / 2$$

$$\hat{f}_9 = (\bar{f}_{14} - \bar{f}_{17} - \bar{f}_{20} + \bar{f}_{23} + \bar{f}_{16} - \bar{f}_{19} - \bar{f}_{22} + \bar{f}_{25}) / 2$$

$$\hat{f}_{10} = (\bar{f}_{14} - \bar{f}_{16} - \bar{f}_{20} + \bar{f}_{22} + \bar{f}_{17} - \bar{f}_{19} - \bar{f}_{23} + \bar{f}_{25}) / 2$$

$$\hat{f}_{11} = (16\bar{f}_2 - 16\bar{f}_5 - 32\bar{f}_3 + 32\bar{f}_4) / 3$$

$$\hat{f}_{12} = (16\bar{f}_6 - 16\bar{f}_9 - 32\bar{f}_7 + 32\bar{f}_8) / 3$$

$$\hat{f}_{13} = (16\bar{f}_{10} - 16\bar{f}_{13} - 32\bar{f}_{11} + 32\bar{f}_{12}) / 3$$

$$\hat{f}_{14} = \bar{f}_{14} - 2\bar{f}_{15} + \bar{f}_{16} - \bar{f}_{17} + 2\bar{f}_{18} - \bar{f}_{19}$$
$$+ \bar{f}_{20} - 2\bar{f}_{21} + \bar{f}_{22} - \bar{f}_{23} + 2\bar{f}_{24} - \bar{f}_{25}$$

$$\hat{f}_{15} = \bar{f}_{14} - 2\bar{f}_{15} + \bar{f}_{16} - \bar{f}_{20} + 2\bar{f}_{21} - \bar{f}_{22}$$
$$+ \bar{f}_{17} - 2\bar{f}_{18} + \bar{f}_{19} - \bar{f}_{23} + 2\bar{f}_{24} - \bar{f}_{25}$$

$$\hat{f}_{16} = \bar{f}_{14} - 2\bar{f}_{26} + \bar{f}_{17} - \bar{f}_{16} + 2\bar{f}_{27} - \bar{f}_{19}$$
$$+ \bar{f}_{20} - 2\bar{f}_{28} + \bar{f}_{23} - \bar{f}_{22} + 2\bar{f}_{29} - \bar{f}_{25}$$

$$\hat{f}_{17} = \bar{f}_{14} - 2\bar{f}_{26} + \bar{f}_{17} - \bar{f}_{20} + 2\bar{f}_{28} - \bar{f}_{23}$$
$$+ \bar{f}_{16} - 2\bar{f}_{27} + \bar{f}_{19} - \bar{f}_{22} + 2\bar{f}_{29} - \bar{f}_{25}$$

$$\hat{f}_{18} = \bar{f}_{14} - 2\bar{f}_{30} + \bar{f}_{20} - \bar{f}_{16} + 2\bar{f}_{31} - \bar{f}_{22}$$
$$+ \bar{f}_{17} - 2\bar{f}_{32} + \bar{f}_{23} - \bar{f}_{19} + 2\bar{f}_{33} - \bar{f}_{25}$$

$$\hat{f}_{19} = \bar{f}_{14} - 2\bar{f}_{30} + \bar{f}_{20} - \bar{f}_{17} + 2\bar{f}_{32} - \bar{f}_{23}$$
$$+ \bar{f}_{16} - 2\bar{f}_{31} + \bar{f}_{22} - \bar{f}_{19} + 2\bar{f}_{33} - \bar{f}_{25}$$

$$\hat{f}_{20} = \bar{f}_{14} - \bar{f}_{16} - \bar{f}_{17} + \bar{f}_{19} - \bar{f}_{20} + \bar{f}_{22} + \bar{f}_{23} - \bar{f}_{25}$$

$$\hat{f}_1 = \bar{f}_1 + (\hat{f}_5 + \hat{f}_6 + \hat{f}_7) / 12$$

$$\hat{f}_2 = \bar{f}_2 - \bar{f}_5 - \frac{\hat{f}_{11}}{10} + (\hat{f}_{16} + \hat{f}_{18}) / 12$$

$$\hat{f}_3 = \bar{f}_6 - \bar{f}_9 - \frac{\hat{f}_{12}}{10} + (\hat{f}_{14} + \hat{f}_{19}) / 12$$

$$\hat{f}_4 = \bar{f}_{10} - \bar{f}_{13} - \frac{\hat{f}_{13}}{10} + (\hat{f}_{15} + \hat{f}_{17}) / 12$$

The nodal to modal transcription of the fluxes at $\tau>0$ is given by

$$\hat{f}_{24} = (-9\overline{f}_2 + 9\overline{f}_5 + 12\overline{f}_{35} - 12\overline{f}_{36} - 3\overline{f}_{49} + 3\overline{f}_{50}) / 2$$

$$\hat{f}_{25} = (-9\overline{f}_6 + 9\overline{f}_9 + 12\overline{f}_{37} - 12\overline{f}_{38} - 3\overline{f}_{51} + 3\overline{f}_{52}) / 2$$

$$\hat{f}_{26} = (-9\overline{f}_{10} + 9\overline{f}_{13} + 12\overline{f}_{39} - 12\overline{f}_{40} - 3\overline{f}_{53} + 3\overline{f}_{54}) / 2$$

$$\hat{f}_{27} = 9(\overline{f}_{49} - \overline{f}_{50} - 2\overline{f}_{35} + 2\overline{f}_{36} + \overline{f}_2 - \overline{f}_5) / 2$$

$$\hat{f}_{28} = 9(\overline{f}_{51} - \overline{f}_{52} - 2\overline{f}_{37} + 2\overline{f}_{38} + \overline{f}_6 - \overline{f}_9) / 2$$

$$\hat{f}_{29} = 9(\overline{f}_{53} - \overline{f}_{54} - 2\overline{f}_{39} + 2\overline{f}_{40} + \overline{f}_{10} - \overline{f}_{13}) / 2$$

$$\hat{f}_{30} = 6(\overline{f}_{35} - 2\overline{f}_{34} + \overline{f}_{36}) - 3\hat{f}_5$$

$$\hat{f}_{31} = 6(\overline{f}_{37} - 2\overline{f}_{34} + \overline{f}_{38}) - 3\hat{f}_6$$

$$\hat{f}_{32} = 6(\overline{f}_{39} - 2\overline{f}_{34} + \overline{f}_{40}) - 3\hat{f}_7$$

$$\hat{f}_{33} = 3(\overline{f}_{41} - \overline{f}_{42} - \overline{f}_{43} + \overline{f}_{44} + \overline{f}_{45} - \overline{f}_{46} - \overline{f}_{47} + \overline{f}_{48}) / 2 - 3\hat{f}_8$$

$$\hat{f}_{34} = 3(\overline{f}_{41} - \overline{f}_{43} - \overline{f}_{45} + \overline{f}_{47} + \overline{f}_{42} - \overline{f}_{44} - \overline{f}_{46} + \overline{f}_{48}) / 2 - 3\hat{f}_9$$

$$\hat{f}_{35} = 3(\overline{f}_{41} - \overline{f}_{42} - \overline{f}_{45} + \overline{f}_{46} + \overline{f}_{43} - \overline{f}_{44} - \overline{f}_{47} + \overline{f}_{48}) / 2 - 3\hat{f}_{10}$$

$$\hat{t}_1 = \overline{f}_{34} - \overline{f}_1 + (\hat{f}_{30} + \hat{f}_{31} + \hat{f}_{32}) / 36$$

$$\hat{t}_2 = (\overline{f}_{49} + \overline{f}_{50} + \overline{f}_{51} + \overline{f}_{52} + \overline{f}_{53} + \overline{f}_{54}) / 6 - \overline{f}_1$$
$$- (\hat{f}_5 + \hat{f}_6 + \hat{f}_7) / 12$$

$$\hat{t}_3 = \overline{f}_{55} - \overline{f}_1 + (\hat{f}_{30} + \hat{f}_{31} + \hat{f}_{32}) / 12$$

$$\hat{f}_{21} = (18\hat{t}_1 - 9\hat{t}_2 + 2\hat{t}_3) / 2$$

$$\hat{f}_{22} = (-45\hat{t}_1 + 36\hat{t}_2 - 9\hat{t}_3) / 2$$

$$\hat{f}_{23} = (27\hat{t}_1 - 27\hat{t}_2 + 9\hat{t}_3) / 2$$

where $\hat{t}_1$, $\hat{t}_2$ and $\hat{t}_3$ are temporary variables.

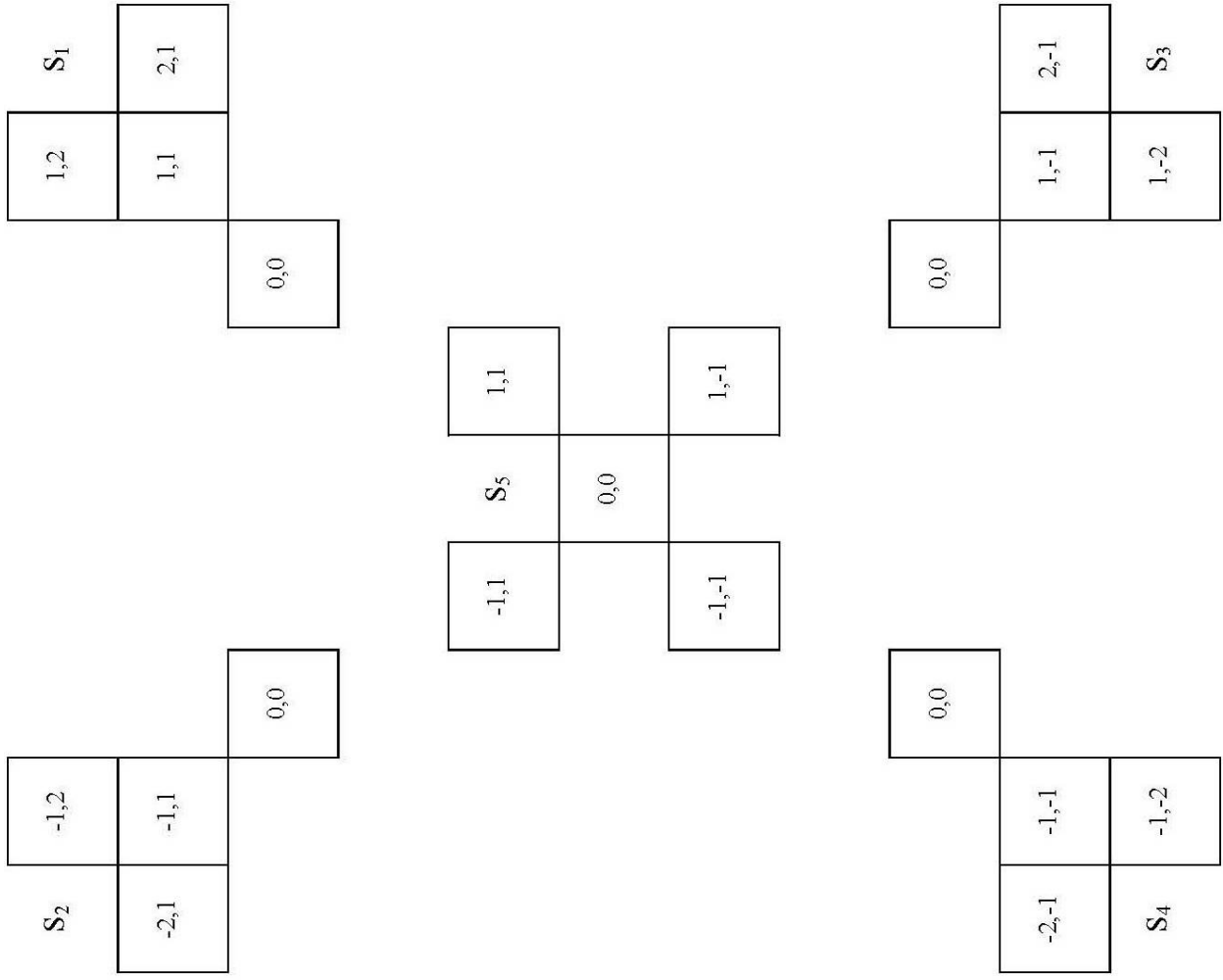

Fig. 1) Shows the five stencils that are used for evaluating the cross-terms in the fourth order WENO reconstruction.

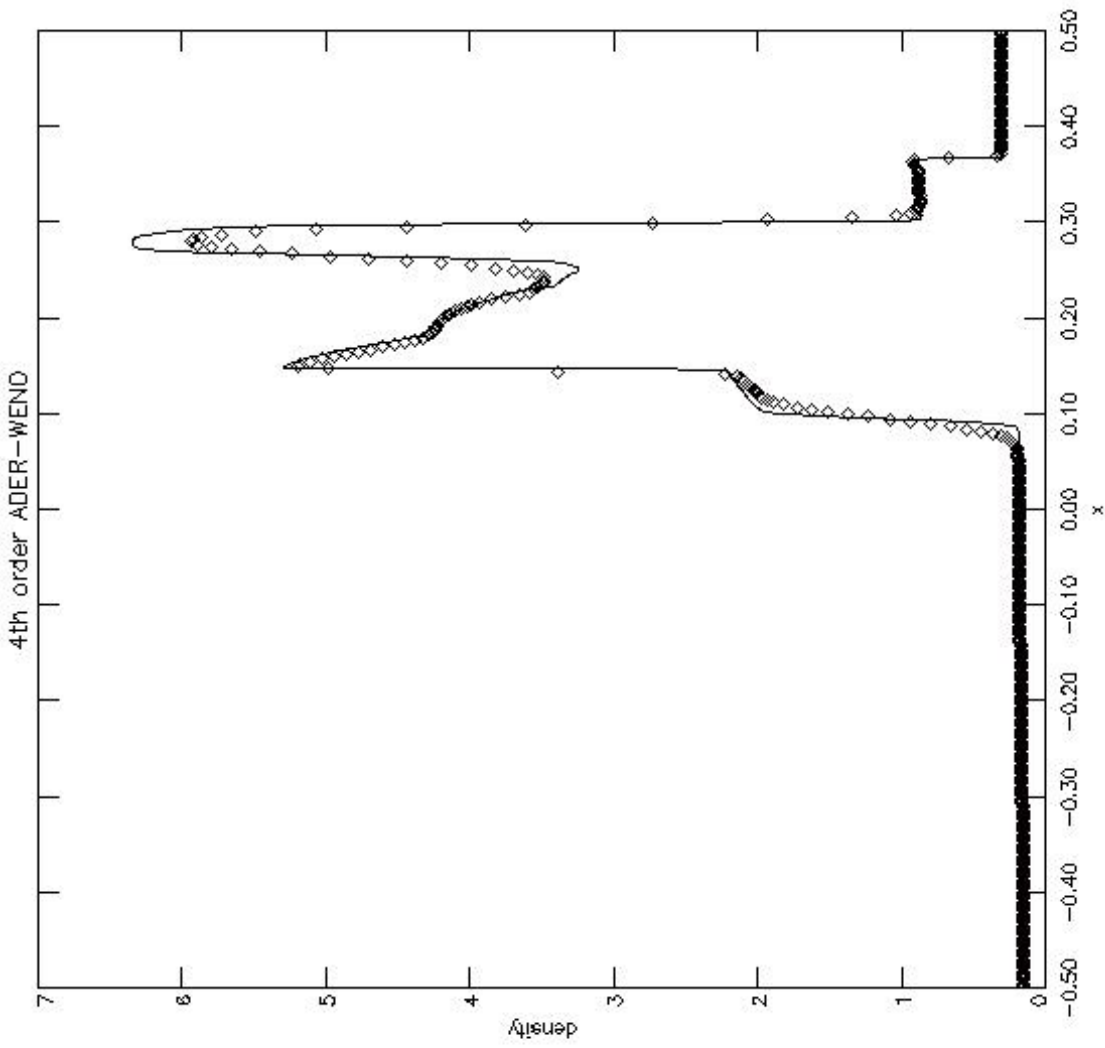

Fig 2) The density profile for the interacting blast problem. The diamonds show the results of a 400 zone simulation. The solid line is a converged density profile obtained from a 1600 zone simulation. The fourth order ADER-WENO scheme with a linearized Riemann solver was used for both simulations.

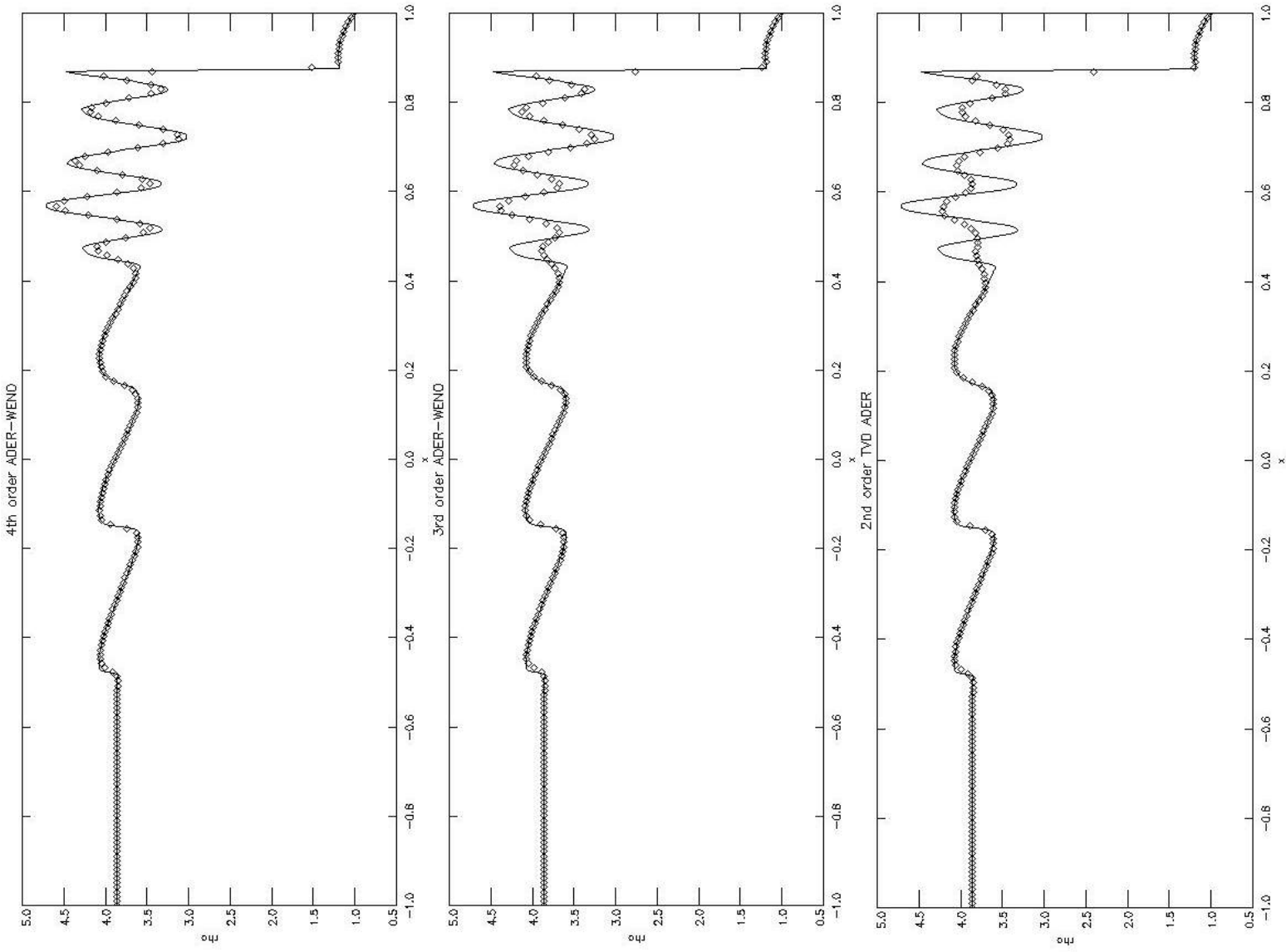

Fig. 3) Density from the shock entropy problem. The first and second panels show the 4th order and 3rd order ADER-WENO schemes. The third panel shows the TVD scheme with an MC limiter. The converged solution was obtained from an 800 zone simulation and is shown with a solid line. Notice the better convergence of the higher order schemes.

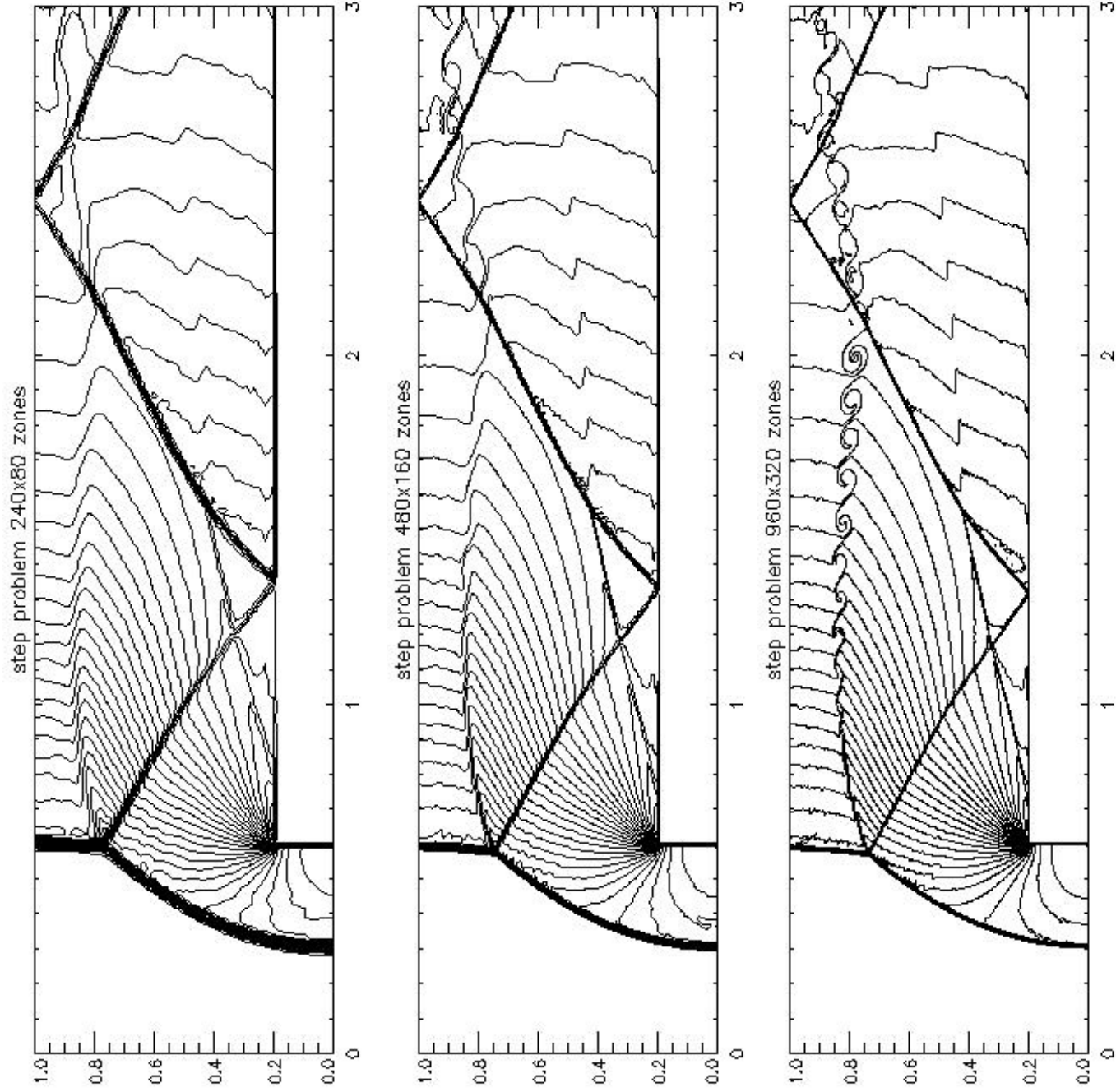

Fig. 4) This resolution study shows the density variable from the forward facing step problem at resolutions of 240X80, 480X160 and 960X320 zones at a time of 4 units. Thirty equally spaced contours are shown in the density variable ranging from 0.090338 to 6.2365. The fourth order scheme with a linearized Riemann solver was used. We see the beginnings of the vortex sheet roll-up at a resolution of 480X160 zones and the 960X320 zone simulation captures the roll-up very clearly.

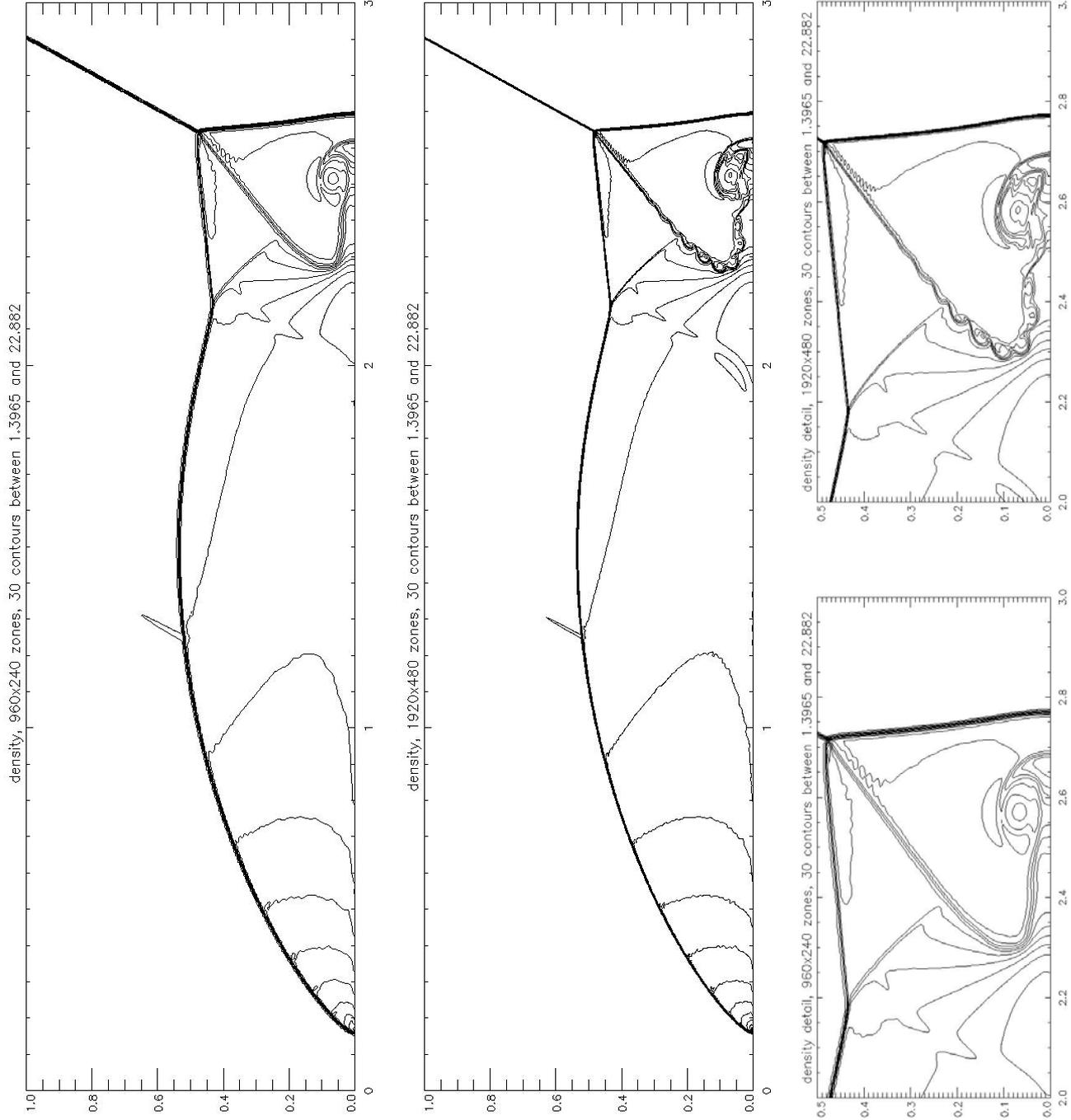

Fig. 5) Shows a resolution study of the double Mach reflection of a strong shock. The 1st and 2nd panels show the density from 960X240 and 1920X480 zone simulations. The lowest two panels show details at the Mach stem for each of those two simulations with the lowest left panel corresponding to the lower resolution simulation. The 4rd order ADER-WENO scheme with an HLL Riemann solver was used. 30 contours were fit between a range of 1.3965 and 22.882. We clearly see the roll up of the Mach stem due to Kelvin-Helmholtz instability in the higher resolution simulation.

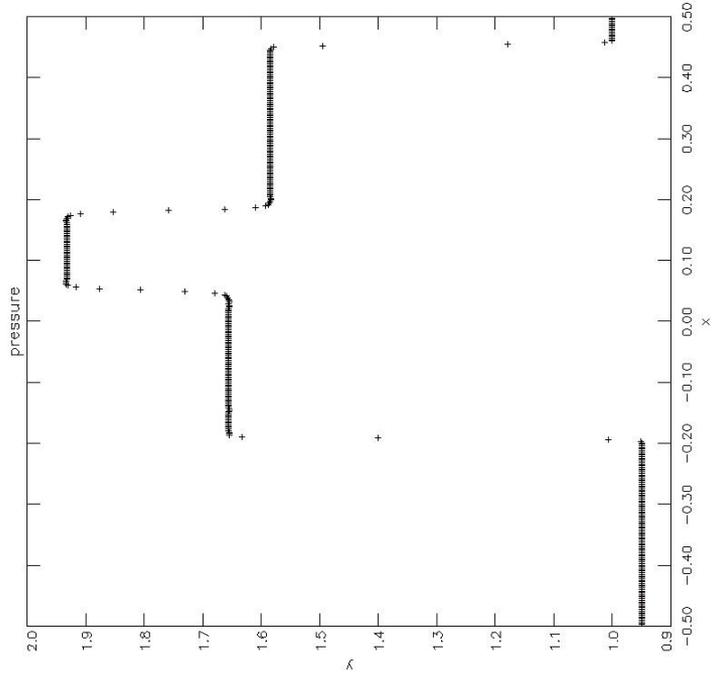
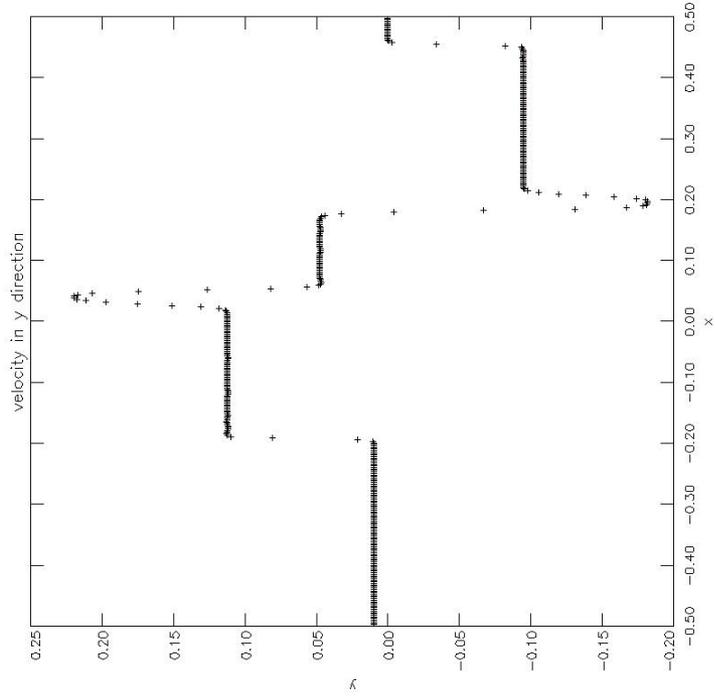
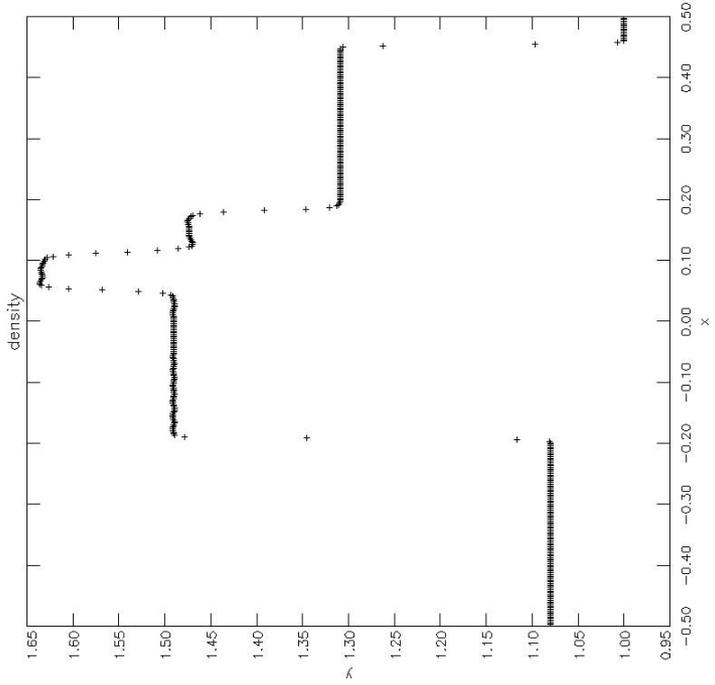
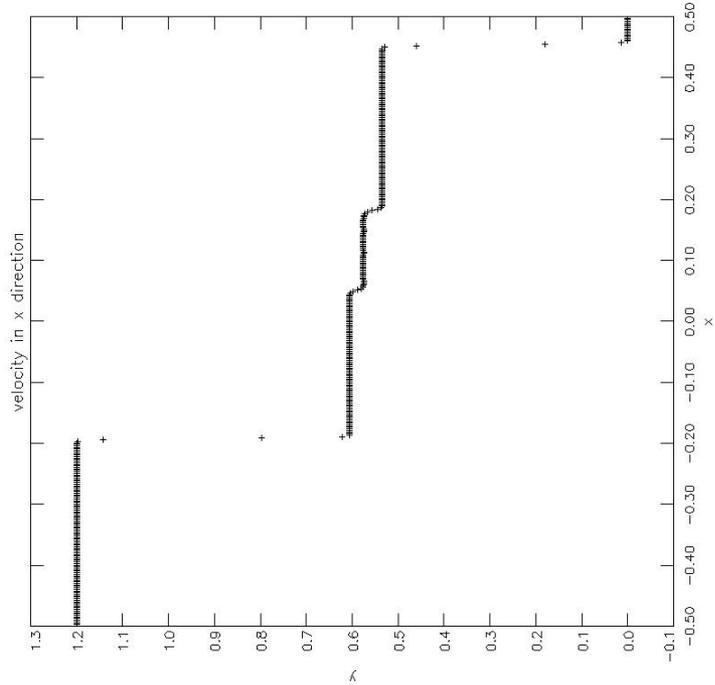

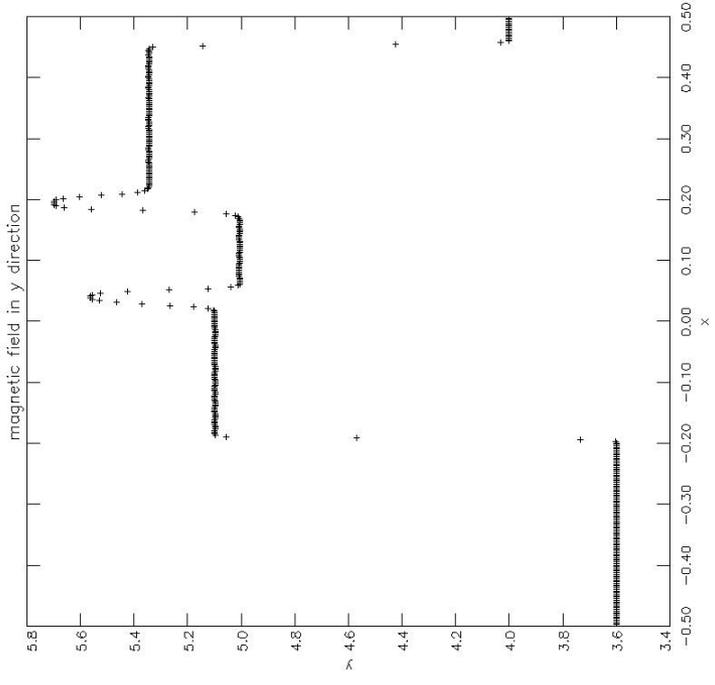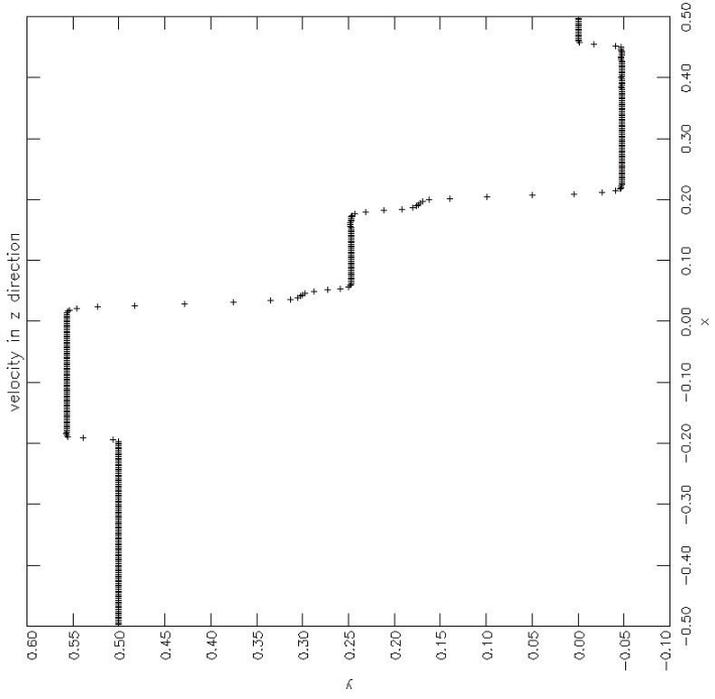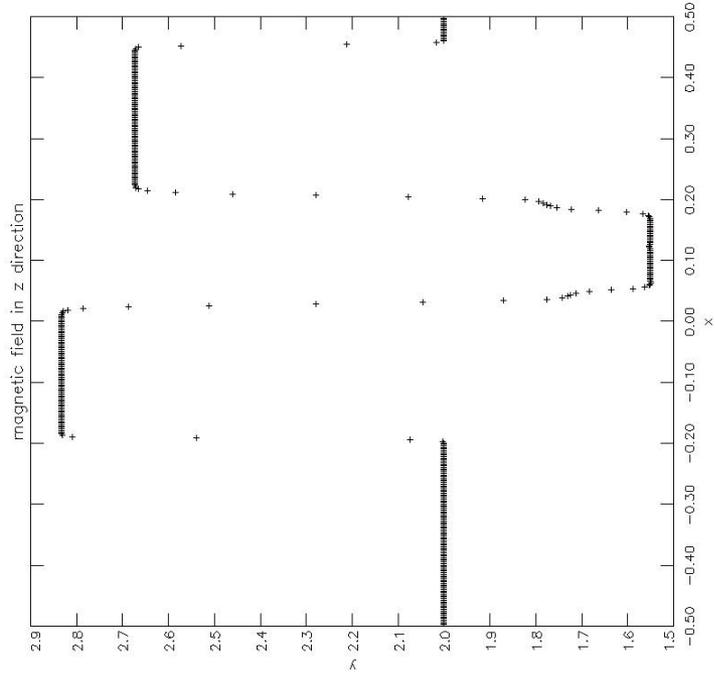

Fig 6) One-Dimensional Riemann problem showing all seven waves. The 4th order ADER-WENO scheme was used with an HLLE Riemann solver. Notice that the Alfven waves as well as the slow shocks are captured with very few zones.

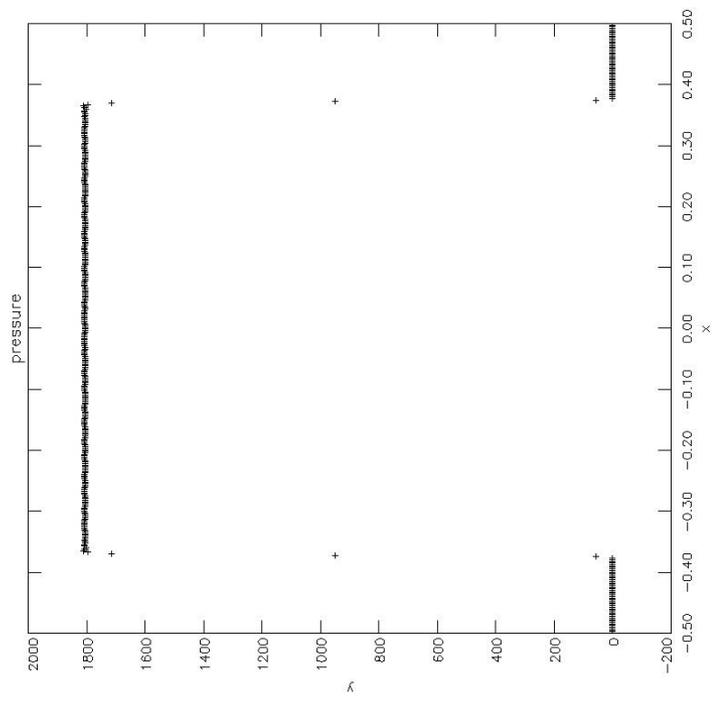
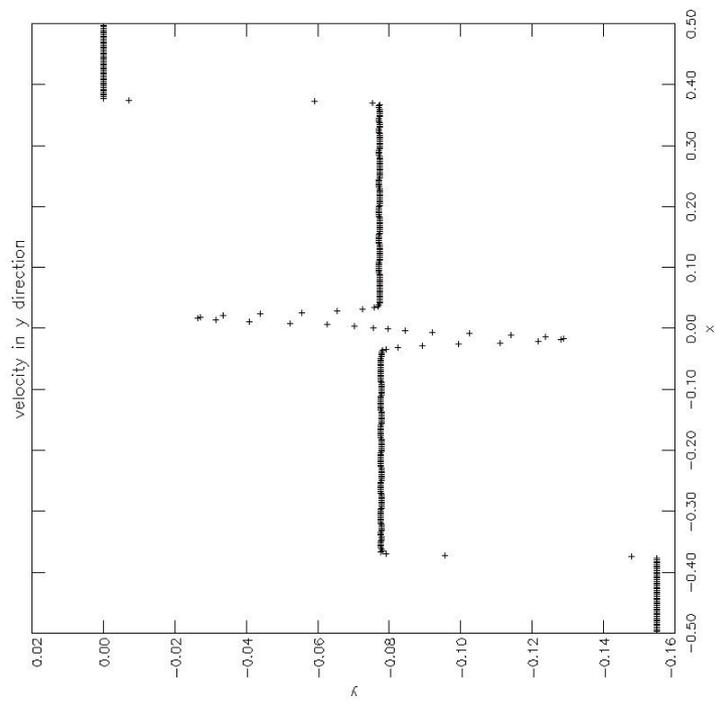
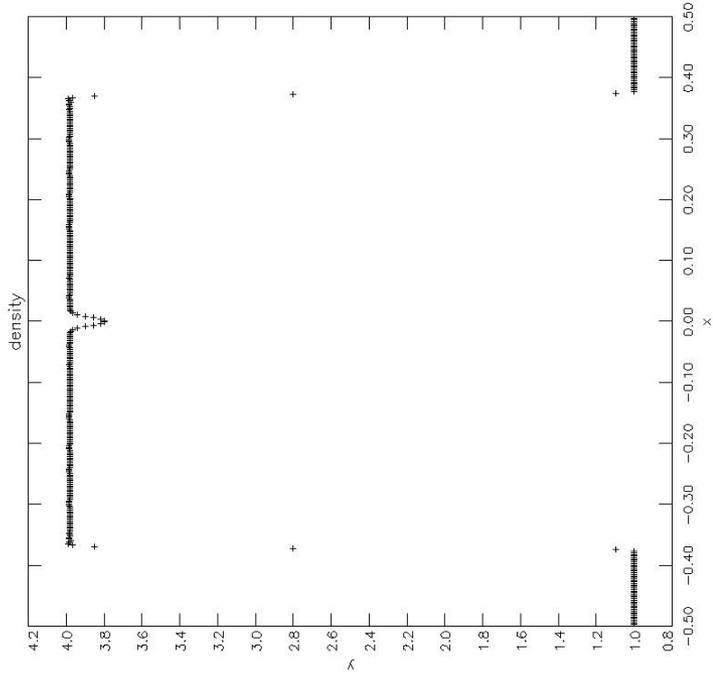
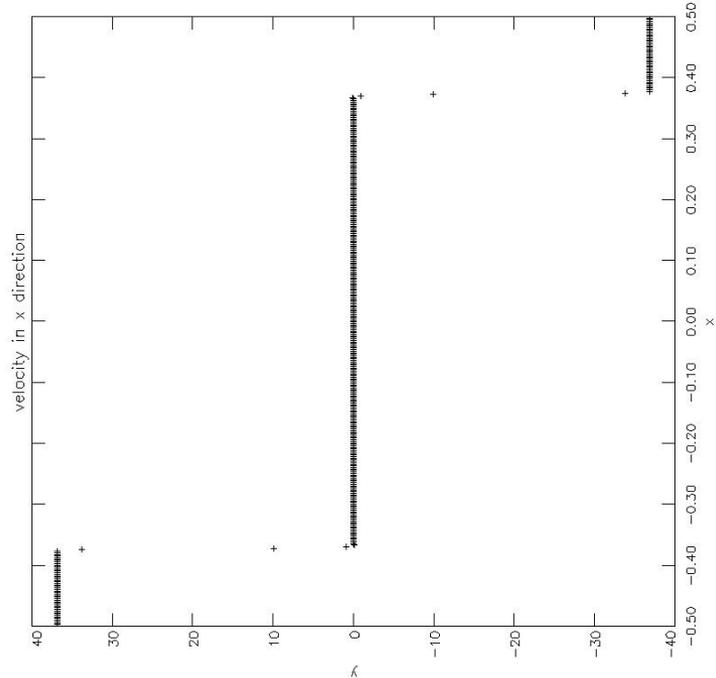

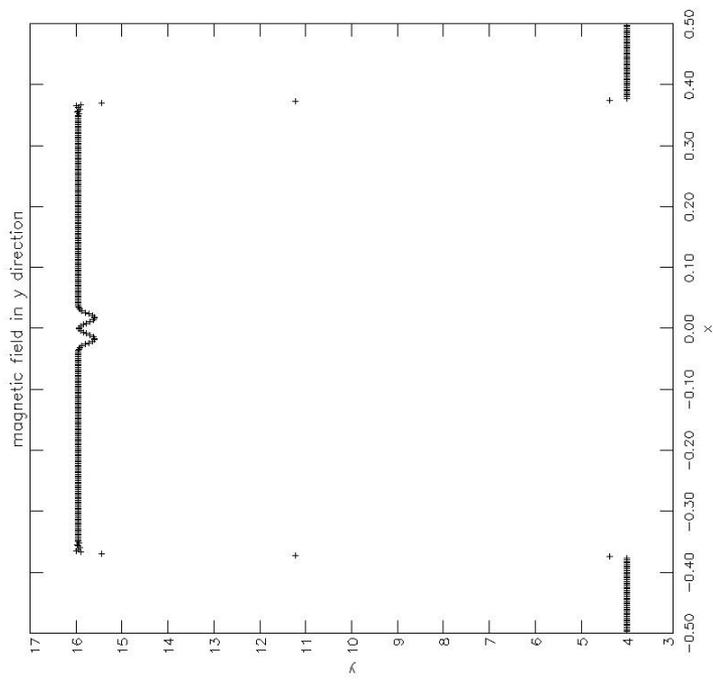
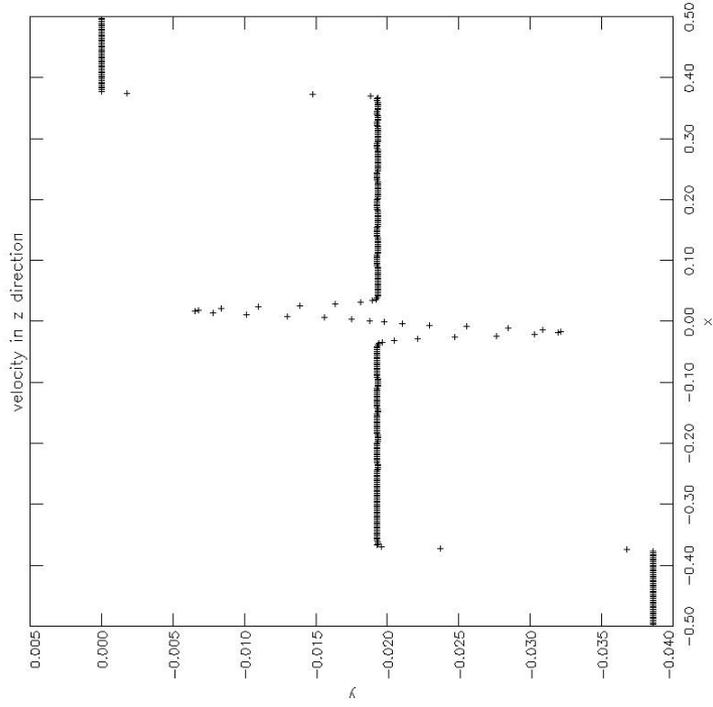
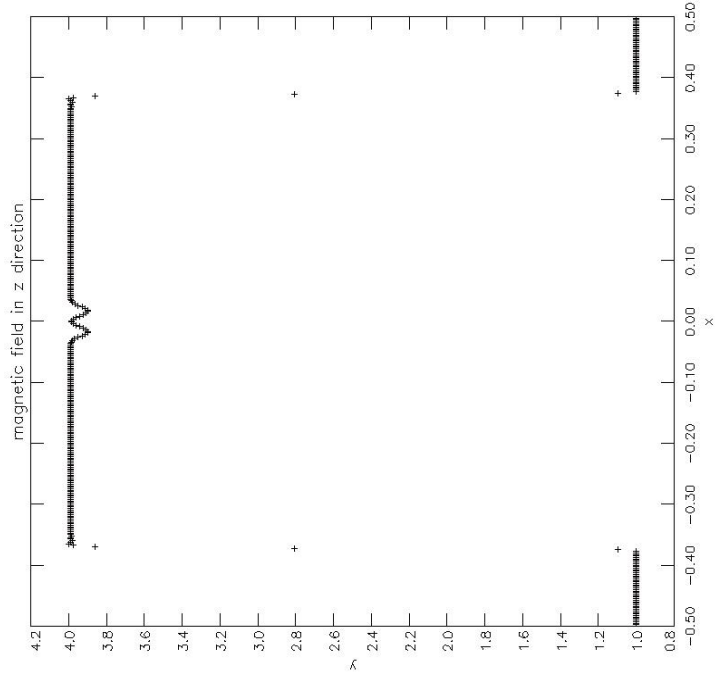

Fig 7) One-Dimensional Riemann problem showing the interaction of two very high Mach number streams of magnetized fluid. The 4th order ADER-WENO scheme was used with an HLLE Riemann solver. Notice that despite the near-infinite shocks that are generated in this problem, the flow variables are virtually free of any post-shock oscillations.

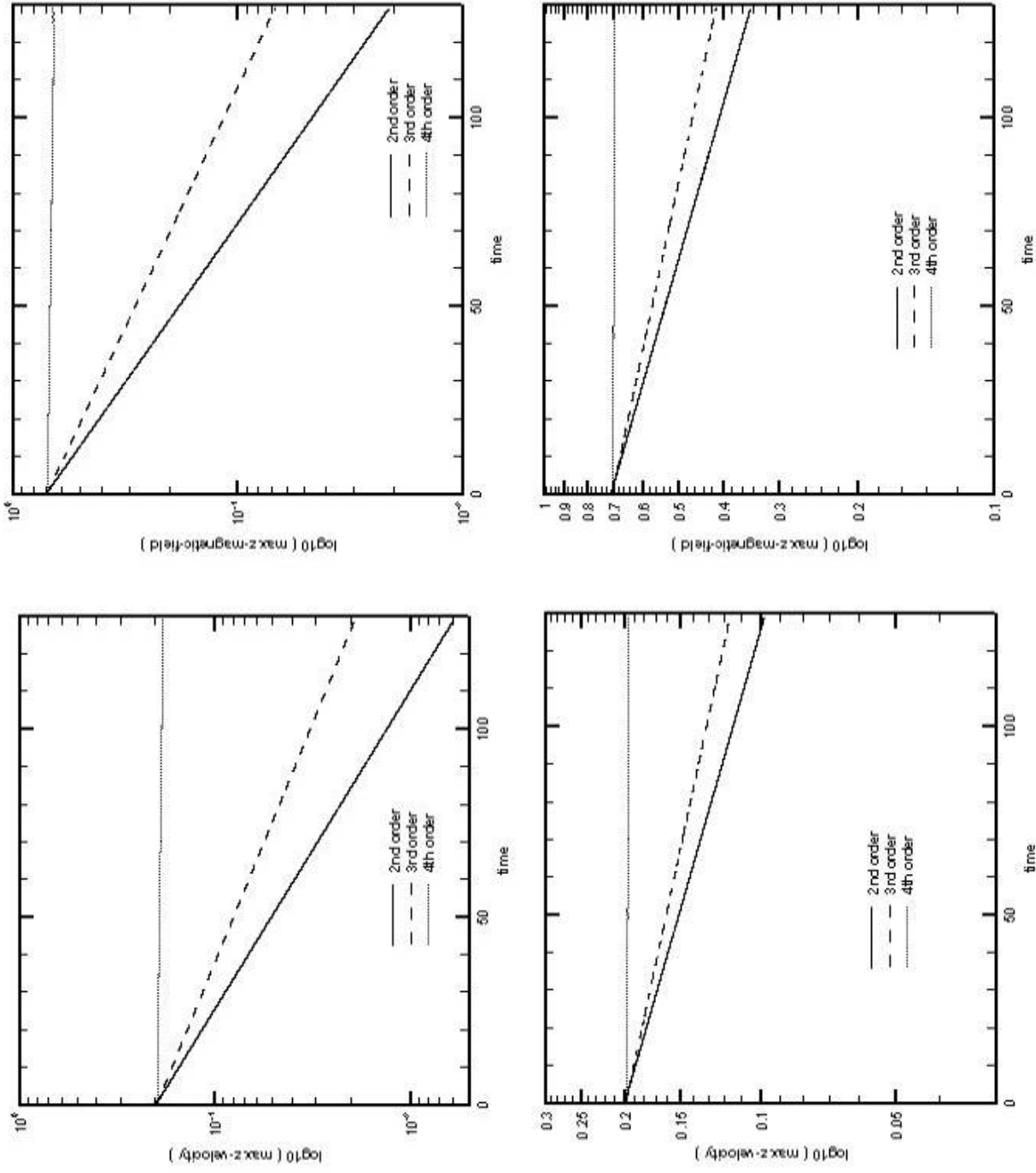

Fig 8) The log-linear plots show the decay of torsional Alfven waves that are made to propagate obliquely on a two dimensional square. The above two panels show the decay of the maximum z-velocity and the maximum z-component of the magnetic field when second, third and fourth order schemes are used with an HLLE Riemann solver. The lower two panels show the same information when a linearized Riemann solver is used. Notice that the decay is substantially reduced with increasing order. Notice too that the linearized Riemann solver provides a substantial improvement to the solution, especially at lower orders.

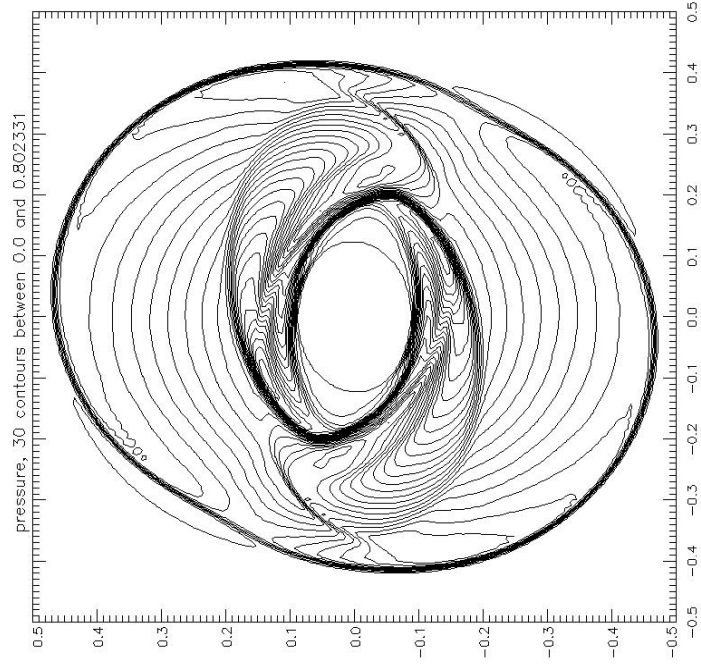
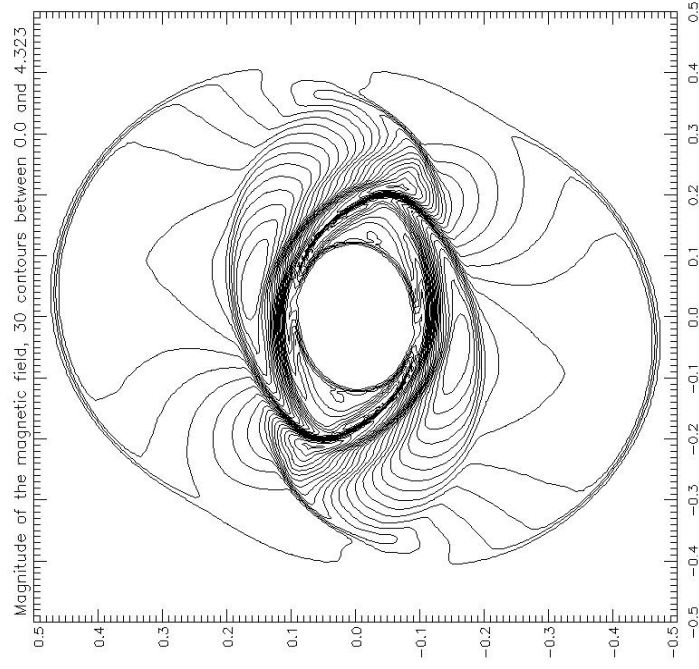
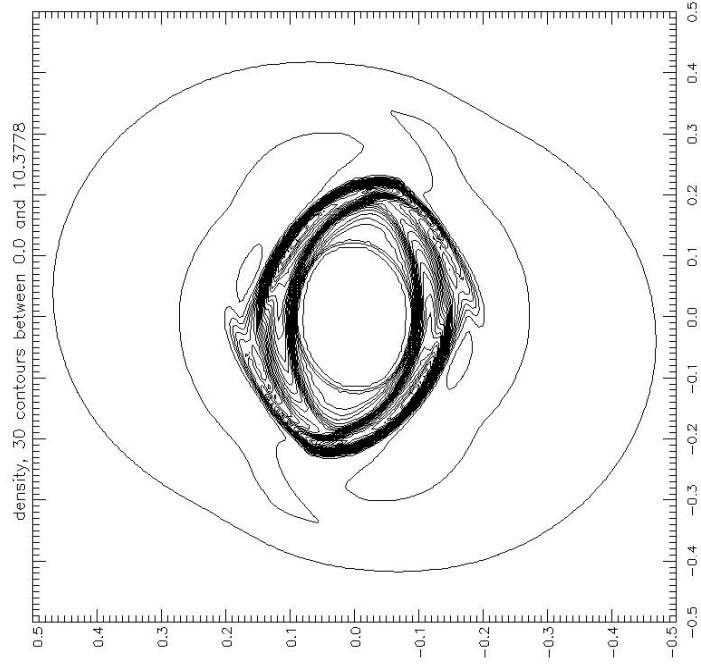
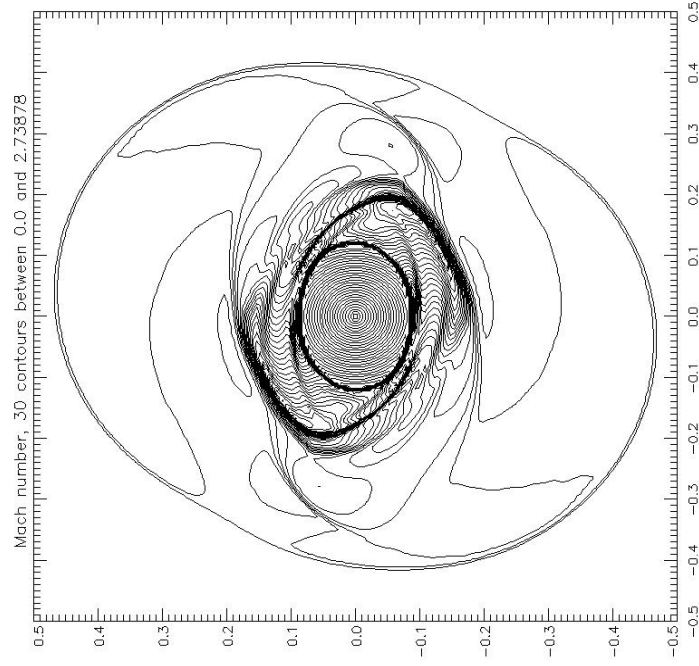

Fig 9) Rotor problem where the top two panels show the density and pressure and the bottom two panels show the Mach number and the magnitude of the magnetic field. The fourth order ADER-WENO scheme with a linearized Riemann solver was used. 30 contours are shown for each figure with the min and max values catalogued above the panels.

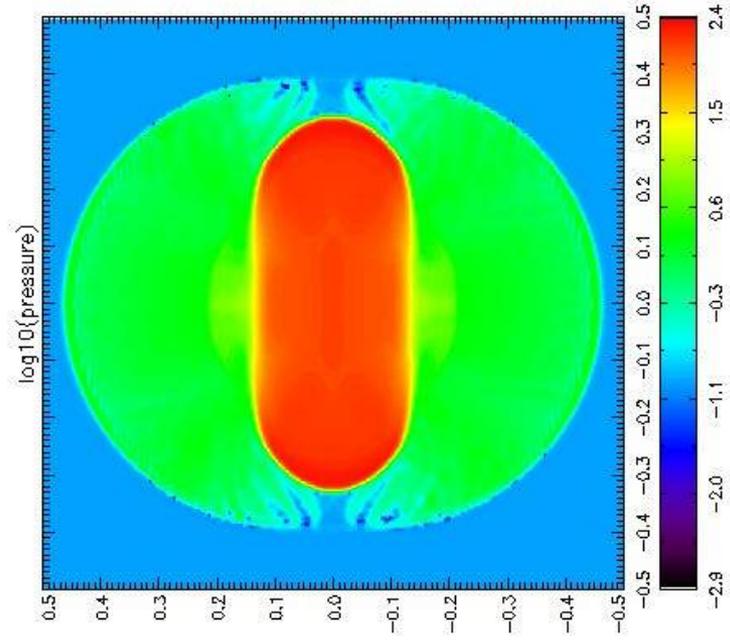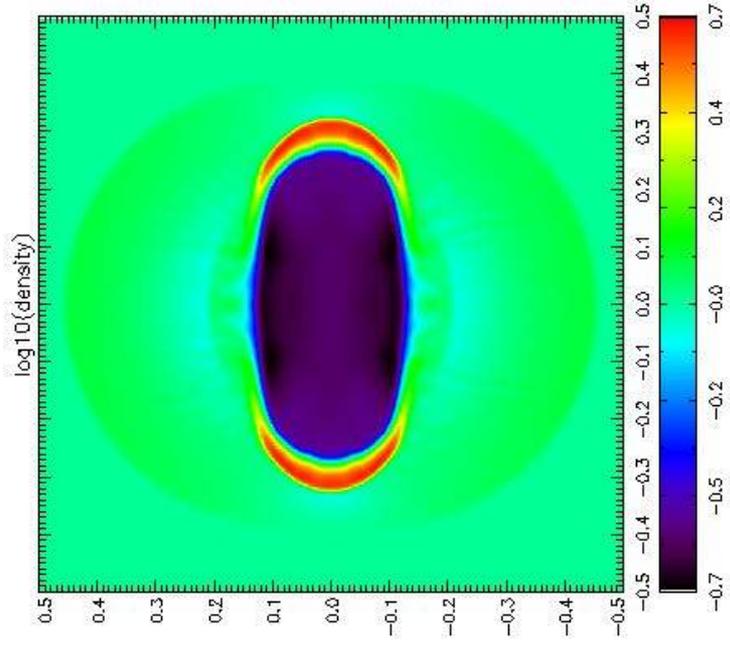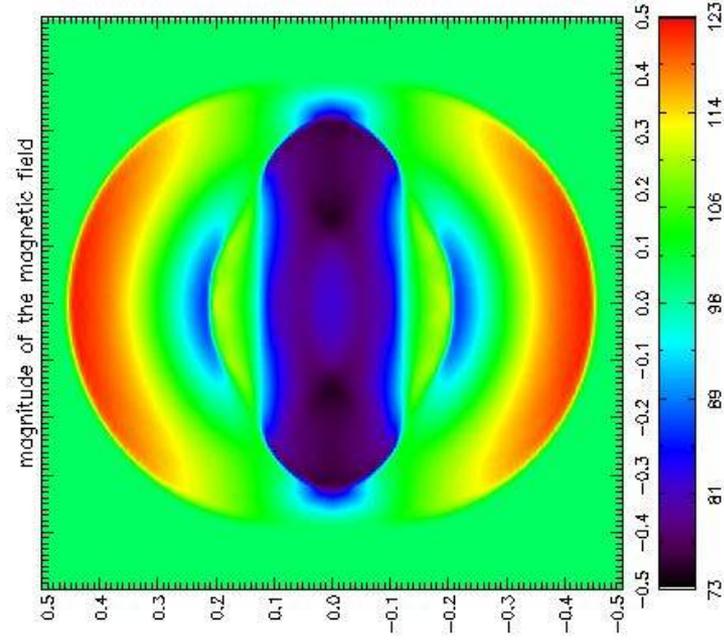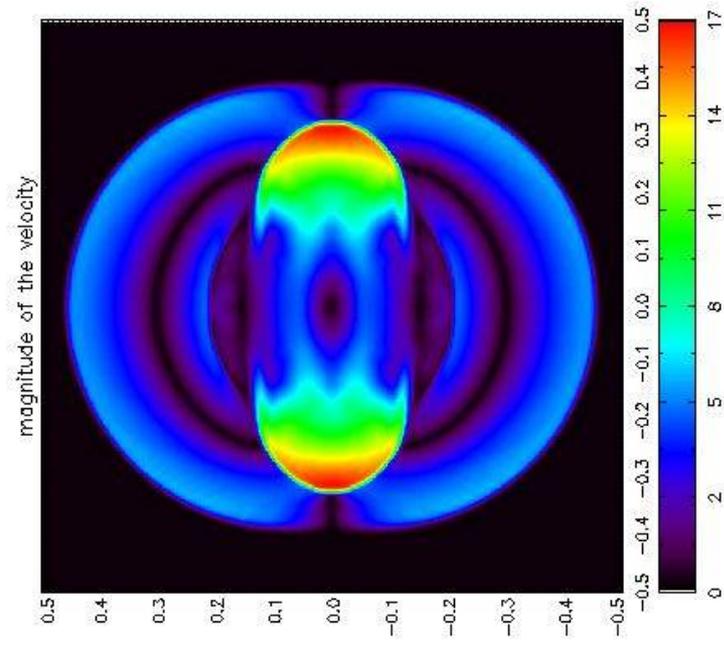

Fig 10) Blast problem in 2d where the top two panels show the $\log_{10}$ of the density and pressure and the bottom two panels show the magnitudes of the velocity and the magnetic field. The fourth order ADER-WENO scheme with an HLLE Riemann solver was used.

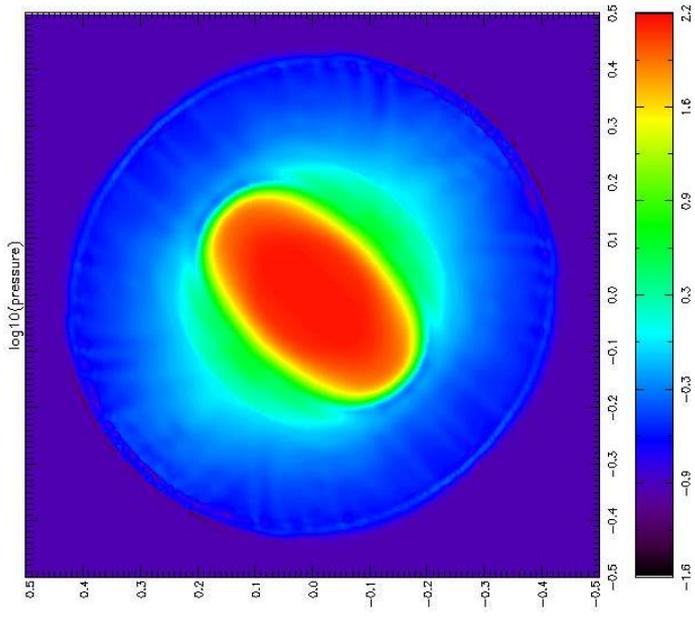 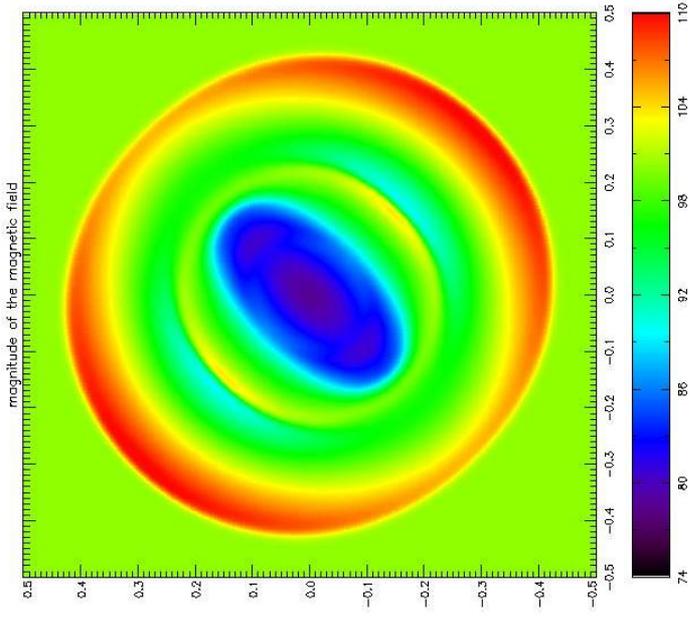
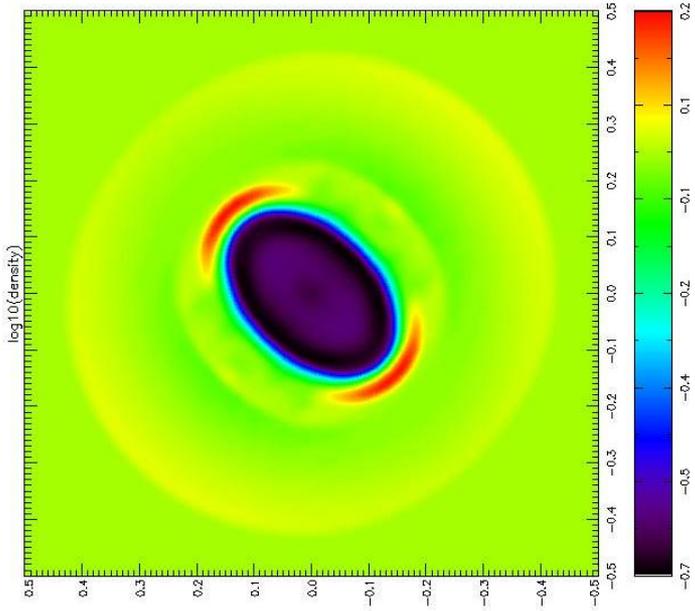 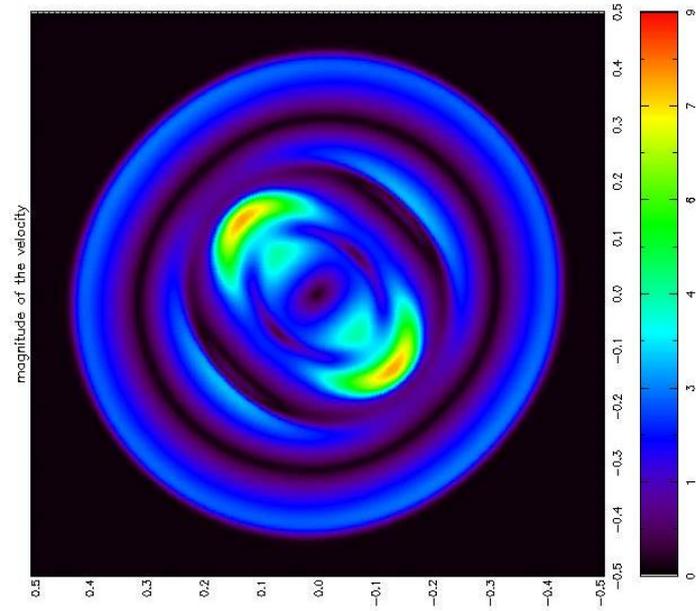

Fig. 11) Shows the $\log_{10}$ (density), the $\log_{10}$ (pressure), the magnitude of the velocity and the magnitude of the magnetic field for the three dimensional blast problem. The flow variables are shown in the mid-plane of the computational domain. Even though a very strong shock propagates through an ambient medium with very low plasma-$\beta$, all densities and pressures remain positive.